%% file: paper.tex
\documentclass[a4paper,USenglish,cleveref]{lipics-v2021}

\hideLIPIcs
\nolinenumbers

\newcommand{\psfrage}[1]{{\color{blue}{\sf[PS: #1]}}}
\newcommand{\hpfrage}[1]{{\color{violet}\sf[HP: #1]}}
\newcommand{\swfrage}[1]{{\color{teal}\sf[SW: #1]}}
 \renewcommand{\psfrage}[1]{} \renewcommand{\hpfrage}[1]{} \renewcommand{\swfrage}[1]{}

\newcommand{\myparagraph}[1]{\subparagraph*{#1}}
\newcommand{\Bin}{\mathrm{Bin}}
\newcommand{\Oh}{\mathcal{O}}

\input{pgf_header}

\input{unicode}

\usepackage{dsfont}
\usepackage{pgfplots}
\usepackage{caption}
\usepackage{subcaption}
\usepackage{hyperref}
\usepackage{booktabs}
\usepackage[nocompress]{cite}
\usepackage{listings}
\usepackage{mathtools}
\crefname{listing}{Algorithm}{Algorithms}
\def\refRelObs#1#2{\stackrel{\mathclap{\text{Obs. \ref{#1}}}}{#2}\quad}

\def\refRelLem#1#2{\stackrel{\mathclap{\text{Lem. \ref{#1}}}}{#2}\quad}
\def\refRelLemT#1#2#3{\stackrel{\mathclap{\text{Lem. \ref{#1},\ref{#2}}}}{#3}\quad\:\:}
\def\refRelEq#1#2{\stackrel{\mathclap{\text{(\ref{#1})}}}{#2}\quad}
\def\cond{\hspace{0.8mm}\middle\vert\hspace{0.8mm}}
\def\ShockHashRS{ShockHash\nobreakdash-RS}

\newcommand{\mytitle}{ShockHash: Near Optimal-Space Minimal Perfect Hashing Beyond Brute-Force}
\title{\mytitle}
\titlerunning{ShockHash: Near Optimal-Space MPH Beyond Brute-Force}

\author{Hans-Peter Lehmann}{Karlsruhe Institute of Technology, Germany}{hans-peter.lehmann@kit.edu}{https://orcid.org/0000-0002-0474-1805}{}
\author{Peter Sanders}{Karlsruhe Institute of Technology, Germany}{sanders@kit.edu}{https://orcid.org/0000-0003-3330-9349}{}
\author{Stefan Walzer}{Karlsruhe Institute of Technology, Germany}{stefan.walzer@kit.edu}{https://orcid.org/0000-0002-6477-0106}{}

\newcommand{\myauthorrunning}{H.-P. Lehmann, P. Sanders, S. Walzer}
\authorrunning{\myauthorrunning}
\Copyright{Hans-Peter Lehmann, Peter Sanders, and Stefan Walzer}
\hypersetup{
  colorlinks=true,
  pdftitle={\mytitle},
  pdfauthor={\myauthorrunning},
  pdfsubject={}
}


\supplementdetails[subcategory={Source code}]{Software}{https://github.com/ByteHamster/ShockHash}
\supplementdetails[subcategory={Comparison with competitors}]{Software}{https://github.com/ByteHamster/MPHF-Experiments}

\relatedversion{
  This preprint for the ShockHash journal version is based on a conference paper \cite{lehmann2024shockhash} published at ALENEX 2024.
  The basic idea of bipartite ShockHash was described in an earlier version \cite{lehmann2023bipartiteV1} of this arXiv paper without an analysis.
}

\funding{
\flag[2cm]{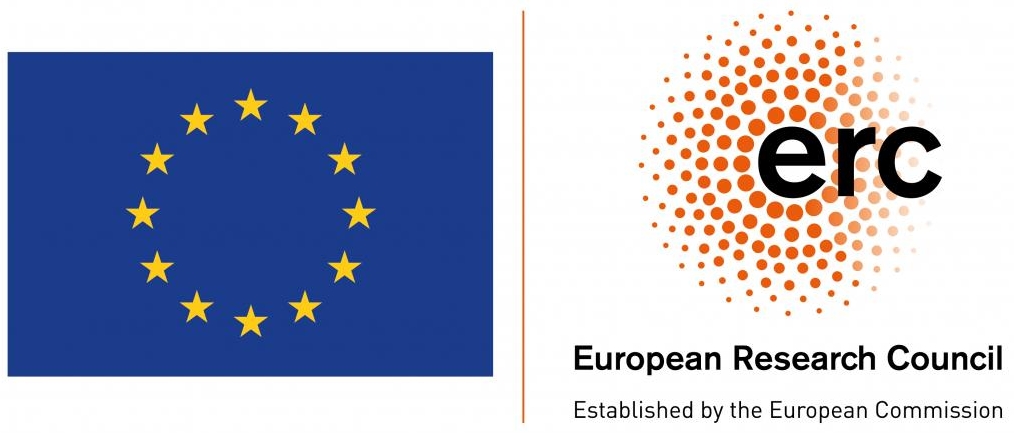}
This project has received funding from the European Research Council (ERC) under the European Union’s Horizon 2020 research and innovation programme (grant agreement No. 882500).
}

\begin{document}
\maketitle

\begin{abstract}
  A minimal perfect hash function (MPHF) maps a set~$S$ of $n$ keys to the first $n$ integers without collisions.
  There is a lower bound of $n\log_2e-\Oh(\log n) \approx 1.44n$ bits needed to represent an MPHF.
  This can be reached by a \emph{brute-force} algorithm that tries $e^n$ hash function seeds in expectation and stores the first seed that leads to an MPHF.
  The most space-efficient previous algorithms for constructing MPHFs all use such a brute-force approach as a basic building block.

  In this paper, we introduce ShockHash -- {\bf S}mall, {\bf h}eavily {\bf o}verloaded cu{\bf ck}oo {\bf hash} tables for minimal perfect hashing.
  ShockHash uses two hash functions $h_0$ and $h_1$, hoping for the existence of a function $f : S \rightarrow \{0,1\}$ such that $x \mapsto h_{f(x)}(x)$ is an MPHF on $S$.
  It then uses a 1-bit retrieval data structure to store $f$ using $n + o(n)$~bits.

  In graph terminology, ShockHash generates $n$-edge random graphs until stumbling on a \emph{pseudoforest} -- where each component contains as many edges as nodes.
  Using cuckoo hashing, ShockHash then derives an MPHF from the pseudoforest in linear time.
  We show that ShockHash needs to try only about $(e/2)^n \approx 1.359^n$ seeds in expectation.
  This reduces the space for storing the seed by roughly $n$ bits (maintaining the asymptotically optimal space consumption) and speeds up construction by almost a factor of $2^n$ compared to brute-force.
  \emph{Bipartite} ShockHash reduces the expected construction time again to about $1.166^n$ by maintaining a pool of candidate hash functions and checking all possible pairs.

  Using ShockHash as a building block within the RecSplit framework we obtain \ShockHashRS{}, which can be constructed up to 3 orders of magnitude faster than competing approaches.
  \ShockHashRS{} can build an MPHF for $10$ million keys with $1.489$ bits per key in about half an hour.
  When instead using ShockHash after an efficient $k$-perfect hash function, it achieves space usage similar to the best competitors, while being significantly faster to construct and query.
\end{abstract}

\newpage

\section{Introduction}
A perfect hash function (PHF) maps a set of $N$ keys to the first $M$ integers without collisions.
If $M=N$, the hash function is called \emph{minimal} perfect (MPHF) and is a bijection between the keys and the first $N$ integers $[N]$.
Minimal perfect hashing has many applications.
For example, it can be used to implement static hash tables with guaranteed constant access time \cite{fredman1984storing}.
Storing only payload data in the hash table cells, we obtain an updatable retrieval data structure \cite{MSSZ14}, and storing only fingerprints \cite{fan2014cuckoo,bender2018bloom}, we obtain an approximate membership data structure.
The hashes can also be used as small identifiers of the input keys \cite{botelho2007perfect}, which are more efficient to deal with than large and complex keys.
Finally, there is a range of applications in bioinformatics \cite{chapman2011meraculous,chen2013hybrid,chikhi2016compacting,almodaresi2018space} and text indexing \cite{belazzougui2014alphabet,belazzougui2010fast,Witten:1999}.

\myparagraph{Related Work.}
There is a lower bound of about $N\log_2 e\approx 1.44N$ bits needed to represent an MPHF.
There is also a theoretical construction matching this bound that runs in linear time and allows constant query time \cite{HagTho01}.
However, this construction does not work for realistic $N<2^{150}$ \cite{botelho2013practical}, so that (minimal) perfect hashing remains an interesting topic for algorithm engineering.
A long sequence of previous work has developed a range of practical approaches with different space-time tradeoffs.

Many approaches first construct an outer hash function $g : S \rightarrow [b]$ that partitions the input set $S$ into small subsets $S_1, S_2, \ldots, S_b$ of sizes $s_1 \approx s_2 \approx \ldots \approx s_b$ and then construct a perfect hash function $h_i : S_i \rightarrow \{1,\ldots, s_i\}$ for $i \in [b]$.
Given $(s_i)_{i \in [b]}$, or better yet the prefix sums $p_i = s_1 + \cdots + s_{i}$ for $i \in [b]$, an MPHF on $S$ is given as $x \mapsto p_{g(x)} + h_{g(x)}(x)$.

On the one hand, there are holistic methods where such a partitioning step is not essential (though still possibly useful).
These are (so far) all a constant factor away from the space lower bound (e.g. \cite{chapman2011meraculous,MSSZ14,lehmann2023sichash,beling2023fingerprinting}).
One of the most space-efficient approaches among these is \emph{SicHash} \cite{lehmann2023sichash} that maps $N$ keys to $N(1+\varepsilon)$ unique table entries using a generalization of cuckoo hashing \cite{pagh2004cuckoo,fotakis2005space}.
The choice of hash function for each key is then stored in a retrieval data structure and a small fallback data structure turns the constructed PHF into an MPHF.
SicHash is not space-optimal partly because the cuckoo table tends to admit many valid placements of its keys, meaning a single input set is redundantly handled by many distinct states of the PHF data structure.

On the other hand, there are methods that use brute-force trial-and-error of hash functions on subsets of size $n$ \cite{esposito2020recsplit,bez2023high,fox1992faster,belazzougui2009hash,pibiri2021pthash}, which takes roughly $e^n \approx 2.718^n$ trials.
Hence, an aggressive partitioning step is required to obtain an acceptable overall running time.
The previously most space efficient approach, \emph{RecSplit} \cite{esposito2020recsplit}, is of this kind and recursively splits the input set into very small ($n\approx 16$) \emph{leaf} subsets.
Surprisingly, when tuned thoroughly, this enables higher construction throughput than the best holistic methods even when being fairly far away from the space lower bound \cite{bez2023high}.

\myparagraph{Contribution.}
In this paper, we introduce \emph{ShockHash} -- {\bf S}mall, {\bf h}eavily {\bf o}verloaded cu{\bf ck}oo {\bf hash} tables, which can be seen as an extreme version of SicHash where we use two hash functions for each key and retry construction until we can completely fill the cuckoo hash table.
That way, we achieve an MPHF without an intermediate non-minimal PHF.
In graph terminology, ShockHash repeatedly generates an $n$-edge random graph where each key corresponds to one edge, connecting the candidate positions of the key.
The table can be filled if and only if the graph is a \emph{pseudoforest} -- a graph where no component contains more edges than nodes.
While the ShockHash idea is straightforward in principle, we can \emph{prove} that when using binary cuckoo hashing with two choices (and thus 1-bit retrieval) there is only an insignificant amount of redundancy.
Therefore, ShockHash approaches the information theoretic lower bound for large $n$ and has running time $(e/2)^n\cdot\textrm{poly}(n) \approx 1.359^n$ (nearly a factor $2^n$ faster than brute-force).

In \emph{bipartite} ShockHash, further exponential improvements are possible.
Instead of using a pair of fresh hash functions for each construction attempt, we build a growing pool of hash functions and consider all pairs that can be formed from this pool.
Also, we let the two hash functions hash to disjoint ranges, meaning we effectively sample a bipartite graph where each edge has one endpoint in both partitions.
In this bipartite setting, the hash functions of both partitions need to be \emph{individually} surjective.
We can therefore filter the set of candidate hash functions in each partition individually -- before testing all combinations.
This improves the construction time by an additional exponential factor, to about $1.166^n\cdot\textrm{poly}(n)$.

\myparagraph{Evaluation.}
Still being an exponential time algorithm, we use ShockHash as a building block after partitioning the input.
We obtain ShockHash-RS by using ShockHash instead of brute-force as a base case within the RecSplit framework.
Though there is a small penalty in query time due to the additional access to a retrieval data structure, ShockHash-RS construction is about two orders of magnitude faster than tuned RecSplit \cite{bez2023high} for space efficient configurations, using the same architecture.
Bipartite ShockHash-RS improves this by a factor of $20$ again.
An important step in this harmonization of theory and practice is the observation that only an exponentially small fraction of the hash functions tried by ShockHash require the construction of a cuckoo hash table.
The other cases can be covered with a simple bit-parallel filter that checks whether all entries of the cuckoo table are hit by some key.
This removes much of the time overhead which made brute-force seemingly superior \cite{bez2023high}.

On large instances and for a space consumption of $1.56$ bits per key, the most space efficient competitor, tuned RecSplit \cite{bez2023high}, requires $137\mu$s per key.
In a similar amount of time, $1.499$ bits per key can be achieved by massive parallelization of the approach using a GPU \cite{bez2023high}.
Bipartite ShockHash-RS now achieves a new record of just $1.489$ bits per key with a similar construction time, while using only a single CPU thread (and efficient GPU parallelization seems not too difficult in future work).

We also demonstrate that ShockHash is useful outside the RecSplit framework.
When using $k$-perfect hashing for partitioning the input, we obtain ShockHash-Flat, which achieves a space usage similar to the most space efficient competitors.
At the same time, it is significantly faster to construct and reduces the query time by about 30\%, which brings it closer to the query performance of way less space efficient approaches.

\myparagraph{Outline.}
In \cref{s:prelim}, we explain preliminary ingredients needed to understand ShockHash, and in \cref{s:related}, we discuss related work.
We then explain the ShockHash algorithm in \cref{s:shockhash} and a bipartite variant in \cref{s:shockhash2}.
In \cref{s:analysis}, we then analyze both approaches.
To make ShockHash feasible for large input sizes, we show how to use it as a building block in different partitioning schemes in \cref{s:partitioning}.
In \cref{s:refinements}, we give additional variants and refinements that improve the construction in practice.
We conduct detailed experiments in \cref{s:experiments} and conclude the paper in \cref{s:conclusion}.

\section{Preliminaries}\label{s:prelim}
In this section, we explain basic ingredients of ShockHash.
This also includes the two perfect hash function constructions SicHash \cite{lehmann2023sichash} and RecSplit \cite{esposito2020recsplit} that ShockHash-RS is based on.
Finally, we describe pairing functions that we use to encode the seeds in bipartite ShockHash.

\myparagraph{Cuckoo Hashing.}
Cuckoo Hashing \cite{pagh2004cuckoo} is a well known approach to handle collisions in hash tables.
Each object gets two candidate cells via two hash functions.
A query operation looks at the two cells and searches for the object.
If an insertion operation tries to insert an object into a cell that is already full, the object already stored in the cell is taken out and recursively inserted using its other candidate position.
Cuckoo hashing can be extended to use more than two hash functions \cite{fotakis2005space}, or cells with more than one object in them \cite{dietzfelbinger2007balanced}.
In this paper, we are only interested in the basic version with two hash functions and one object per cell.
\label{s:loadThresholds}
The load threshold of a cuckoo hash table \cite{L:A_New_Approach:2012,FKP:The_Multiple:2011,fountoulakis2012sharp} is the percentage of cells that can be filled before insertion likely fails.
For cuckoo hashing with two candidate cells, the load threshold is $c=0.5$.
Despite this, in ShockHash we consider cuckoo hash tables that are filled completely using many retries.

\myparagraph{Cuckoo Graph and Pseudoforests.} \label{s:pseudotrees}
Cuckoo hashing can be modeled as a random graph $G$, where each node represents a table cell and each edge represents one object, connecting its two candidate cells.
It is easy to see that a cuckoo hash table can be constructed successfully if and only if the edges of $G$ can be directed such that the indegree of each node is $\leq 1$.
In the following, we call this a 1-orientation.
A 1-orientation exists if and only if $G$ is a \emph{pseudoforest}, i.e.\ every connected component of $G$ is a pseudotree.
A pseudotree is either a tree or a cycle with trees branching from it.
A way to check whether a graph is a pseudoforest is to check whether each component contains at most as many edges as nodes.

\myparagraph{Retrieval Data Structures.}\label{s:retrieval}
For a given set $S$ of $N$ keys and an integer $r$, a retrieval data structure (or \emph{static function} data structure) stores a function $f: S\rightarrow\{0, 1\}^r$ that maps each key to a specific $r$-bit value.
Because it may return arbitrary values for keys not in $S$, it is possible to represent the function without representing $S$ itself.
Representing a retrieval data structure needs at least $rN$ bits of space and there are practical data structures that need $rN+o(rN)$ bits allowing linear construction time and constant query time.
In particular, for $r = 1$, \emph{Bumped Ribbon Retrieval (BuRR)} \cite{dillinger2022burr} reduces function evaluation to \texttt{XOR}ing a hash function value with a segment of a precomputed table and reporting the parity of the result.
This table can be determined by solving a nearly diagonal system of linear equations (a ``ribbon'').
In practice, BuRR has a space overhead below $1\%$.

\myparagraph{SicHash.}
{\bf S}mall {\bf i}rregular {\bf c}uckoo tables for perfect {\bf Hash}ing \cite{lehmann2023sichash} constructs perfect hash functions through cuckoo hashing.
It constructs a cuckoo hash table and then uses a retrieval data structure to store which of the hash function choices was finally used for each key.
The main innovation of SicHash is using a careful mix of \mbox{1--3} bit retrieval data structures, corresponding to 2, 4, or 8 choices for the keys.
It achieves a favorable space-performance tradeoff when being allowed 2--3 bits of space per key.
It cannot go below this because using only 1-bit retrieval seems to lead far from minimality while using 2 or more bits for retrieval allows redundant choices that cannot achieve space-optimality.
SicHash achieves a rather limited gain in space efficiency by \emph{overloading} the table beyond the load thresholds and trying multiple hash functions.
This mainly exploits the variance in the number of keys that can fit.
SicHash leaves the success probability of constructing overloaded tables as an open question.
ShockHash drives the idea of overloading to its extreme and gives a formal analysis for this case.

\myparagraph{RecSplit.}\label{s:recsplit}
RecSplit \cite{esposito2020recsplit} is a minimal perfect hash function that is mainly focused on space efficiency.
First, all keys are hashed to buckets of constant expected size~$b$.
A bucket's set of keys is partitioned into different subsets recursively in a tree-like structure by \emph{splitting} hash functions that are chosen by brute-force to ensure that all leaves except the last have size exactly $n$ (in Ref. \cite{esposito2020recsplit}, the leaf size is called~$\ell$).
Within the leaves, RecSplit then performs brute-force search for a minimal perfect hash function (also called \emph{bijection}).
The tree structure is based only on the size of the current bucket.
This makes it possible to store only the seed values for the hash functions without storing structural information.
Apart from encoding overheads for the seeds, this makes RecSplit information theoretically optimal within a bucket.
The number of child nodes (\emph{fanout}) in the two lowest levels is selected such that the amount of brute-force work is balanced between splittings and bijections.

There also is a parallel implementation using multi-threading and SIMD instructions or the GPU \cite{bez2023high}.
The paper also proposes a new technique for searching for bijections called \emph{rotation fitting}.
Instead of just applying hash functions on the keys in a leaf directly, rotation fitting splits up the keys into two sets using a 1-bit hash function.
It then hashes each of the two sets individually, forming two words where the bits indicate which hash values are occupied.
Then it tries to find a way to cyclically rotate the second word, such that the empty positions left by the first set are filled by the positions of the second set.
The paper shows that each rotation has a similar success probability as a new hash function seed, so it is a way to quickly evaluate additional hash function seeds.

\begin{figure}[t]
  \centering
  \begin{subfigure}[b]{0.3\textwidth}
    \centering
    \includegraphics[page=1]{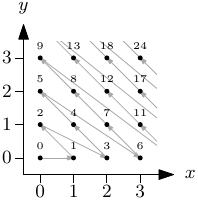}
    \caption{Cantor pairing function}
    \label{fig:pairingFunctions:cantor}
  \end{subfigure}
  \begin{subfigure}[b]{0.3\textwidth}
    \centering
    \includegraphics[page=2]{fig/pairingFunctions}
    \caption{Szudzik's pairing function}
    \label{fig:pairingFunctions:szudzik}
  \end{subfigure}
  \begin{subfigure}[b]{0.3\textwidth}
    \centering
    \includegraphics[page=3]{fig/pairingFunctions}
    \caption{Triangular pairing function}
    \label{fig:pairingFunctions:triangular}
  \end{subfigure}
  \caption{Illustrations of different pairing functions.}
  \label{fig:pairingFunctions}
\end{figure}

\myparagraph{Pairing Functions.}\label{s:pairing}
A pairing function encodes two natural numbers in a single natural number.
More precisely, a pairing function is a bijection between the grid $\mathbb{N}_0^2$ and $\mathbb{N}_0$.
We are interested in pairing functions that can be inverted efficiently.
The most popular pairing function is the Cantor pairing function, which enumerates the 2D grid diagonally (see \cref{fig:pairingFunctions:cantor}).
It can be calculated by $\textrm{pair}_{\textrm{c}}(x, y) = (x + y)(x + y + 1)/2 + y$.
Another pairing function is the one by Szudzik \cite{szudzik2006elegant}, which enumerates the 2D grid following the edges of a square (see \cref{fig:pairingFunctions:szudzik}).
The pairing function can be calculated by $\textrm{pair}_{\textrm{s}}(x, y) = y^2+x$ if $x \leq y$ and $\textrm{pair}_{\textrm{s}}(x, y) = x^2+x+y$ otherwise.

In this paper, we require a function that enumerates only those coordinates of the 2D grid with $x > y$. We will still call it a pairing function in slight abuse of traditional terminology.
Our \emph{triangular pairing function} (see \cref{fig:pairingFunctions:triangular}) can be calculated by $\textrm{pair}_{\textrm{t}}(x, y) = x(x - 1)/2 + y$ with the intuition stemming from the \emph{Little Gauss} formula.
The basic idea for inverting our function $(x', y') = \textrm{pair}_{\textrm{t}}^{-1}(z)$ is to set $y=0$ in the definition and solve for $x$.
This gives $x'=\lfloor1/2 + \sqrt{1/4 + 2z}\rfloor$ and $y' = z - \textrm{pair}_{\textrm{t}}(x', 0)$.
In our bipartite implementation, we use both our triangular pairing function and Szudzik's pairing function, depending on the distribution of the numbers we want to encode.

While pairing uses only integer operations, all three pairing functions rely on the square root operation and rounding for inverting.
This means that inverting the functions in practice can lead to problems due to floating point inaccuracies.
Whether inverting $z$ succeeded can easily be checked by verifying that $\textrm{pair}(x', y') = z$.
In our implementation, we check invertibility at construction time, so we do not get a run-time overhead during queries.

\section{More Related Work}\label{s:related}
In addition to RecSplit and ShockHash, which we describe in the preliminaries, there is a range of other minimal perfect hash functions.
We give an overview over the approaches in the following paragraphs.

\myparagraph{Hash-and-Displace.}
Perfect hashing with Hash-and-Displace \cite{fox1992faster,belazzougui2009hash,pibiri2021pthash} allows fast queries and asymptotically optimal space consumption.
Each key $x$ is first hashed to a small bucket $b(x)$ of keys.
For each bucket $b$, an index $i(b)$
of a hash function $f_{i(b)}$ is stored such $x \mapsto f_{i(b(x))}(x)$ is an injective function.
For a particular bucket, this index is searched in a brute-force way.
To accelerate the search, buckets are first sorted by their size.
FCH \cite{fox1992faster} uses an asymmetric bucket assignment: it hashes 60\% of the keys to 30\% of the buckets.
This is efficient because when placing the largest (and therefore hardest to place) buckets, most positions are still available.
CHD \cite{belazzougui2009hash} uses buckets of expected equal size but compresses the stored hash function seeds.
PTHash \cite{pibiri2021pthash} combines the ideas by using an asymmetric bucket assignment and storing the seeds in compressed form.

\myparagraph{Fingerprinting.}
Perfect hashing through fingerprinting \cite{chapman2011meraculous,MSSZ14} hashes the $N$ keys to $\gamma N$ positions using an ordinary hash function, where $\gamma$ is a tuning parameter.
The most space efficient choice $\gamma=1$ leads to a space consumption of $e$ (not $\log_2 e$) bits per key.
A bit vector of length $\gamma N$ indicates positions to which exactly one key was mapped.
Keys that caused collisions are handled recursively in another layer of the same data structure.
At query time, when a key is the only one mapping to its location, a rank operation on the bit vector gives the MPHF value.
Publicly available implementations include BBHash \cite{limasset2017fast} and the significantly faster FMPH \cite{beling2023fingerprinting}.
\mbox{FMPHGO} \cite{beling2023fingerprinting} combines the idea with a few brute-force tries to select a hash function that causes fewer collisions.

\myparagraph{Table Lookup.}
A tempting way to replace expensive brute-force search is precomputation of solutions with subsequent table lookup -- a standard technique used in many compressed data structures.
For a rough idea, suppose for a subproblem with $n$ keys, we first map them injectively to a range of size $U' \in \Omega(n^2)$ using an intermediate hash function (less would lead to collisions -- birthday paradox).
Then, using a lookup table of size $2^{U'}$, we can find precomputed perfect hash functions in constant time.
However, polynomial running time limits the subproblem size to $n \in \Oh(\sqrt{\log N})$, where $N$ is the size of the overall input set.
Computing concrete values for realistic values of $N$, one gets subproblem size much smaller than what can be easily handled even with plain RecSplit.
Nevertheless, Hagerup and Tholey \cite{HagTho01} develop this approach to a comprehensive theoretical solution of the perfect hashing problem yielding linear construction time, constant query time, and space $1+o(1)$ times the lower bound.
However, this method is not even well-defined for $N<2^{150}$ \cite{botelho2013practical}.
A variant of RecSplit with rotation fitting \cite{bez2023high} uses lookup tables of size $2^{n}$ to find feasible rotations in constant time.
Unfortunately, this turns out to be slower than trying all rotations directly.

\begin{figure}
  \centering
  \includegraphics[scale=0.8]{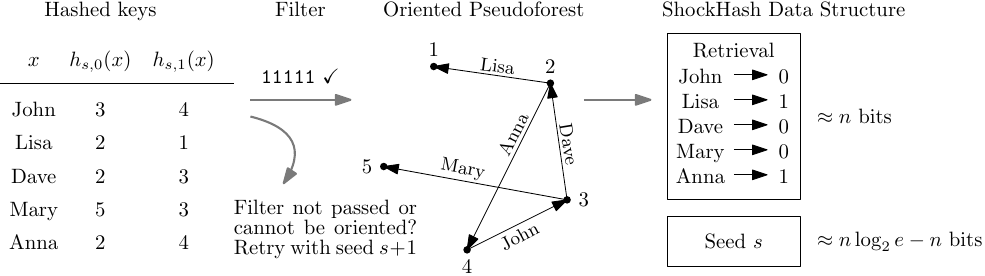}
  \caption{\label{fig:illustration} Illustration of the ShockHash construction.
    Functions $h_0$ and $h_1$ are randomly sampled hash functions using a seed $s$.
    Here, $s$ is a seed value where the resulting graph is a pseudotree.
    During construction, many seeds need to be tried.
  }
\end{figure}

\section{ShockHash}\label{s:shockhash}
We now introduce the main idea of this paper, ShockHash.
The asymptotic load threshold of a binary cuckoo hash table is $c=0.5$ (see \cref{s:loadThresholds}), so the success probability of constructing a table with $n$ cells and more than $n/2$ keys tends to zero.
ShockHash overloads a cuckoo hash table far beyond its asymptotic load threshold -- it inserts $n$ keys into a binary cuckoo hash table of size $n$.
As we will see in \cref{lem:constructionTries}, the construction succeeds after $(e/2)^{n} \textrm{poly}(n)$ tries in expectation.
We then record the successful seed
\begin{equation*}\label{eq:ShockHash}
  s=\min\{s ∈ ℕ \mid ∃f ∈ \{0,1\}^S\!: x \!↦\! h_{s,f(x)}(x) \text{ is MPHF}\} \hspace{8mm} (h_{i,0})_{i \in ℕ}, (h_{i,1})_{i \in ℕ} : S\rightarrow[n]
\end{equation*}
and a successful choice $f$ between the two candidate positions of each key.
The seed needs $n \cdot \log(e/2) + o(n) \approx 0.44n+o(n)$ bits in expectation using Golomb-Rice codes \cite{golomb1966run, rice1979some}.
The choices are stored in a $1$-bit retrieval data structure, requiring $n+o(n)$ bits.
This means that the majority of the MPHF description is not stored in the seed, like with the brute-force construction, but in the retrieval data structure.
A query for key $x$ retrieves $f(x)$ from the retrieval data structure and returns $h_{s,f(x)}(x)$.
\Cref{fig:illustration} gives an illustration of the ShockHash construction.

The beauty of ShockHash is that it can check $2^n$ different possible hash functions (determined by the $2^n$ different functions representable by the retrieval data structure) in $\Oh(n)$ time.
This enables significantly faster construction than brute-force while still consuming the same amount of space up to lower order terms.
\label{s:sccFilter}
As discussed in \cref{s:pseudotrees}, a seed leads to a successful cuckoo hash table construction if and only if the corresponding random (multi)graph with edges $\{\{h_{s,0}(x), h_{s,1}(x)\} \mid x \in S\}$ forms a pseudoforest.
Each component of size $c$ is a pseudotree if and only if it contains no more than $c$ edges.
This can be checked in linear time using connected components algorithms, or in close to linear time using an incremental construction of an ordinary cuckoo hash table.
However, compared to the simple bit-parallel perfectness test of brute-force \cite{esposito2020recsplit}, each individual check is slower by a large constant factor.
In the following paragraph, we discuss a way to address this bottleneck.

\myparagraph{Filter by Bit Mask.}\label{s:bitmaskFilter}
To reduce the time spent checking if a graph is a pseudoforest, we use a filter to quickly reject most seeds.
More specifically, we reject seeds for which some table cell is not a candidate position of any of the keys.
If there is such a cell, we already know that cuckoo hashing cannot succeed and we can skip the full test.
Otherwise, cuckoo hashing might succeed.
The filter can be implemented using simple shift and comparison operations.
Also, the filter can use registers, in contrast to the more complex full construction.
It is one of the main ingredients for making ShockHash practical and is easily proven to be very effective:

\begin{lemma}
  \label{lem:filterProbability}
  The probability for a seed to pass the filter, i.e.\ for every table cell to be hit by at least one key, is at most $(1-e^{-2}+o(1))^n ≈ 0.864^n$.
\end{lemma}
\begin{proof}
  Let $X_i$ denote the number of times that cell $i ∈ [n]$ is hit. Then $(X₁,…,Xₙ)$ follows a multinomial distribution.
  The variables $X₁,…,Xₙ$ are \emph{negatively associated} in the sense introduced in \cite{JDP:NegativeAssociation:1983} and satisfy
  \[
    ℙ(∀i∈[n]:X_i ≥ 1) ≤ \prod_{i = 1}^n ℙ(X_i ≥
    1),
  \]
  the intuition being that since the sum $X₁+…+Xₙ= 2n$ is fixed, the events $\{X_i ≥ 1\}$ for $i ∈ [n]$ are less likely to co-occur compared to corresponding independent events.
  Since $X_i \sim \Bin(2n,\frac 1n)$ for all $i ∈ [n]$ we have
  \[
    ℙ(X_i ≥ 1) = 1- (1-\tfrac 1n)^{2n} = 1-e^{-2}
    + o(1) ≈ 0.864
  \]
  and the claim follows.
  \hfill
\end{proof}
A more careful analysis \cite{walzer2024probability} reveals that the probability to pass the filter is around $b^n$ where $b = 2e^λ / (λe^2) ≈ 0.836$ and where $λ ≈ 1.597$ is the solution to $2 = λ/(1-e^{-λ})$.

\myparagraph{Enhancements.}
In \cref{s:refinements}, we explain additional enhancements that improve the construction performance in practice.
This includes applying the idea of \emph{rotation fitting} \cite{bez2023high} to ShockHash, as well as faster orientability checks.

\section{Bipartite ShockHash}\label{s:shockhash2}
\begin{figure}[t]
  \centering
  \includegraphics[width=0.9\textwidth]{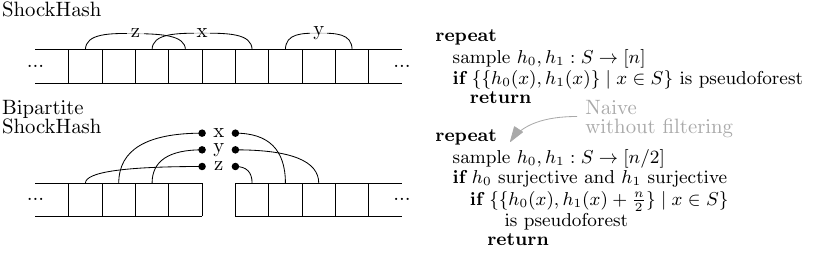}
  \caption{ShockHash and bipartite ShockHash. The pseudocode illustrates the overall idea but does not lead to any performance improvements yet.}
  \label{fig:shockhash1vs2}
\end{figure}

Bipartite ShockHash is an extension of the ShockHash idea.
It enables significantly faster construction compared to plain ShockHash.
In turn, this enables more aggressive parameter choices, thereby leading to improved space-efficiency.
While ShockHash samples random graphs, \emph{bipartite} ShockHash now samples \emph{bipartite} random graphs.
\Cref{fig:shockhash1vs2} gives an illustration and very simple pseudocode.
In plain ShockHash, each edge is connected to two nodes using two independent hash functions.
In bipartite ShockHash, the hash functions have a range of $[n/2]$, but we shift the hashes of one of the hash functions by $n/2$, meaning each edge gets one endpoint in $[n/2]$ and one in $n/2 + [n/2]$.
This is similar to the original implementation of cuckoo hashing using two independent hash tables \cite{pagh2004cuckoo}.
The idea might sound not very helpful at first, but opens up several ways of pruning the search space.
In the following, we assume that $n$ is an even number.
We give an extension to uneven numbers n \cref{ss:unevenN}.

\myparagraph{Filtering Seed Candidates.}
We show in \cref{s:analysis} that testing about $(e/2)^n \approx 1.359^n$ pairs of hash functions is sufficient for plain ShockHash.
The idea of bipartite ShockHash is that it is almost as good to consider roughly $\sqrt{(e/2)^n} = (e/2)^{n/2}$ hash functions and all pairs that can be formed from them.
This already improves the practical running time because fewer hash functions need to be evaluated.
However, the asymptotic construction time is not improved much because we still need to test all combinations.
A key realization is that in the bipartite case, a pair $(h_0, h_1)$ of hash functions can only work if \emph{both} $h_0$ and $h_1$ (both with range $[n/2]$) are \emph{individually} surjective.
In the non-bipartite case, in contrast, the check was that $h_0$ and $h_1$ (both with range $[n]$) \emph{together} hit each position in $[n]$ at least once.
This means that we can filter the list of hash functions \emph{before} pairing them up.

\label{s:filterEffectiveness}
In each of the partitions, we look at $n$ keys mapping to $n/2$ positions.
Similar to \cref{lem:filterProbability}, the probability of passing the filter is $0.836^{n/2}$ \cite{walzer2024probability}.
This suggests that if we pair up only the hash functions passing the filter then we will be considering at most $((e/2)^{n/2} \cdot 0.836^{n/2})^2 \approx 1.136^n$ pairs.
Refer to \cref{s:analysis} for details.

\begin{figure}[t]
      \begin{lstlisting}
Function construct$(S)$
  surjectiveCandidates $\leftarrow$ $\emptyset$
  for $s_0 = 0$ to $\infty$
    if $h_{s_0}$ is surjective on $S$
      for $s_1$ $\in$ surjectiveCandidates
        if $\exists f \in \{0,1\}^S: x \mapsto \frac{n}{2} \cdot f(x)+h_{s_{f(x)}}(x)$ is a bijection
          return $f$ as retrieval data structure, $s_0$, $s_1$
      surjectiveCandidates $\leftarrow$ surjectiveCandidates $\cup$ $\{s_0\}$
Function evaluate$(x)$
  return $\frac{n}{2} \cdot f(x)+h_{s_{f(x)}}(x)$
      \end{lstlisting}
  \caption{Pseudocode of bipartite ShockHash.}
  \label{fig:pseudocode}
\end{figure}

\begin{figure}[t]
  \centering
  \begin{subfigure}[b]{0.38\textwidth}
    \centering
    \includegraphics[scale=0.8,page=1,trim={0 0 65mm 0},clip]{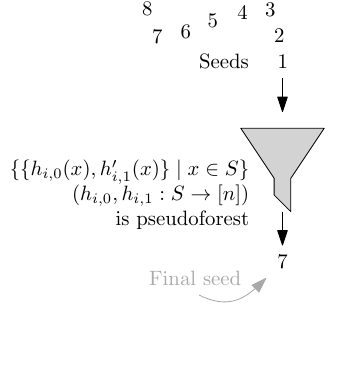}
    \caption{ShockHash.}
  \end{subfigure}%
  \begin{subfigure}[b]{0.62\textwidth}
    \centering
    \includegraphics[scale=0.8,page=2,trim={20mm 0 0 0},clip]{fig/shockhash1vs2filters.pdf}
    \caption{Bipartite ShockHash.}
  \end{subfigure}
  \caption{
    Illustration of the filtering involved in ShockHash and bipartite ShockHash.
    The construction is complete if we find one final seed.
    ShockHash determines both hash functions from the same seed.
    Bipartite ShockHash uses independent seeds for the two hash functions and filters the seeds before combining them.
  }
  \label{fig:shockhash1vs2filters}
\end{figure}

\myparagraph{The Bipartite ShockHash Algorithm.}
The following paragraph describes our new bipartite ShockHash algorithm.
We maintain a pool of seed candidates that are surjective on $[n/2]$.
To find a new candidate, we linearly check hash function seeds until we find a seed $s_0$ that gives a surjective hash function.
Given that new candidate, we try to combine it with all previous candidates $s_1$ from the pool.
More precisely, we check if the graph defined by the nodes $[n]$ and the edges $\{\{h_{s_0}(x), h_{s_1}(x) + n/2\} \mid x \in S\}$ is a pseudoforest.
If it is a pseudoforest, we have found a perfect hash function.
We only need to store the assignment from keys to their candidate hash function ($h_{s_0}$ or $h_{s_1}$) in a retrieval data structure, as well as the two seeds $s_0$ and $s_1$.
If the combination with none of the previous seed candidates leads to a pseudoforest, we add the new candidate $s_0$ to the set of surjective candidates and search for the next one.
\Cref{fig:pseudocode} gives a pseudocode for this algorithm and \cref{fig:shockhash1vs2filters} illustrates the idea of filtering hash functions before putting them in the pool.

Note that it does not matter which of the two hash functions we use for which partition of the graph.
Switching the partitions just gives an isomorphic graph and does not influence orientability.
We therefore always use the newly determined candidate directly and shift the old candidate by $n/2$ to be in the second partition.
Also, we neglect the possibility that a hash function combined with itself on both partitions leads to successful construction.%
\footnote{The function would have to map exactly two keys to each of the $n/2$ positions, which happens with probability $\binom{n}{2\,2\,\cdots\,2}(\frac{n}{2})^{-n} = e^{-n}2^{n/2}\mathrm{poly}(n)$.}
This allows us to store the two seeds $s_0$ and $s_1$, knowing that $s_1 < s_0$.
We do so in one integer using our triangular pairing function that we explain in \cref{s:pairing}.
Note that the pairing function enumerates the seed pairs in exactly the same order that we test them in.
Compared to storing two variable-length integers, pairing reduces constant overheads in the encoding.

\myparagraph{Enhancements.}
In \cref{s:refinements}, we give additional enhancements that improve the construction performance significantly in practice.
This includes other ways of coming up with a stream of hash function candidates, bit-parallel filtering, and support for uneven input sizes.

\section{Analysis}\label{s:analysis}
In this section, we analyze the space usage and construction time of ShockHash.
The main challenge is to lower bound the probability that a hash function seed enables successful construction of the heavily overloaded cuckoo hash table.
First, in \cref{ss:simpleProof}, we give a very simple analysis of the success probability of plain ShockHash.
It is less tight than our more complex proof, but it is significantly shorter.
In \cref{ss:analysisPreliminaries}, we explain tools used in the analysis and prove small building blocks of the full proof.
In \cref{ss:successProbability,ss:successProbabilityBipartite}, we then analyze the success probability of plain and bipartite ShockHash, respectively.
We then show in \cref{ss:pooling} that a pool containing about $(\sqrt{e/2})^{n}$ hash function candidates is usually sufficient.
Finally, we give the construction time and space consumption of ShockHash and bipartite ShockHash in \cref{ss:analysisShockHash}.
In the following we assume that a seed is given.
We suppress it in notation.

It will be useful to consider the graph
\[
G = ([n], \{\{h₀(x),h₁(x)\} \mid x ∈ S\}).
\]
While similar to an Erdős-Renyi random graph, $G$ may have self-loops\footnote{Graphs with self-loops are easier to analyze here but we avoid them in practice for better performance.} and multi-edges.
Our model matches Model A in \cite{frieze2016introduction}.

\subsection{A Simple Proof}\label{ss:simpleProof}
First, in \cref{lem:simpleProof}, we give a simple combinatorial argument showing that the probability for $G$ to be a pseudotree is at least $(e/2)^{-n}\sqrt{\pi/(2n)}$.
This lower bounds the probability of $G$ being a pseudoforest consisting of potentially more than one tree.
\Cref{ss:successProbability} then shows that the probability is at least $(e/2)^{-n} \pi/e$.
Therefore, the simple argument is only a factor of $\Oh(\sqrt{n})$ less tight than the much more complex proof.

\begin{theorem}
  \label{lem:simpleProof}
  Let $G$ be a multigraph with $n$ nodes and $n$ edges.
  The probability space underlying $G$ is that of sampling $2n$ vertices (with replacement) and creating an edge from the samples $2i-1$ and $2i$ for each $i ∈ [n]$.
  Then the probability that $G$ is a pseudotree is at least $(e/2)^{-n}\sqrt{\pi/(2n)}$.
\end{theorem}
\begin{proof}
  For $G$ to be a pseudotree it is sufficient (though not necessary) that the first $n-1$ created edges form a tree.
  There are $n^{n-2}$ labeled $n$-node trees (Cayley's Formula \cite{cayley1878theorem}).
  Since the ordering of the edges and the order of the two samples forming an edge does not matter, each of the trees can be generated in $2^{n-1}(n-1)!$ ways.
  The last two samples can be anything, giving us $n^2$ choices.
  By applying Stirling's approximation, namely
  \begin{align*}
    n! ∈ \left[\big(\tfrac{n}{e}\big)^{n}\sqrt{2 \pi n}\cdot e^{1/(12n+1)}, \big(\tfrac{n}{e}\big)^{n}\sqrt{2 \pi n} \cdot e^{1/(12n)}\right],
  \end{align*}
  we can show that the total probability to draw a pseudotree is at least
  \begin{align*}
    \frac{n^{n-2}2^{n-1}(n-1)!n^2}{n^{2n}} \geq \left(\frac{e}{2}\right)^{-n}\sqrt{\pi/(2n)}.
    \tag*{\qedhere}
  \end{align*}
\end{proof}

\subsection{Tools}\label{ss:analysisPreliminaries}
In this section, we explain tools that are later needed in the analysis, such as the configuration model and graph peeling.
We start with proving two small lemmas that are used in the remaining analysis but are very generic in nature.

\begin{lemma}
  \label{lem:expectation}
  Let $X \in \mathbb{N}_0$ be a random variable.
  Then the probability that $X$ is \emph{at least} 1 is
  \[
    ℙ(X > 0) = \frac{𝔼(X)}{𝔼(X \mid X > 0).}
  \]
\end{lemma}
\begin{proof}
  For any non-negative random variable $X$ we can apply the law of total expectation to get
  $𝔼(X) = ℙ(X = 0)·𝔼(X \mid X = 0) + ℙ(X > 0)·𝔼(X \mid X > 0) = ℙ(X > 0)·𝔼(X \mid X > 0)$.
  Rearranging this for $ℙ(X > 0)$ yields the desired result.
\end{proof}

\begin{lemma}
  \label{lem:binomial}
  For $n,c\in\mathbb{N}$, it holds that ${cn \choose n} \leq \left(ec\right)^n$
\end{lemma}
\begin{proof}
  We first upper bound all factors $(cn-k)$ by $cn$ and then apply Stirling's approximation.
  \begin{align*}
    {cn \choose n}
      & = \frac{(cn)(cn-1)(cn-2)\ldots(cn-n+1)}{n!}
      \leq \frac{(cn)^n}{n!}
      \leq \frac{(cn)^n}{\sqrt{2 \pi n}\left(n/e\right)^n}
      \leq \left(ec\right)^n
    \tag*{\qedhere}
  \end{align*}
\end{proof}

\myparagraph{Configuration Model.}\label{ss:configurationModel}
The \emph{configuration model} \cite{newman2010networks} is a way to describe distributions of random graphs.
In the model, we can fix the exact degree of every node in the graph by giving each node a number of edge \emph{stubs}.
The graph is obtained by repeatedly sampling, uniformly at random, two unconnected stubs and connecting them.
In other words, if we take one stub and look at its partner, all other stubs are equally likely.

\myparagraph{Graph Peeling.}\label{ss:graphPeelingPrelim}
In several sections of this paper, we are interested in \emph{peeling} \cite{walzer2021peeling,Molloy05:Cores-in-random-hypergraphs,Luczak:A-simple-solution} graphs.
For this, we iteratively take any node of degree $1$ and remove it together with its corresponding edge.
The process continues until all nodes have degree $>1$.
A graph is $1$-orientable or a pseudoforest, if all nodes in the remaining graph have degree $2$, meaning that the graph consists of only cycles.

\myparagraph{Graph Peeling in the Configuration Model.}
To analyze the peeling process, it will be useful to reveal $G$ in two steps.
First the degree of each node is revealed by randomly distributing $2n$ \emph{stubs} (or half-edges) among the $n$ nodes.
This yields a configuration model from which the edges are then obtained by randomly matching the stubs.
The following lemma should clarify what exactly we need.

\begin{lemma}
  \label{lem:probability-spaces}
  Let $x_1,…,x_{2n} ∈ [n]$ be independent and uniformly random. The graphs $G₁,G₂,G₃$ defined in the following have the same distribution as $G$.
  \begin{enumerate}
    • $G₁ = ([n],\{\{x_{2i-1},x_{2i}\} \mid i ∈ [n]\})$.
    • $G₂ = ([n],\{\{x_i,x_j\} | \{i,j\} ∈ M\})$ where $M$ is a uniformly random perfect matching of $[2n]$, i.e.\ a partition of $[2n]$ into $n$ sets of size $2$.
    • $G₃$ is defined like $G₂$, except that $M$ is obtained in a sequence of $n$ rounds. In each round an unmatched number $i ∈ [2n]$ is chosen \emph{arbitrarily} and matched to a distinct unmatched number $j$, chosen uniformly at random. The choice of $i$ may depend on $x₁,…,x_{2n}$ and on the set of numbers matched previously.
  \end{enumerate}
\end{lemma}
The reason for considering these alternative probability spaces for $G$ is that they permit conditioning on partial information about $G$ (such as its degree sequence implicit in $x₁,…,x_{2n}$) but retaining a clean probability space for the remaining randomness.
\begin{proof}
  Compared to $G$, the definition of $G₁$ simply collects the $2n$ relevant hash values in a single list.%
\footnote{Here, we assume that $h₀$ and $h₁$ are fully random hash functions and given for free, which is common in previous papers (Simple Uniform Hashing Assumption) \cite{dietzfelbinger1990new,pagh2007linear,pagh2008uniform,dietzfelbinger2009applications}.}
  Concerning $G₂$, imagine that $M$ is revealed first. Conditioned on $M$, $G₂$ is composed of $n$ uniformly random edges like $G₁$.
  Concerning $G₃$, the key observation is that $M$ is a uniformly random matching even if the number to be randomly matched in every round is chosen by an adversary. A formal proof could consider any arbitrary adversarial strategy and use induction.
  \hfill
\end{proof}

Therefore, when peeling in the configuration model, we can interleave the peeling process and the process of uncovering the sampled graph.
To peel, we take a node with degree $1$ and look at the other endpoint of its adjacent edge, which is uniformly distributed between all other stubs.
If the node connected to it has degree $2$, we have found a new degree-$1$ node that we can directly continue peeling.
Otherwise, we have to start with a new node of degree $1$ in a next iteration.

\subsection{Success Probability in Plain ShockHash}\label{ss:successProbability}
In this section, we give the tighter analysis of the success probability of ShockHash.
\def\orient{\mathrm{ori}}
\def\PF{\mathrm{PF}}
Given two hash functions $h₀,h₁: S → [n]$ and a function $f: S → \{0,1\}$, let $\orient(f)$ be the event that $x ↦ h_{f(x)}(x)$ is bijective.
We are now interested in the probability that there exists such a function $f$ that leads to a bijective function, namely $\Pr(\exists f : \orient(f))$.
There is a one-to-one correspondence between functions $f$ with $\orient(f)$ and $1$-orientations of $G$, i.e.\ ways of directing $G$ such that each node has indegree at most $1$.%
\footnote{Note that in our model, there are two ways of directing a self-loop.}

We write $\PF(G)$ for the event that $G$ is a pseudoforest.
As pointed out in \cref{s:pseudotrees}:
\begin{equation}
  \PF(G) ⇔ ∃f: \orient(f).
  \label{eq:pseudoforests-orientable}
\end{equation}
In our case with $n$ nodes and $n$ edges, $\PF(G)$ implies that $G$ is a \emph{maximal} pseudoforest, where every component is a pseudotree and not a tree.
Note that a pseudotree that is not a tree admits precisely two $1$-orientations because the unique cycle can be directed in two ways and all other edges must be directed away from the cycle.
A useful observation is therefore
\begin{equation}
  \PF(G) ⇒ \# \{f : \orient(f) \} = 2^{c(G)}
  \label{eq:orientations-cycles}
\end{equation}
where $c(G)$ is the number of connected components of $G$.

The basic idea of our proof is as follows.
The probability that a random function is minimal perfect is $e^{-n}\textrm{poly}(n)$ (see \cref{lem:bruteForce}).
Each of the $2^n$ functions $f: S → \{0,1\}$ has that chance of satisfying $\orient(f)$ and yielding an MPHF.
However, simply multiplying $e^{-n}\textrm{poly}(n)$ by $2^n$ does not necessarily yield an approximation for the probability that such an $f$ exists.
The key point here is that the $2^n$ functions $(x \mapsto h_{f(x)}(x))_{f \in \{0,1\}^S}$ determined by the $2^n$ different options for $f$ are correlated.
If there are some graphs that permit many different $1$-orientations, we may find many MPHFs at once and the probability that at least one $1$-orientation exists is reduced.
This intuition is stated more formally in \cref{lem:expectation} using $X=\#\{f : \orient(f)\}$:
\[
  ℙ(∃f: \orient(f)) = \frac{𝔼(\#\{f : \orient(f)\})}{𝔼(\#\{f : \orient(f)\} \mid ∃f: \orient(f)).}
\]

As discussed above, a key step in the analysis is to show that we usually find only a few MPHFs at once.
This amounts to analyzing the distribution of the number of components in random maximal pseudoforests, which we do in \cref{lem:numberOfOrientations}.
The main proof in \cref{lem:constructionTries} then formally bounds the probability that a random graph is a pseudoforest, juggling different probability spaces.

\begin{figure}[t]
  \centering
  \includegraphics[scale=0.85]{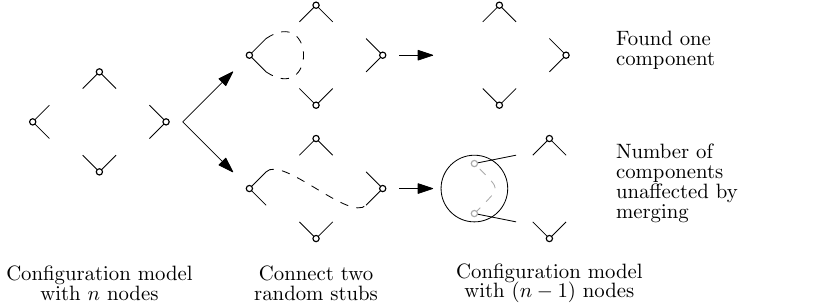}
  \caption{Number of components in the configuration model.}
  \label{fig:orientations}
\end{figure}

\begin{lemma}
  \label{lem:numberOfOrientations}
  Let $G_n$ be the random graph sampled from the configuration model with $n$ nodes of degree $2$, i.e.\ the $2$-regular graph obtained by randomly joining $2n$ stubs that are evenly distributed among $n$ nodes.
  Then the number $c(G_n)$ of components of $G$ satisfies $𝔼(2^{c(G_n)}) ≤ e·\sqrt{2n}$.
\end{lemma}
We remark that a similar proof shows that $\mathds{E}(c(G_n)) \in \Oh(\log n)$.
Note also the similarity to the locker puzzle, which analyzes the length of the largest cycle in a random permutation \cite{stanley2011enumerative}.
\begin{proof}
  We will find a recurrence for $d_n := 𝔼(2^{c(G_n)})$.
  Consider an arbitrary node $v$ of $G_n$ and one of the stubs at $v$.
  This stub forms an edge with some other stub.
  We have $n-1$ other nodes, each with $2$ stubs, and we have the second stub at $v$. Each of these $2n-1$ stubs is matched with $v$ with equal probability.
  Therefore, the probability that $v$ has a self-loop is $\frac{1}{2n-1}$.

  (1) Conditioned on $v$ having a self-loop, we have found an isolated node.
  The distribution of the remaining graph is that of $G_{n-1}$ and the conditional expectation of $2^{c(G)}$ is therefore $𝔼(2^{1+c(G_{n-1})}) = 2d_{n-1}$.

  (2) Now condition on the formed edge connecting $v$ to $w ≠ v$.
  We can now merge the nodes to a single one without affecting the number of components. The merged node inherits two unused stubs, one from $v$ and one from $w$.
  The distribution of the remaining graph is that of $G_{n-1}$.
  Therefore, in this case, the conditional expectation of $2^{c(G)}$ is simply $d_{n-1}$.

  These two cases are illustrated in \cref{fig:orientations} and lead us to the following recurrence:
  \begin{align*}
    d_{n}=\tfrac{1}{2n-1}2d_{n-1}+\big(1-\tfrac{1}{2n-1}\big)d_{n-1}=\big(1+\tfrac{1}{2n-1}\big)d_{n-1}.
  \end{align*}
  With the base case $d_0=1$, we can solve the recurrence and bound its value as follows, using that $\ln(1+x) \leq x$ for $x \geq 0$ as well as $H_n := \sum_{i = 1}^n \frac 1i ≤ 1 + \ln n$:
  \begin{align*}
    d_n
    &= \prod_{i=1}^n \left( 1 + \frac{1}{2i-1} \right)
    = \textrm{exp}\left(\sum_{i=1}^n \ln\left(1 + \frac{1}{2i-1}\right)\right)
    \leq \textrm{exp}\left(\sum_{i=1}^n \frac{1}{2i-1}\right)\\
    &= \textrm{exp}\left(1 + \sum_{i=2}^n \frac{1}{2i-1}\right)
    = \textrm{exp}\left(1 + \frac{1}{2} \sum_{i=2}^n \left(\frac{1}{2i-1} + \frac{1}{2i-1}\right)\right)\\
    &\leq \textrm{exp}\left(1 + \frac{1}{2} \sum_{i=2}^n \left(\frac{1}{2i-1} + \frac{1}{2i-2}\right)\right)
    = \textrm{exp}\left(1 + \frac{1}{2} \sum_{i=2}^{2n-1} \frac{1}{i}\right)\\
    &= \textrm{exp}\left((1+H_{2n-1})/2\right)
    \leq \textrm{exp}\left(1 + \ln(2n)/2\right)
    \leq e·\sqrt{2n}.
    \tag*{\qedhere}
  \end{align*}
\end{proof}

Let us now re-state the known result proven in RecSplit \cite{esposito2020recsplit}, which bounds the probability that a random function is minimal perfect.
We will use this later when proving \cref{lem:constructionTries}.

\begin{lemma}[see \cite{esposito2020recsplit}]
  \label{lem:bruteForce}
  A random function $h: S \rightarrow [n]$ on a set $S$ of $n$ keys, is minimal perfect
  (i.e.\ is a bijection)
  with probability $e^{-n}\sqrt{2 \pi n}\cdot (1+o(1))$.
\end{lemma}
\begin{proof}
  Given $S$, there are $n^n$ possible functions from $S$ to~$[n]$ and $n!$ of them are bijective.
  Therefore, the probability that a randomly selected function is minimal perfect is $n!/n^n$.
  The claim is obtained by applying Stirling's approximation.
  \hfill
\end{proof}

Finally, we can derive the success probability of the ShockHash search as follows.

\begin{theorem}
  \label{lem:constructionTries}
  Let $h₀,h₁ : S → [n]$ be uniformly random functions. The probability that there exists $f : S → \{0,1\}$ such that $x ↦ h_{f(x)}(x)$ is bijective is at least $(e/2)^{-n} e^{-1} \sqrt{\pi}$.
\end{theorem}
\def\perf{\mathrm{perf}}
\begin{proof}
  Recall our shorthand $\orient(f)$ for the event that $x ↦ h_{f(x)}(x)$ is bijective.
  As given in \cref{lem:expectation}, we can calculate the success probability as follows.
  \[
    ℙ(∃f: \orient(f)) = \frac{𝔼(\#\{f : \orient(f)\})}{𝔼(\#\{f : \orient(f)\} \mid ∃f: \orient(f)).}
  \]
  We will consider the numerator and denominator in turn.

  \myparagraph{Numerator: Expectation.}
  Linearity of expectation (holding even for dependent variables) yields
  \begin{gather*}
    𝔼(\#\{f : \orient(f)\}) = \sum_{f} ℙ(\orient(f))
    = \sum_{f} ℙ(x ↦ h_{f(x)}(x) \text { is bijective on $S$}).
  \end{gather*}
  For any fixed $f$, the function $x ↦ h_{f(x)}(x)$ assigns independent random numbers to each $x ∈ S$, i.e.\ is a random function as considered in \cref{lem:bruteForce} and hence bijective with probability $e^{-n}\sqrt{2 \pi n}·(1+o(1))$. We therefore get
  \begin{align}
    \label{eq:numOfOrientations}
    \mathds{E}\left(\# \{f : \orient(f)\} \right) ≥ 2^n \cdot e^{-n}\sqrt{2 \pi n}.
  \end{align}

  \myparagraph{Denominator: Conditional Expectation.}
  Using observations (\ref{eq:pseudoforests-orientable}) and (\ref{eq:orientations-cycles}) we can shift our attention to the graph $G$:
  \[
    𝔼(\#\{f : \orient(f)\} \mid ∃f: \orient(f))
    = 𝔼(2^{c(G)} \mid \PF(G)).
  \]
  By virtue of \cref{lem:probability-spaces} we can moreover move to a configuration model à la $G₃$.
  We first reveal the locations $x₁,…,x_{2n}$ of the $2n$ stubs (hence the degree sequence of $G₃$) and then consider the following peeling process (see \cref{ss:graphPeelingPrelim}) that reveals edges of $G₃$ and simplifies $G₃$ in a step-by-step fashion.

  As long as there exists a node $v$ with only one stub, firstly, match it to a random stub to form a corresponding edge $\{v,w\}$ (consuming the two stubs) and, secondly, remove the node $v$ and the newly formed edge $\{v,w\}$. These removals do not affect the number of components of the resulting graph (since $v$ was connected to $w$), nor whether the resulting graph is a pseudoforest (since the component of $w$ lost one node and one edge).

  Let $n'$ be the number of nodes that remain after peeling and let $G'$ be the graph obtained by matching the remaining stubs. As discussed we have $\PF(G₃) ⇔ \PF(G')$ and $c(G') = c(G₃)$. Since the average degree of $G₃$ is $2$ and since we removed one node and one edge in every round, the average degree of $G'$ is also $2$. There are two cases.

  (1) Some node of $G'$ has degree $0$. Then $¬\PF(G')$ because some component of $G'$ must have average degree $> 2$.

  (2) No node of $G'$ has degree $0$. Since we ran the peeling process, there is also no node of $G'$ with degree $1$. Hence, every node of $G'$ has degree $2$. This makes $G'$ a collection of cycles. In particular $\PF(G')$ holds. Moreover, the generation of $G'$ is precisely the situation discussed in \cref{lem:numberOfOrientations}.

  Because the two cases imply opposite results on $G'$ being a pseudoforest, we know that $\PF(G')$ holds if \emph{and only if} we arrive in Case 2.
  While we have no understanding of the distribution of $n'$, we can nevertheless compute:
  \begin{align}
    \label{eq:conditionalPseudotree}
    𝔼(2^{c(G)} &\mid \PF(G)) = 𝔼(2^{c(G₃)} \mid \PF(G₃))
    = 𝔼(2^{c(G')} \mid \PF(G')) = 𝔼(2^{c(G')} \mid \text{Case 2})\notag\\
    &≤ \max_{1 ≤ i ≤ n} 𝔼(2^{c(G')} \mid \text{Case 2 with $n' = i$})
    ≤ \max_{1 ≤ i ≤ n} e \sqrt{2i} = e \sqrt{2n}.
  \end{align}
  \myparagraph{Putting the Observations Together.}
  Combining our bounds on the numerator (\ref{eq:numOfOrientations}) and the denominator (\ref{eq:conditionalPseudotree}) gives the final result
  \begin{align*}
    ℙ(∃f: \orient(f))
      &≥ 2^n e^{-n}\sqrt{2 \pi n} / (e \sqrt{2n})
      = (e/2)^{-n}e^{-1}\sqrt{\pi}.
    \tag*{\qedhere}
  \end{align*}
\end{proof}

\subsection{Success Probability in Bipartite ShockHash}\label{ss:successProbabilityBipartite}
In the bipartite case, we can basically perform the same steps as in the non-bipartite version.
For simplicity, we restrict ourselves to even $n$ in the analysis.
While our implementation does support uneven $n$ (see \cref{ss:unevenN}), this complicates the analysis and can be largely avoided when ShockHash is integrated into a partitioning framework like RecSplit (see \cref{s:partitioning}).
In this section, we again suppress the seed of the hash functions.
Let $h_0,h_1 : S \rightarrow [n/2]$ be hash functions and $f: S \rightarrow \{0, 1\}$ be a function that selects between the two hash functions.

We now look at the effect of testing all correlated choices of the function $f$.
As argued in \cref{ss:successProbability}, we start with peeling the corresponding graph until there are no nodes with degree $1$ left.
Conditioned on the graph being a pseudoforest, this leaves us with a graph where each node has degree $2$.
The distribution of this graph is captured again by a configuration model (see \cref{ss:configurationModel}), namely giving a random bipartite matching between the stubs.
Additionally, remember that we started with a bipartite graph, so both partitions have the same size.
Similar to \cref{lem:numberOfOrientations}, we can now show in \cref{lem:components} that the number of components $c$ in the remaining graph satisfies $\mathds{E}(2^{c}) \leq e\cdot\sqrt{n}$.
Because the peeling process does not change the number of components, the same applies also to the original bipartite graph.
This gives us a bound for the expected number of orientations of the graph, e.g., the number of different functions $f$ that all make the hash function pair $(h_0,h_1)$ minimal perfect.
\begin{lemma}
  Let $n$ be an even number, and let $G_n$ be a random bipartite graph with $n/2$ nodes in each partition, where all nodes have degree $2$, sampled from the corresponding \emph{bipartite} configuration model.
  Hence, the stubs from one partition are matched to the stubs of the other partition uniformly at random.
  Then the number $c(G_n)$ of components of $G_n$ satisfies $\mathds{E}(2^{c(G_n)}) \leq e\cdot\sqrt{n}$.
  \label{lem:components}
\end{lemma}
\begin{proof}
  We will find a recurrence for $d_n := \mathds{E}(2^{c(G_n)})$.
  Consider an arbitrary node $v$ from the first partition of $G_n$ and one of the stubs at $v$.
  Because the graph is bipartite, this stub forms an edge to a node $r$ from the other partition of the graph.
  The node $r$ has a second stub that is connected back to the first partition.
  We now have $n/2-1$ other nodes in the first partition, each with $2$ stubs, and we have the second stub at $v$.
  Each of these $n-1$ stubs is matched with equal probability.
  Therefore, the probability that this edge closes a cycle is $\frac{1}{n-1}$.

  (1) Conditioned on closing the cycle, the distribution of the remaining graph is that of $G_{n-2}$ and the conditional expectation of $2^{c(G_n)}$ is therefore $\mathds{E}(2^{1+c(G_{n-2})}) = 2d_{n-2}$.

  (2) Now condition on the edge not closing a cycle.
  We can now merge the three considered nodes to a single one without affecting the number of components.
  The merged node inherits two unused stubs, and the graph is now bipartite with $n/2-1$ nodes in each partition.
  The distribution of the remaining graph therefore is that of $G_{n-2}$.
  Therefore, the conditional expectation of $2^{c(G)}$ is simply $d_{n-2}$.

  These two cases are similar to the non-bipartite case illustrated in \cref{fig:orientations} and lead us to the following recurrence:
  \begin{align*}
    d_{n}=\tfrac{1}{n-1}2d_{n-2}+\big(1-\tfrac{1}{n-1}\big)d_{n-2}=\big(1+\tfrac{1}{n-1}\big)d_{n-2}.
  \end{align*}
  With the base case $d_0=1$, we can solve the recurrence and bound its value as follows, using that $\ln(1+x) \leq x$ for $x \geq 0$ as well as $H_n := \sum_{i = 1}^n \frac 1i \leq 1 + \ln n$:
  \begin{align*}
    d_{n}
    &= \prod_{i=1}^{n/2} \left( 1 + \frac{1}{2i-1} \right)
    = \textrm{exp}\left(\sum_{i=1}^{n/2} \ln\left(1 + \frac{1}{2i-1}\right)\right)
    \leq \textrm{exp}\left(\sum_{i=1}^{n/2} \frac{1}{2i-1}\right)\\
    &= \textrm{exp}\left(1 + \sum_{i=2}^{n/2} \frac{1}{2i-1}\right)
    = \textrm{exp}\left(1 + \frac{1}{2}\sum_{i=2}^{n/2} \left(\frac{1}{2i-1} + \frac{1}{2i-1}\right)\right)\\
    &\leq \textrm{exp}\left(1 + \frac{1}{2}\sum_{i=2}^{n/2} \left(\frac{1}{2i-1} + \frac{1}{2i-2}\right)\right)
    = \textrm{exp}\left(1 + \frac{1}{2}\sum_{i=2}^{n-1} \frac{1}{i}\right)\\
    &= \textrm{exp}\left((1+H_{n-1})/2\right)
    \leq \textrm{exp}\left(1 + \ln(n)/2\right)
    \leq e\cdot\sqrt{n}.
    \tag*{\qedhere}
  \end{align*}
\end{proof}

We can lower bound the success probability by applying \cref{lem:expectation} similarly as in \cref{lem:constructionTries}.
In contrast to \cref{lem:constructionTries}, we now use $\orient(f)$ adapted for the bipartite case.
Given two hash functions $h₀,h₁: S → [n/2]$ and a function $f: S → \{0,1\}$, let $\orient(f)$ be the event that $x ↦ h_{f(x)}(x) + f(x)\cdot\frac{n}{2}$ is bijective.
We write $\PF(h_0, h_1)$ for the event that the graph defined by the two hash functions $h_0$ and $h_1$ is a pseudoforest, and again get $\PF(h_0, h_1) ⇔ ∃f: \orient(f)$.

\begin{theorem}
  \label{lem:constructionTriesBipartite}
  Let $h₀,h₁ : S → [n/2]$ be uniformly random functions. The probability that there exists $f : S → \{0,1\}$ such that $x ↦ h_{f(x)}(x) + f(x)\cdot\frac{n}{2}$ is bijective is at least $(e/2)^{-n}\sqrt{n}/e$.
\end{theorem}
\begin{proof}
  All bipartite ShockHash functions have the form $(x \mapsto h_{f(x)}(x) + f(x)\cdot(n/2))$.
  While it is clear that the results of different $x$ are independent, let us first justify why the function is uniform.
  For this, let $c \in [n]$ be a constant, $x$ an input value, and $g: S \rightarrow \{0,1\}$ a uniform random function.
  Then we get
  \begin{align*}
    \hspace{-3mm}
    \Pr&(h_{g(x)}(x) + g(x)\cdot\tfrac{n}{2} = c)
      = \left\{\begin{array}{lr}
          \Pr(h_0(x) = c \wedge g(x) = 0),     & c < \frac{n}{2} \\
          \Pr(h_1(x) = c-\frac{n}{2} \wedge g(x) = 1) & c \geq\frac{n}{2}
        \end{array}\right\}
      = \frac{1}{n/2} \cdot \frac{1}{2}
      = \frac{1}{n}.
  \end{align*}

  For uniform random functions $g$ it holds that $𝔼(\#\{f : \orient(f)\})=2^n \cdot \Pr(\orient(g))$.
  Because we now also know that bipartite ShockHash with random $g$ gives a uniform random function, we know that $\Pr(\orient(g))$ matches the probability that a random function is a bijection.
  Applying \cref{lem:bruteForce}, this gives $𝔼(\#\{f : \orient(f)\}) \geq 2^n \cdot e^{-n}\sqrt{2 \pi n}$.

  To determine the overall success probability, we can now continue similar to \cref{lem:constructionTries}.
  Therefore, a random pair of hash functions $h_0,h_1$ permits at least one valid placement $f$ with at least the following probability.
  \begin{align*}
    \Pr(\exists f: \orient(f))\quad
      &\refRelLem{lem:expectation}{=} \frac{𝔼(\#\{f : \orient(f)\})}{𝔼(\#\{f : \orient(f)\} \mid ∃f: \orient(f))}
      \geq \frac{2^n \cdot e^{-n}\sqrt{2 \pi n}}{𝔼(\#\{f : \orient(f)\} \mid ∃f: \orient(f))}\\
      &\refRelLem{lem:components}{\geq} e^{-n}\sqrt{2 \pi n} \cdot 2^n / (e\cdot\sqrt{n})
      = (e/2)^{-n}\sqrt{n}/e
    \tag*{\qedhere}
  \end{align*}
\end{proof}

Our bound is a factor of $\sqrt{2}$ better than with plain ShockHash, which reduces our bound on the reduced expected space usage by $\log_2(\sqrt{2})=0.5$ bits.
It might be possible to save a few additional bits in the retrieval data structure because $f$ is known to map exactly half of the keys to $1$, i.e., only a $1/\Theta(\sqrt{n})$ fraction of all functions can occur as $f$.
However, we do not consider this in more detail here.

\subsection{Hash Function Pools}\label{ss:pooling}
An important ingredient for making bipartite ShockHash so much more efficient than the plain version is the fact that we can combine hash functions from a pool of candidates.
This makes it possible to filter the candidates \emph{before} combining them.
In the following section, we show why testing all combinations of two hash functions from a pool of candidates has a similar success probability as always sampling two fresh functions.
For this analysis, let us form \emph{two} pools of hash functions of size $k=(e/2)^{n/2}$.
Note that this is slightly different to the construction explained before, which uses a single, growing pool.
However, this does not influence the asymptotic construction time.%
\footnote{%
  One could view the two pools as one single pool of twice the size, where we only test a subset of the combinations.
  The time for testing more seed combinations would then be a constant factor larger.
  In practice, however, we test all combinations within the single pool, so it does not actually need to be twice as large.
}
The first pool $(l_i)_{i\in[k]}$ contains uniformly drawn hash functions $l_1,\ldots,l_k$ and the second pool $(r_i)_{i\in[k]}$ contains uniformly drawn hash functions $r_1,\ldots,r_k$.
Our algorithm tests all combinations between two hash functions $l_i$ and $r_j$ $(i,j \in [k])$ from the pools.
We are therefore interested in the probability that the pools contain two hash functions that are \emph{compatible}, meaning that their combined graph is a pseudoforest.
For easier presentation, we suppress polynomial factors in this section and assume large $n$.

\def\PF{\mathrm{PF}}
\def\uniform{\mathcal{U}}
\def\expect{\mathbb{E}}
\def\H{\mathcal{H}}
\def\HProm{\mathcal{H}_{\textrm{promisc}}}
  \def\B{\mathcal{B}}

\subsubsection{Notation and Overview}
Let $\H := [n/2]^S$ be the set of all (hash) functions from $S$ to $[n/2]$.
Sampling two hash functions $l,r \sim \uniform(\H)$ uniformly at random gives $\Pr(\PF(l, r)) = (e/2)^{-n}$, as shown in \cref{lem:constructionTriesBipartite} (ignoring polynomial factors).
In the following, it will be useful to also consider $\HProm$ containing \emph{promiscuous} functions.
The intuition is that a promiscuous function has a very large number of compatible partners, which is very unlikely.
For reasons we will discuss later, $\HProm$ is defined as the set of functions that hit more than $90\%$ of the hash values exactly two times:
\begin{align*}
  \HProm \coloneqq \left\{ f \in \H \cond \left|\left\{i \in [n/2] \cond |f^{-1}(i)| = 2\right\}\right| \geq 0.9n/2\right\}.
\end{align*}

For a specific function $l$, we define the set $C(l)$ as all hash functions that are compatible with $l$.
We also define $C^*(l)$ as only those compatible functions that are not promiscuous:
\begin{align*}
  C(l) \coloneqq \{ r \in \H \mid \PF(l, r)\},
  \hspace{5mm}
  C^*(l) \coloneqq C(l) \setminus \HProm.
\end{align*}

For a set $X \subseteq \H$ of hash functions, let $\|X\| \coloneqq |X|/|\H|$.
This is the probability that a randomly sampled hash function is one of the hash functions in $X$.
Then $\|C(l)\|$ is the probability that a randomly sampled hash function is compatible with $l$.
If $l$ is a random variable, this probability is a random variable as well.
The expected value of this variable for uniform random $l$ is $\expect_{l \sim \uniform(\H)}(\|C(l)\|) = \Pr_{l,r \sim \uniform(\H)}(\PF(l, r)) = (e/2)^{-n}$ (ignoring polynomial factors).

The main random variable we are interested in is $Z \coloneqq \|\bigcup_{i \in [k]} C(l_i)\|$.
It depends on the hash functions that we sampled for the pool.
$Z$ is the probability that a randomly sampled hash function is compatible with at least one function from our pool $(l_i)_{i\in[k]}$.

If $Z$ was small, it would mean that the hash functions in our pool need a very specific set of partners.
Therefore, in this case, it would be unlikely that one of the compatible partners was drawn for the pool $(r_i)_{i\in[k]}$.
However, we will show that $Z$ is large enough that $(r_i)_{i\in[k]}$ likely contains a compatible partner.
More specifically, we will show that $Z$ is closely concentrated around $(e/2)^{-n/2}$.
Since we have $k=(e/2)^{n/2}$ hash functions in each pool, we get a constant probability that a compatible function is in $(r_i)_{i\in[k]}$.

We start our proof in \cref{s:promiscuous}, showing that it is unlikely that a hash function is promiscuous.
We then show in \cref{sec:peelingBipartite} that functions $\not\in\HProm$ do not have too many compatible partners.
This is a key ingredient for providing concentration bounds on $Z$ in \cref{s:concentrationBounds}.
Finally, we combine this with the pool $(r_i)_{i\in[k]}$ in \cref{s:combining}.
\Cref{fig:proofStructure} gives an overview over the proof structure.

\def\constantMaxC{c_1}
\def\constantD{c_2}
\def\constantZHat{c_3}
\def\constantZ{c_4}

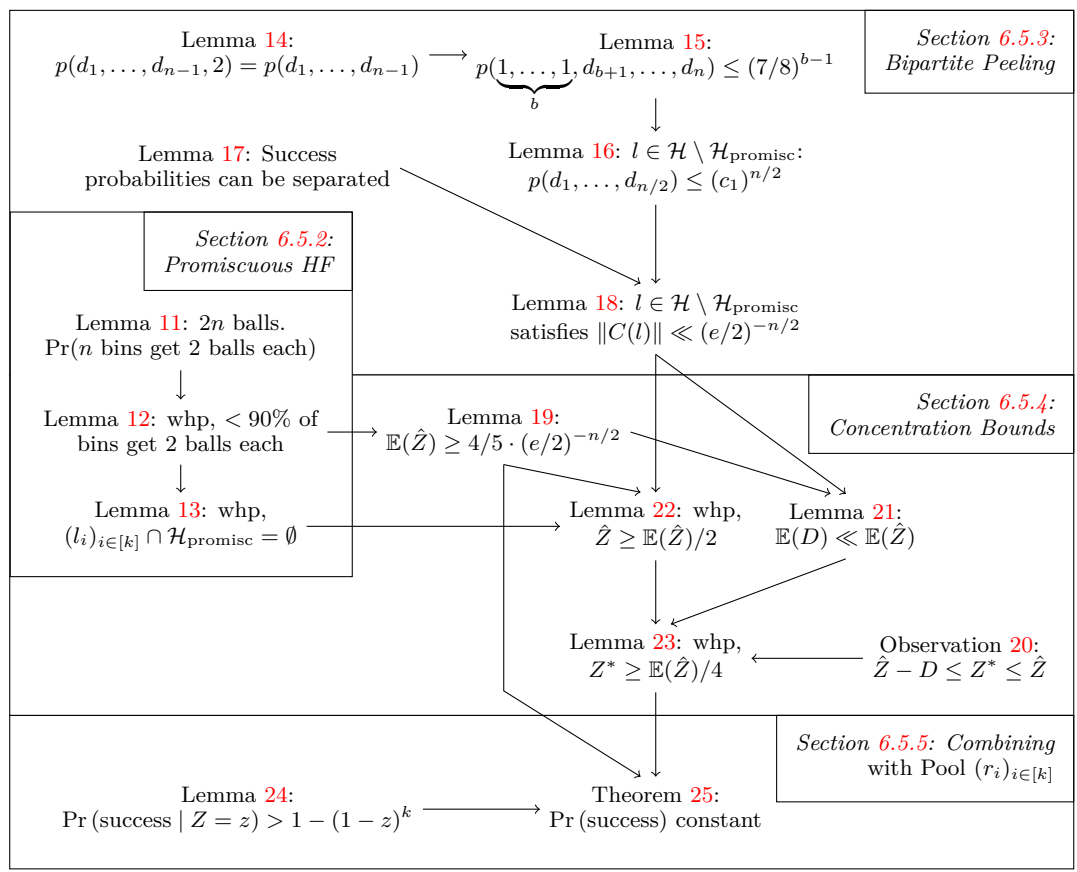
\begin{figure}[t]
  \centering
  \usetikzlibrary{positioning}
  \begin{tikzpicture}[every node/.style={font={\footnotesize}}]

    \node[style=rectangle,minimum width=14cm,text depth=4.3cm,text height=3mm,anchor=south,align=center,draw=black,text=black,fill=white] (largeBoxPeeling) at (4.5cm, 1cm) {~};
    \node[anchor=north east,align=right,text=black,draw=black,inner sep=0.25cm] (e) at (largeBoxPeeling.north east) {\it\cref{sec:peelingBipartite}:\\\it Bipartite Peeling};
    \node[align=center] (peelingDegree2) at (0.5cm, 5.25cm) {\cref{lem:peelingDegree2}:\\$p(d_1, \ldots, d_{n-1},2) = p(d_1,\ldots,d_{n-1})$};
    \node[align=center] (peelingDegree1) at (6cm, 5.25cm) {~\\~\\\cref{lem:peelingDegree1}:\\$p(\underbrace{1, \ldots, 1}_b, d_{b+1}, \ldots, d_{n}) \leq (7/8)^{b-1}$};
    \node[align=center] (nodeDegree1) at (6cm, 3.75cm) {\cref{lem:nodeDegree1}: $l \in \H\setminus\HProm$:\\$p(d_1, \ldots, d_{n/2}) \leq (\constantMaxC)^{n/2}$};
    \node[align=center] (peelingSeparate) at (0.5cm, 3.75cm) {\cref{lem:peelingSeparate}: Success\\probabilities can be separated};
    \node[align=center] (maxC) at (6cm, 1.75cm) {\cref{lem:maxC}: $l \in \H\setminus\HProm$\\satisfies $\|C(l)\| \ll (e/2)^{-n/2}$};

    \node[style=rectangle,minimum width=14cm,text depth=3.98cm,text height=3mm,anchor=north,align=center,draw=black,text=black,fill=white] (largeBoxConcentration) at ($(largeBoxPeeling.south)+(0, \the\pgflinewidth)$) {~};
    \node[anchor=north east,text=black,align=right,draw=black,inner sep=0.25cm] (e) at (largeBoxConcentration.north east) {\it\cref{s:concentrationBounds}:\\\it Concentration Bounds};
    \node[align=center] (upperLower) at (10cm, -2.75cm) {\cref{cor:upperLower}:\\$\hat Z-D \leq Z^* \leq \hat Z$};
    \node[align=center] (smallD) at (8.5cm, -1cm) {\cref{lem:smallD}:\\$\expect(D) \ll \expect(\hat Z)$};
    \node[align=center] (sConcentration) at (6cm, -1cm) {\cref{lem:sConcentration}: whp,\\$\hat Z \geq \expect(\hat Z)/2$};
    \node[align=center] (largeZ) at (6cm, -2.75cm) {\cref{lem:largeZ}: whp,\\$Z^* \geq \expect(\hat Z)/4$};
    \node[align=center] (largeZ2) at (4.0cm, 0.25cm) {\cref{lem:largeZ2}:\\$\expect(\hat Z) \geq 4/5 \cdot (e/2)^{-n/2}$};

    \node[style=rectangle,minimum width=14cm,text depth=1.5cm,text height=3mm,anchor=north,align=center,draw=black,text=black,fill=white] (largeBoxPool) at ($(largeBoxConcentration.south)+(0, \the\pgflinewidth)$) {~};
    \node[anchor=north east,align=right,text=black,draw=black,inner sep=0.25cm] (e) at (largeBoxPool.north east) {\it\cref{s:combining}: Combining\\with Pool $(r_i)_{i\in[k]}$};
    \node[align=center] (poolSuccessConditioned) at (0.5cm, -4.75cm) {\cref{lem:poolSuccessConditioned}:\\$\Pr\left(\textrm{success} \cond Z = z \right) > 1-(1-z)^k$};
    \node[align=center] (poolSuccess) at (6cm, -4.75cm) {\cref{thm:poolSuccess}:\\$\Pr\left(\textrm{success}\right)$ constant};

    \node[style=rectangle,minimum width=4.5cm,text depth=4.0cm,text height=6mm,anchor=west,align=center,draw=black,text=black,fill=white] (largeBoxPromisc) at (-2.5cm, 0.75cm) {~};
    \node[anchor=north east,align=right,text=black,draw=black,inner sep=0.25cm] (e) at (largeBoxPromisc.north east) {\it\cref{s:promiscuous}:\\\it Promiscuous HF};
    \node[align=center] (perfectBallsBins) at (-0.25cm, 1.5cm) {\cref{lem:perfectBallsBins}: $2n$ balls.\\$\Pr(n$ bins get $2$ balls each$)$};
    \node[align=center] (nodeDegree2) at (-0.25cm, 0.25cm) {\cref{lem:nodeDegree2}: whp, $<90\%$ of\\bins get $2$ balls each};
    \node[align=center] (eventPoolNodeDegree2) at (-0.25cm, -1.0cm) {\cref{lem:eventPoolNodeDegree2}: whp,\\$(l_i)_{i\in[k]} \cap \HProm = \emptyset$};

    \draw[->] (perfectBallsBins.south) -- (nodeDegree2.north);
    \draw[->] (nodeDegree2.east) -- (largeZ2.west);
    \draw[->] (nodeDegree2.south) -- (eventPoolNodeDegree2.north);
    \draw[->] (nodeDegree1.south) -- (maxC);
    \draw[->] (peelingDegree2.east) -- (peelingDegree1.west);
    \draw[->] ($(peelingDegree1.south)+(0,0.3)$) -- (nodeDegree1.north);
    \draw[->] (peelingSeparate.east) -- ($(maxC.north)-(0.2,0)$);
    \draw[->] (eventPoolNodeDegree2.east) -- (sConcentration.west);
    \draw[->] (maxC.south) -- (sConcentration.north);
    \draw[->] (maxC.south) -- (smallD.north);
    \draw[->] (smallD.south) -- ($(largeZ.north)+(0.2,0)$);
    \draw[->] (sConcentration.south) -- (largeZ);
    \draw[->] (upperLower.west) -- (largeZ.east);
    \draw[->] (poolSuccessConditioned.east) -- (poolSuccess.west);
    \draw[->] (largeZ.south) -- (poolSuccess.north);
    \draw[->] (largeZ2.south) -- ($(largeZ2.south)-(0,3)$) -- ($(poolSuccess.north)-(0.2,0)$);
    \draw[->] (largeZ2.east) -- ($(smallD.north)-(0.2,0)$);
    \draw[->] (largeZ2.south) -- ($(sConcentration.north)-(0.2,0)$);
  \end{tikzpicture}
  \caption{%
    Illustration of the proof structure showing the success probability of sampling from hash function pools instead of independent hash functions.
    Variables $\hat Z$, $D$ and $Z^*$ are defined in \cref{s:concentrationBounds}.
    Functions $p$ and $q$ are defined in \cref{sec:peelingBipartite}.
  }
  \label{fig:proofStructure}
\end{figure}

\subsubsection{Promiscuous Hash Functions}\label{s:promiscuous}
Before being able to give concentration bounds, we have to rule out a special case of hash functions with too many compatible partners.
As stated before, we call these \emph{promiscuous}.
In this section, we show that these functions are very rare.
To show this, we interpret the output values of our hash functions as a balls-into-bins process.
In the following, we show a general property of balls-into-bins processes that we then later apply to our hash functions.

\begin{lemma}
  When throwing $2n$ balls into $n$ bins independently and uniformly at random, the probability that each bin receives exactly two balls is $p_n \leq 5 \sqrt{n}(2e^{-2})^n$.
  \label{lem:perfectBallsBins}
\end{lemma}
\begin{proof}
  To assign the balls to the bins such that each bin receives exactly two balls, we choose $n$ subsets of size $2$ from the set of $2n$ balls.
  The number of ways we can do this is given by the multinomial coefficient ${2n \choose {2,2,\ldots,2}}$.
  Each of these combinations has a probability of $n^{-2n}$.
  Using Stirling's approximation in the step annotated with $*$, this gives
  \begin{align*}
    p_n
      &= {2n \choose \underbrace{2,2,\ldots,2}_\textrm{$n$ times}} \cdot n^{-2n}
       = \frac{(2n)!}{\underbrace{2 \cdot 2 \cdot \ldots \cdot 2}_\textrm{$n$ times}}n^{-2n}
       \overset{*}{\leq} \sqrt{2\pi 2n}\left(\frac{2n}{e}\right)^{2n}e^{\frac{1}{24n}}\cdot 2^{-n} \cdot n^{-2n}\\
      &=\sqrt{4\pi n} \cdot (2e^{-2})^n \cdot e^{\frac{1}{24n}}
      \leq 5 \sqrt{n}(2e^{-2})^n.
    \tag*{\qedhere}
  \end{align*}
\end{proof}

\begin{lemma}
  When throwing $n$ balls into $n/2$ bins independently and uniformly at random, the probability $p_{90\%}$ that more than $90\%$ of the bins receive exactly $2$ balls is $p_{90\%} \leq 0.66^n$.
  \label{lem:nodeDegree2}
\end{lemma}
\begin{proof}
  Let us consider a set $A$ of $0.9 \cdot n/2$ bins that will each receive exactly $2$ balls each.
  We do not care \emph{which} bins receive exactly $2$ balls, so there are ${n/2 \choose 0.9 \cdot n/2} = {n/2 \choose 0.1 \cdot n/2}$ ways of selecting the set $A$.
  The probability that exactly $2|A|$ balls land in a bin in $A$ is ${n \choose 0.1n} 0.1^{0.1n} 0.9^{0.9n}$.
  Conditioned on this, the probability that these $2|A|$ balls are evenly distributed among the $|A|$ bins is $p_{|A|} \leq 5 \sqrt{0.9 \cdot n/2} \left(2e^{-2}\right)^{0.9 \cdot n/2}$ by \cref{lem:perfectBallsBins}.
  Bringing this together we can bound $p_{90\%}$ as follows, applying \cref{lem:binomial}.
  \begin{align*}
    p_{90\%}
      &\leq {n \choose 0.1n} 0.1^{0.1n} 0.9^{0.9n} \cdot {n/2 \choose 0.1 \cdot n/2} \cdot \left(2e^{-2}\right)^{0.9 \cdot n/2} 5 \sqrt{0.9 \cdot n/2} \\
      &\leq (10e)^{0.1n} \cdot 0.1^{0.1n} 0.9^{0.9n} \cdot (10e)^{0.1\cdot n/2} \cdot \left(2e^{-2}\right)^{0.9 \cdot n/2} 5 \sqrt{0.45n}  \\
      &= \left((10e)^{0.1} \cdot 0.1^{0.1} 0.9^{0.9} \cdot (10e)^{0.05} \cdot \left(2e^{-2}\right)^{0.45} \right)^{n} 5 \sqrt{0.45n}\\
      &< 0.66^{n} \textrm{ for $n$ large enough.}
    \tag*{\qedhere}
  \end{align*}
\end{proof}

\begin{lemma}
  With a probability of $1 - 0.77^n$, none of the hash functions in the pool $(l_i)_{i\in[k]}$ is promiscuous.
  \label{lem:eventPoolNodeDegree2}
\end{lemma}
\begin{proof}
  Each of our hash functions maps the $n$ keys to $n/2$ nodes in the graph.
  Because the hash functions are sampled at random, this can be modeled as a balls-into-bins process.
  We can apply \cref{lem:nodeDegree2}, telling us that the probability that a function is promiscuous is $p_{90\%} \leq 0.66^n$.
  We sample a pool of $k=(e/2)^{n/2}$ hash functions.
  Therefore, the probability that at least one of our hash functions is promiscuous can be upper bounded as follows using a union bound.
  \begin{align*}
    \Pr\left(\bigvee_{i\in[k]} l_i \in \HProm\right)
      \leq k \cdot p_{90\%}
      = ((e/2)^{1/2} \cdot 0.66)^n
      < 0.77^n
    \tag*{\qedhere}
  \end{align*}
\end{proof}

\subsubsection{Peeling Bipartite Graphs} \label{sec:peelingBipartite}
For plain ShockHash, we have looked at the event $\PF(G)$ that the resulting graph $G$ is a pseudoforest, and have given bounds for $\Pr(\PF(G))$.
In \cref{ss:graphPeelingPrelim}, we have shown that we can uncover the graph in two steps: we first reveal the degree of each node and then match these stubs at random.
Staying with plain ShockHash for the moment, we now condition on the degrees of the nodes.
For this, let $(d_1,\ldots,d_n)$ be the degrees of each node, and let $p(d_1, \ldots, d_n) \coloneqq \Pr(\PF(G) \mid \textrm{nodes of $G$ have degrees $d_1, \ldots, d_n$})$ be the conditioned success probability.
Note that the order of the function arguments does not matter to the success probability because the stubs are matched randomly.
Let $\B_n$ be the distribution of balls in the balls-into-bins process with $n$ balls and $n/2$ bins.
Before we give additional properties of $p(d_1, \ldots, d_n)$ in the following lemmas, let us look at a simple observation about the function:
\begin{align}
  \expect_{(d_1, \ldots, d_n) \sim \B_{2n}}(p(d_1, \ldots, d_n))
    = \Pr(\PF(G))
    \quad\refRelLem{lem:constructionTries}{=} (e/2)^{-n}.
  \label{obs:expectP}
\end{align}

The graph $G$ is a pseudoforest if we can repeatedly peel away nodes of degree $1$ and end up with a graph that consists of only nodes of degree $2$ (forming cycles).
When peeling, we repeatedly take a node of degree $1$ and follow its edge to a random stub.
There are now three possible outcomes.

(1) The edge leads to a node with degree $1$.
Then we have found a component with more nodes than edges, meaning that the remaining graph cannot be a pseudoforest.

(2) The edge leads to a node with degree $\geq 3$.
Then we have peeled away a subtree of a component that is potentially still a pseudotree.
We now have one less node of degree $1$ in our graph.
In that case, we continue the peeling process with another node of degree $1$.

(3) The edge leads to a node with degree $2$.
Then it generates a new node with degree $1$, and we can continue the peeling process with that same node.

\begin{lemma}
  Nodes of degree $2$ do not influence the success probability.
  More formally, $p(d_1, \ldots, d_{i},2) = p(d_1,\ldots,d_{i})$.
  \label{lem:peelingDegree2}
\end{lemma}
\begin{proof}
  If the peeling process arrives in case (3) where it connects to a node of degree $2$, we can immediately continue peeling with that node.
  This always succeeds and does not cause an abort of the peeling process.
  Afterwards, we have one less node but did not influence the degrees of all other nodes.
  Therefore, the probabilities to arrive in the more interesting cases (1) and (2) stay the same, relative to each other.
\end{proof}

\begin{lemma}
  If the graph corresponding to our hash function has $b$ nodes of degree $1$, $p$ is exponentially small in $b$.
  More formally, $p(\underbrace{1, \ldots, 1}_b, d_{b+1}, \ldots, d_{n}) \leq (7/8)^{b-1}$.
  \label{lem:peelingDegree1}
\end{lemma}
\begin{proof}
  During the peeling process, we can ignore all nodes of degree $2$ because they do not influence the success probability (see \cref{lem:peelingDegree2}).
  Let us therefore now look at the remaining nodes.
  If we connect to one of the nodes of degree $1$, we have found a tree.
  This means that the construction fails because we need each component to be a pseudotree and not a tree.
  Because the average degree is $2$, we have at most $3b$ stubs at nodes with degree $\geq 3$.
  Therefore, the probability that we connect to a node of degree $1$ (possibly indirectly through nodes of degree $2$) is at most $(b-1)/(4b)$.
  We can now apply this iteratively until all nodes of degree $1$ are peeled away.
  This gives the following probability that we never connect to a node of degree~1 (and therefore fail):
  \begin{align*}
    p(\underbrace{1, \ldots, 1}_b, d_{b+1}, \ldots, d_{n})
      &\leq \prod_{j=1}^b\left(1 - \frac{j-1}{4j}\right)
      = \prod_{j=1}^b\left(\frac{3}{4} + \frac{1}{4j}\right) \\
      &= \prod_{j=2}^b\left(\frac{3}{4} + \frac{1}{4j}\right)
      \leq \prod_{j=2}^b\left(\frac{7}{8}\right)
      = \left(\frac{7}{8}\right)^{b-1}.
    \tag*{\qedhere}
  \end{align*}
\end{proof}

We call a hash function promiscuous if more than 90\% of the nodes in its corresponding (stub) graph have degree 2.
This means that it cannot have too many nodes of degree $1$.
Promiscuous hash functions are rare, as we have seen in \cref{s:promiscuous}.
In the following, we bound the number of compatible hash functions when our function is \emph{not} promiscuous.

\begin{lemma}
  Let $l$ be a hash function from $\H\setminus\HProm$ and $(d_1, \ldots, d_{n/2})$ be the corresponding degree sequence.
  Then $p(d_1, \ldots, d_{n/2}) \leq (\constantMaxC)^{n/2}$ where $\constantMaxC$ is a constant $\in (0, 1)$ and $n$ large enough.
  \label{lem:nodeDegree1}
\end{lemma}
\begin{proof}
  Let $X_{\not=2}$ be the set of nodes with degree $\not=2$.
  Because the functions considered here are not promiscuous, we have $|X_{\not=2}| \geq 0.1 \cdot n/2$.
  Because there are two times more stubs than nodes, the nodes in $X_{\not=2}$ receive $2$ stubs on average.
  If one of the nodes receives $0$ stubs, the success probability is $0$.
  Otherwise, at least half of the nodes in $X_{\not=2}$ have to receive exactly $1$ stub to satisfy the average.
  Therefore, $l$ has at least $b=0.05n/2$ nodes with degree~$1$.
  Applying \cref{lem:peelingDegree1} and selecting $\constantMaxC > (7/8)^{0.05} \in (0, 1)$ then concludes the proof.
\end{proof}

Now let us turn back to bipartite ShockHash.
Here we have two random hash functions $l$ and $r$ that give us a bipartite graph $G$.
Let $(d_1, \ldots, d_{n/2})$ and $(d'_1, \ldots, d'_{n/2})$ be the degrees of the nodes in the two partitions and remember that $\B_n$ is the distribution of balls in the balls-into-bins process with $n$ balls and $n/2$ bins.
Then we can define $q$ as the following probability conditioned on the degree sequence of $G$:
\begin{align*}
  q&((d_1, \ldots, d_{n/2}), (d'_1, \ldots, d'_{n/2}))\\
    &\coloneqq \Pr\left(\PF(G) \cond G \textrm{ has degree sequence } (d_1, \ldots, d_{n/2}), (d'_1, \ldots, d'_{n/2}) \right) \\
    &= \Pr\left(\PF(l, r) \cond l,r \textrm{ give degree sequence } (d_1, \ldots, d_{n/2}), (d'_1, \ldots, d'_{n/2}) \right)
\end{align*}

\begin{lemma}
  For any $d₁,…,d_{n/2},d₁',…,d_{n/2}'$ we have $q((d₁,…,d_{n/2}),(d₁',…,d_{n/2}')) ≤ p(d₁,…,d_{n/2})·p(d₁',…,d_{n/2}')$.
  \label{lem:peelingSeparate}
\end{lemma}
\begin{proof}
  Recall that the peeling process in the configuration model iteratively matches stubs that are the only remaining stub of their node.
  It can therefore be understood as the process of alternately removing a lonely stub and a random stub (and emitting a corresponding edge).
  Failure means that a randomly removed stub was lonely.
  If in the bipartite case we alternate between picking the lonely stubs on the left and on the right, then within each of the two partitions we are alternating between removing a lonely stub and a random stub.
  Therefore, we are basically running the peeling process within the two partitions separately.
  This suggests that the probability $q((d₁,…,d_{n/2}),(d₁',…,d_{n/2}'))$ that a bipartite graph is a pseudoforest when sampled from the configuration model with degree sequence $((d₁,…,d_{n/2}),(d₁',…,d_{n/2}'))$ is equal to the product of the probabilities $p(d₁,…,d_{n/2})$ and $p(d₁',…,d_{n/2}')$ that two graphs are pseudoforests, namely those sampled from the configuration model with degree sequences $(d₁,…,d_{n/2})$ and $(d₁',…,d_{n/2}')$, respectively.

  However, there is no strict equality due to a slight asymmetry:
  In the first partition we begin with removing a lonely stub while in the second partition we begin with removing a random stub.
  This slightly increases the failure probability for the second partition because there is always one lonely stub more when selecting a random stub than there would otherwise be.
  Note also that the first partition may run out of lonely stubs while the second partition has at least one lonely stub remaining.
  In that case we would remove a lonely stub from the second partition twice in a row, putting things back on track for the second partition.

  Overall, the increase in failure probability for the second partition during (parts of) the process is accounted for by the “$≤$” in our statement.  
\end{proof}

Both $p(d_1,\ldots,d_{n/2})$ and $\|C(l)\|$ provide a way to rate the quality of a hash function candidate.
However, there is a subtle but important difference between the two.
While $\|C(l)\|$ looks at the compatible partners in the bipartite case, $p(d_1,\ldots,d_{n/2})$ looks at the graph when connected to itself.
For example, a hash function candidate that leads to only nodes of degree $2$ has $p(2,\ldots,2) = 1$.
However, we get $\|C(l)\| < 1$ because $l$ is not compatible with partners that do not hit all output values.

We now formally connect the two probabilities $p(d_1,\ldots,d_{n/2})$ and $\|C(l)\|$.
This gives a bound for the probability of drawing a compatible partner for a function in $\H\setminus\HProm$.

\begin{lemma}
  Each $l \in \H\setminus\HProm$ satisfies $\|C(l)\| < (e/2)^{-n/2} \cdot (\constantMaxC)^{n/2}$ where $\constantMaxC$ is a constant $\in (0, 1)$ and $n$ large enough.
  \label{lem:maxC}
\end{lemma}
\begin{proof}
  Take any hash function $l \in \H\setminus\HProm$ and its corresponding degree sequence $(d_1, \ldots, d_{n/2})$.
  As a reminder, $\B_n$ is the distribution of balls in the balls-into-bins process with $n$ balls and $n/2$ bins.
  Then
  \begin{align*}
    \|C(l)\| \quad
      &=\quad \Pr(\PF(l, r))\\
      &=\quad \expect_{(d'_1, \ldots, d'_{n/2}) \sim \B_n}\left(\Pr\left(\PF(l, r) \cond r \textrm{ has degree sequence } (d'_1, \ldots, d'_{n/2})\right)\right)\\
      &=\quad \expect_{(d'_1, \ldots, d'_{n/2}) \sim \B_n}\left(q((d_1, \ldots, d_{n/2}), (d'_1, \ldots, d'_{n/2}))\right)\\
      &\refRelLem{lem:peelingSeparate}{\leq} \expect_{(d'_1, \ldots, d'_{n/2}) \sim \B_n}\left(p(d_1, \ldots, d_{n/2}) \cdot p(d'_1, \ldots, d'_{n/2})\right)\\
      &=\quad p(d_1, \ldots, d_{n/2}) \cdot \expect_{(d'_1, \ldots, d'_{n/2}) \sim \B_n}\left(p(d'_1, \ldots, d'_{n/2})\right)\\
      &\refRelEq{obs:expectP}{=} p(d_1, \ldots, d_{n/2}) \cdot (e/2)^{-n/2}
      \quad\refRelLem{lem:nodeDegree1}{\leq} \constantMaxC^{n/2} \cdot (e/2)^{-n/2}.
    \tag*{\qedhere}
  \end{align*}
\end{proof}

\subsubsection{Concentration Bounds} \label{s:concentrationBounds}
Remember variable $Z = \|\bigcup_{i \in [k]} C(l_i)\|$, which is the probability that a randomly selected function is compatible with a function in our pool.
Also remember $C^*(l)=C(l)\setminus\HProm$
Unfortunately, we cannot directly give concentration bounds on $Z$ or even calculate $\expect(Z)$ because we do not know how using a pool of hash functions influences the probabilities.
We therefore look at three additional random variables, defined as follows:
\begin{align*}
  Z^* \coloneqq \|\bigcup_{i \in [k]} C^*(l_i)\|,
  \hspace{5mm}
  \hat Z \coloneqq \sum_{i \in [k]} \|C^*(l_i)\|,
  \hspace{5mm}
  D \coloneqq \sum_{1 \leq i < j \leq k} \|C^*(l_i) \cap C^*(l_j)\|
\end{align*}

$Z^*$ considers only partners in $\H\setminus\HProm$.
$\hat Z$ is easier to calculate because it looks at each set separately.
In the remainder of this section, we then determine bounds on $Z^*$, $\hat Z$, and $D$, which we can later use to bound $Z$.

\begin{lemma}
  Let $Z^*$, $\hat Z$ and $D$ be as defined above.
  Then $\expect(\hat Z) > 4/5 \cdot (e/2)^{-n/2}$.
  \label{lem:largeZ2}
\end{lemma}
\begin{proof}
  \begin{align*}
    \expect(\hat Z)
     &= \sum_{i \in [k]}\expect(\|C^*(l_i)\|)
     \geq \sum_{i \in [k]}\left(\expect(\|C(l_i)\|) - \|\HProm\|\right)\\
    &= k \cdot \left((e/2)^{-n} - \|\HProm\|\right)
     \quad\refRelLem{lem:nodeDegree2}{\geq} k \cdot \left((e/2)^{-n} - \frac{0.66^n \cdot |\H|}{|\H|}\right)\\
    &= (e/2)^{-n/2} - 0.66^n (e/2)^{n/2}
     = \left(1 - (0.66 \cdot e/2)^n \right) \cdot (e/2)^{-n/2}\\
    & \geq (1-0.9^n) \cdot (e/2)^{-n/2}
      \geq 4/5 \cdot (e/2)^{-n/2} \textrm{ for $n$ large enough.}
    \tag*{\qedhere}
  \end{align*}
\end{proof}

\begin{observation}
  For $\hat Z$, $D$ and $Z^*$ defined as above, it holds that $\hat Z-D \leq Z^* \leq \hat Z$.
  \label{cor:upperLower}
\end{observation}
\begin{proof}
  To show the bounds, we use the inclusion-exclusion principle, namely that for sets $A$ and $B$, $|A \cup B| = |A|+|B| - |A \cap B| \leq |A|+|B|$.
  We can give an upper bound for the variable $Z^*$ by applying the inequality repeatedly using associativity of the union operation.

  If we take $\hat Z$ and subtract the sizes of all pairwise intersections, we get a lower bound.
  If all partners were compatible with at most two hash functions in our pool, subtracting the pairwise intersections would give $Z^*$.
  However, if a partner is compatible with more than two functions, this subtracts too much, therefore giving the lower bound $\hat Z-D \leq Z^*$.
\end{proof}

We can now show that $\expect(D)$ is exponentially smaller than $\expect(\hat Z)$.
Intuitively, this means that $\expect(\hat Z) \approx \expect(Z^*)$ by \cref{cor:upperLower}.
Our proof idea is to bound the intersections using the bound on $\|C(l_i)\|$ shown in \cref{lem:maxC}.

\begin{lemma}
  Let $D$ and $\hat Z$ be as defined above.
  Then $\expect(D) \leq (\constantD)^{n/2} \cdot \expect(\hat Z)$.
  \label{lem:smallD}
\end{lemma}
\begin{proof}
  \begin{align*}
    \expect(D)
      &= \quad\expect\left(\sum_{1 \leq i < j \leq k} \|C^*(l_i) \cap C^*(l_j)\|\right)
       = {k \choose 2} \expect_{l_1,l_2 \sim \uniform(\H)}( \| C^*(l_1) \cap C^*(l_2) \| )\\
      &= \quad{k \choose 2} \frac{1}{|\H|} \expect_{l_1,l_2 \sim \uniform(\H)}\left( | C^*(l_1) \cap C^*(l_2) | \right)\\
      &= \quad{k \choose 2} \frac{1}{|\H|} \expect_{l_1 \sim \uniform(\H)}\left(\expect_{l_2 \sim \uniform(\H)}( | C^*(l_1) \cap C^*(l_2) | )\right)\\
      &= \quad{k \choose 2} \frac{1}{|\H|} \expect_{l_1 \sim \uniform(\H)}\left(\sum_{r \in C^*(l_1)}\Pr_{l_2 \sim \uniform(\H)}\left(r \in C^*(l_2)\right) \right)\\
      &\leq \quad{k \choose 2} \frac{1}{|\H|} \expect_{l_1 \sim \uniform(\H)}\left(\sum_{r \in C^*(l_1)}\Pr_{l_2 \sim \uniform(\H)}\left(l_2 \in C(r)\right) \right)\\
      &= \quad{k \choose 2} \frac{1}{|\H|} \expect_{l_1 \sim \uniform(\H)}\left(\sum_{r \in C^*(l_1)} \|C(r)\| \right)\\
      &\refRelLem{lem:maxC}{\leq} {k \choose 2} \frac{1}{|\H|} \expect_{l_1 \sim \uniform(\H)}\left(|C^*(l_1)| \cdot (e/2)^{-n/2} \cdot (\constantMaxC)^{n/2} \right)\\
      &\leq \quad(e/2)^{n} \expect_{l_1 \sim \uniform(\H)}(\|C(l_1)\|) \cdot (e/2)^{-n/2} \cdot (\constantMaxC)^{n/2}
      = (\constantMaxC)^{n/2} \cdot (e/2)^{-n/2}\\
      &\refRelLem{lem:largeZ2}{\leq} (\constantD)^{n/2} \cdot \expect(\hat Z) \textrm{ for some $\constantD \in (0, 1)$ and $n$ large enough.}
    \tag*{\qedhere}
  \end{align*}
\end{proof}

Let us now show that $\hat Z$ does not get much smaller than its expected value.

\begin{lemma}
  Let $Z^*$, $\hat Z$ and $D$ be as defined above.
  Then $\Pr(\hat Z \geq \expect(\hat Z)/2) > 1-(\constantZHat)^n$ where $\constantZHat$ is a constant $\in (0, 1)$ and $n$ large enough.
  \label{lem:sConcentration}
\end{lemma}
\begin{proof}
  The Bernstein inequality \cite{bernstein1924modification} states that for independent and zero-mean random variables $V_i$ with $|V_i| \leq M$, it holds that:
  \begin{align*}
    \Pr\left(\sum_{i=1}^n V_i \geq t \right)
      \leq \exp \left( -\frac{\tfrac{1}{2} t^2}{\sum_{i = 1}^n \expect\left(V_i^2 \right)+\tfrac{1}{3} Mt} \right)
  \end{align*}

  To apply the inequality, we now center our variables $\|C^*(l_i)\|$ and mirror them around the value $0$, giving us $V_i=\expect(\|C^*(l_i)\|)-\|C^*(l_i)\|$.
  Centering only makes the maximum smaller.
  Through \cref{lem:eventPoolNodeDegree2}, we can assume that none of our functions $l_i$ is promiscuous.
  This can be formalized by increasing $\constantZHat$ accordingly.
  Therefore, we get $\max_{i \in [k]}(V_i) \leq \max(\|C^*(l_i)\|) \leq \max(\|C(l_i)\|) \leq (\constantMaxC)^{n/2} \cdot (e/2)^{-n/2} =: M$ through \cref{lem:maxC}.
  Before we can get to the Bernstein inequality, we need another ingredient.
  Let us upper bound the value of $\expect\left((V_i)^2 \right)$ as follows:
  \begin{align*}
    \expect\left((V_i)^2 \right)
      &= \expect\left(\left(\|C^*(l_i)\|-\expect\left(\|C^*(l_i)\|\right)\right)^2 \right)
      = \expect\left(\|C^*(l_i)\|^2\right)-\expect\left(\|C^*(l_i)\|\right)^2\\
      &\leq \expect\left(\|C^*(l_i)\|^2\right)
      \leq \expect\left(\max_{j \in [k]}\|C^*(l_j)\| \cdot \|C^*(l_i)\|\right)
      = \max_{j \in [k]}\|C^*(l_j)\| \cdot \expect\left(\|C^*(l_i)\|\right)\\
      &\refRelLem{lem:maxC}{\leq} \constantMaxC^{n/2} \cdot (e/2)^{-n/2} \cdot (e/2)^{-n}
      = \constantMaxC^{n/2} \cdot (e/2)^{-n/2-n}
  \end{align*}
  Setting $t = 4/10 \cdot (e/2)^{-n/2}$ and applying the Bernstein inequality in the step annotated with $*$, we get:
  \begin{align*}
    \Pr&\left(\hat Z \leq \expect(\hat Z)/2\right)
      =\Pr\left(\sum_{i \in [k]} \|C^*(l_i)\| \leq \expect(\hat Z)/2\right)\\
      &=\Pr\left(\sum_{i \in [k]} \expect(\|C^*(l_i)\|) - \sum_{i \in [k]} V_i \leq \expect(\hat Z)/2\right)\\
      &=\Pr\left(\expect(\hat Z) - \sum_{i \in [k]} V_i \leq \expect(\hat Z)/2\right)
      =\Pr\left(\sum_{i=1}^k V_i \geq \expect(\hat Z)/2\right)\\
      &\refRelLem{lem:largeZ2}{\leq} \Pr\left(\sum_{i=1}^k V_i \geq 4/10 \cdot (e/2)^{-n/2} \right)\\
      &\overset{*}{\leq} \exp \left( -\frac{\tfrac{1}{2} \left(4/10 \cdot (e/2)^{-n/2}\right)^2}{k\cdot\constantMaxC^{n/2} \cdot (e/2)^{-n/2-n} + \tfrac{1}{3} \cdot (\constantMaxC^{n/2} \cdot (e/2)^{-n/2}) \cdot ((e/2)^{-n/2}/2)} \right)\\
      &= \exp \left(-\frac{\tfrac{2}{25} (e/2)^{-n}}{\constantMaxC^{n/2} \cdot (e/2)^{-n} + \tfrac{1}{6} \cdot \constantMaxC^{n/2} \cdot (e/2)^{-n}} \right)\\
      &= \exp \left(-12/175 \cdot \constantMaxC^{-n/2}\right)
      =\left(e^{-12/175}\right)^{\left(\left(\constantMaxC^{-1/2}\right)^n\right)}
      \leq (\constantZHat)^n
  \end{align*}
  for some $\constantZHat \in (0, 1)$ and $n$ large enough.
\end{proof}

\Cref{lem:smallD} gives a bound on $\expect(D)$ and \cref{lem:sConcentration} gives a concentration bound on $\hat Z$.
With these insights, we now give a bound on $Z^*$ in terms of $\expect(\hat Z)$.

\begin{lemma}
  Let $Z^*$, $\hat Z$ and $D$ be as defined above.
  Then $\Pr\left(Z^* < \expect(\hat Z)/4\right) \leq (\constantZ)^n$  where $\constantZ$ is a constant $\in (0, 1)$ and $n$ large enough.
  \label{lem:largeZ}
\end{lemma}
\begin{proof}
  We can bound the probability that $Z^*$ deviates too much from $\expect(\hat Z)/4$ as follows.
  In the step annotated with $*$, we use the Markov inequality ($\Pr(D \geq a)\leq \expect(D)/a$).
  \begin{align*}
    \Pr&\left(Z^* < \expect(\hat Z)/4\right)
      \hspace{3mm}\refRelObs{cor:upperLower}{\leq} \Pr\left(\hat Z - D < \expect(\hat Z)/4\right)\\
      &\leq \quad \Pr\left(\hat Z < \expect(\hat Z)/2 \vee D \geq \expect(\hat Z)/4\right)\\
      &\leq \quad \Pr\left(\hat Z < \expect(\hat Z)/2\right) + \Pr\left(D \geq \expect(\hat Z)/4\right)\\
      &\refRelLem{lem:sConcentration}{\leq} (\constantZHat)^n +\Pr\left(D \geq \expect(\hat Z)/4\right)
      \overset{*}{\leq} (\constantZHat)^n + \frac{\expect(D)}{\expect(\hat Z)/4}\\
      &\refRelLem{lem:smallD}{\leq} (\constantZHat)^n + (\constantD)^{n/2}/4
      \leq (\constantZ)^n \textrm{ for some $\constantZ \in (0, 1)$ and $n$ large enough.}
    \tag*{\qedhere}
  \end{align*}
\end{proof}

\subsubsection{Combining with Pool \texorpdfstring{$(r_i)_{i\in[k]}$}{r}} \label{s:combining}
We can now plug together the previous results, giving us the success probability of bipartite ShockHash when using a pool of size $k$.
With $\Pr\left(\textrm{success}\right)$, we denote the probability that there is a pair of compatible hash functions $l_i$ and $r_j$, $i,j\in[n/2]$ in our pools.

\begin{lemma}
  Let $Z$ be defined as above.
  Then $\Pr\left(\textrm{success} \cond Z = z \right) = 1-(1-z)^k$.
  \label{lem:poolSuccessConditioned}
\end{lemma}
\begin{proof}
  \begin{align*}
    \Pr\left(\textrm{success} \cond Z = z\right)
      &= \Pr\left(\exists i \in [k] : r_i \in \bigcup_{j \in [k]} C(l_j) \cond \|\bigcup_{j \in [k]} C(l_j)\| = z\right)\\
      &= \Pr_{p_1,\ldots,p_k \sim \textrm{Ber}(z)}\left(\exists i \in [k] : p_i = 1\right)
      = 1-(1-z)^k.
    \tag*{\qedhere}
  \end{align*}
\end{proof}

We already have most of the proof done.
We just need to factor in the probability that the precondition of the previous lemma holds.

\begin{theorem}
  Let us take two pools $(l_i)_{i\in[k]}$ and $(r_i)_{i\in[k]}$ of size $k=(e/2)^{n/2}$ containing randomly sampled hash functions.
  Then the probability that there are two hash functions $l_i$ and $r_j$ $(i,j \in [k])$ in the pools such that $\PF(l_i, r_j)$ is $> 0.17$ for $n$ large enough.
  \label{thm:poolSuccess}
\end{theorem}
\begin{proof}
  \allowdisplaybreaks
  \begin{align*}
    \Pr&(\textrm{success})
        \geq \Pr\left(\textrm{success} \wedge Z \geq \expect(\hat Z)/4\right)\\
        &= \Pr\left(\textrm{success} \cond Z \geq \expect(\hat Z)/4\right) \cdot \Pr(Z \geq \expect(\hat Z)/4)\\
        &\geq \Pr\left(\textrm{success} \cond Z \geq \expect(\hat Z)/4\right) \cdot \Pr(Z^* \geq \expect(\hat Z)/4)\\
        &= \sum_{z=\expect(\hat Z)/4}^\infty \left(\Pr\left(\textrm{success} \cond Z = z\right) \cdot \Pr(Z = z \mid Z \geq \expect(\hat Z)/4) \right) \cdot \Pr(Z^* \geq \expect(\hat Z)/4)\\
        &\refRelLem{lem:poolSuccessConditioned}{=} \sum_{z=\expect(\hat Z)/4}^\infty \left(\left(1-(1-z)^k\right) \cdot \Pr(Z = z \mid Z \geq \expect(\hat Z)/4) \right) \cdot \Pr(Z^* \geq \expect(\hat Z)/4)\\
        &\geq \sum_{z=\expect(\hat Z)/4}^\infty \left(\left(1-(1-\expect(\hat Z)/4)^k\right) \cdot \Pr(Z = z \mid Z \geq \expect(\hat Z)/4) \right) \cdot \Pr(Z^* \geq \expect(\hat Z)/4)\\
        &= \left(1-(1-\expect(\hat Z)/4)^k\right) \cdot \underbrace{\sum_{z=\expect(\hat Z)/4}^\infty\left( \Pr(Z = z \mid Z \geq \expect(\hat Z)/4) \right)}_{=1} \cdot \Pr(Z^* \geq \expect(\hat Z)/4)\\
        &\refRelLemT{lem:largeZ}{lem:largeZ2}{>} \left(1-\left(1-\frac{(e/2)^{-n/2}}{5}\right)^k\right) \cdot \left(1 - (\constantZ)^n\right)
         = \left(1-\left(1-\frac{1}{5k}\right)^k\right) \cdot \left(1 - (\constantZ)^n\right)\\
        &\geq \left(1-e^{-1/5}\right) \cdot \left(1 - (\constantZ)^n\right)
         > 0.18 \textrm{ for $n$ large enough.}
    \tag*{\qedhere}
  \end{align*}
\end{proof}

This concludes the proof of combining hash functions from a pool of candidates.
We have seen that taking the pools of size $k=(e/2)^{n/2}$ gives us a constant probability that there are two compatible functions in the pools.
Note again that our actual implementation uses one single, growing pool, not two fixed size pools.
Therefore, we do not need to retry the construction if the initial pool size is not enough, but can just continue adding more functions to the pool.

\subsection{ShockHash Construction}\label{ss:analysisShockHash}
ShockHash tries different hash function seeds, which is equivalent to generating random graphs.
Given the probability that a random graph is a pseudoforest, it is easy to determine the expected number of graphs ShockHash needs to try in order to find an MPHF.
This leads directly to the space usage and construction time of ShockHash, which we state in the following Theorem.

\begin{theorem}
  \label{thm:shockHashConstruction}
  A ShockHash minimal perfect hash function mapping $n$ keys to $[n]$ needs $\log_2(e)n + \Oh(\log n)$ bits of space in expectation and can be constructed in expected time $\Oh((e/2)^n \cdot n)$.
\end{theorem}
\begin{proof}
  From \cref{lem:constructionTries}, we know that the probability of the graph being a pseudoforest is $\geq (e/2)^{-n} e^{-1}\sqrt{\pi}$.
  We construct these graphs uniformly at random, so the expected number of seeds to try is $\leq (e/2)^{n} e / \sqrt{\pi}$.
  The space usage is given by the $n + o(n)$ bits for the retrieval data structure, plus the bits to store the hash function index:
  \begin{align*}
    &\mathds{E}(\log_2(\textrm{seed value}))
    \overset{*}{\leq} \log_2(\mathds{E}(\textrm{seed value}))
    \leq \log_2\left((e/2)^{n} e / \sqrt{\pi}\right)
    = \log_2(e)n - n + \Oh(1).
  \end{align*}
  In the step annotated with $*$, we use Jensen's inequality \cite{jensen1906fonctions} and the fact that $\log_2$ is concave.

  For determining if at least one of the $2^n$ functions corresponding to such a seed is valid, we can use an algorithm for finding connected components, as described in \cref{s:sccFilter}.
  This takes linear time for each of the seeds, resulting in an overall construction time of $\Oh((e/2)^n \cdot n)$.
  Constructing the retrieval data structure is then possible in linear time \cite{dillinger2022burr} and happens only once, so it is irrelevant for the asymptotic time here.
  \hfill
\end{proof}

Looking back at the simple brute-force approach, each of its $e^n/\sqrt{2 \pi n}$ expected trials needs $n$ hash function evaluations, leading to a construction time of $\Oh(e^n \sqrt{n})$.
Now, as shown in \cref{thm:shockHashConstruction}, ShockHash needs time $\Oh((e/2)^n \cdot n)$.
This makes ShockHash almost $2^n$ times faster than the previous state of the art.
Given the observations in Ref. \cite{bez2023high}, we conjecture that ShockHash with rotation fitting reduces the number of hash function evaluations by an additional factor of $n$, while the space overhead tends to zero.
In the following Theorem, we now give the resulting construction time and space consumption of the bipartite version.
\begin{theorem}
  \label{thm:bipartiteShockHashConstruction}
  A bipartite ShockHash minimal perfect hash function needs $\log_2(e)n + \Oh(\log n)$ bits of space in expectation and can be constructed in expected time $\Oh(1.166^n)$.
\end{theorem}
\begin{proof}
  We know that we need to test $(e/2)^{n}e/\sqrt{n}$ candidate pairs $(h_0,h_1)$ in expectation before we find a perfect hash function.
  As described in \cref{s:shockhash2}, instead of sampling two hash functions independently, we use a pool of hash functions and test all combinations of them.
  \Cref{thm:poolSuccess} shows that if we use a pool size of $k = (e/2)^{n/2} \approx 1.166^n$, we get a constant success probability.
  Until here, this does not improve the construction time asymptotically because all $k^2$ combinations need to be tested.
  However, instead of combining all of the candidates, we can filter them directly while building the candidate pool.
  The filter, as with plain ShockHash, is very effective:
  The probability that the $n$ hash values in a partition of size $n/2$ hit all output positions is $\Theta(0.836^{n/2})$ \cite{walzer2024probability}.
  This means that we are only considering about $((e/2)^{n/2} \cdot 0.836^{n/2})^2 = (e/2 \cdot 0.836)^n \approx 1.136^n$ pairs of hash functions in expectation.
  The construction time is therefore bounded by $\max \{1.166^n, 1.136^n\}$.
  Note that both of these values were rounded up anyway, so polynomial factors are dominated.

  Looking at the space consumption of bipartite ShockHash, we need to encode two seeds of expected value $\leq k$ each.
  Using Jensen's inequality and the retrieval data structure just like in \cref{thm:shockHashConstruction}, we get the resulting space usage.
  The fact that we suppressed polynomial factors in the analysis disappears in the $\Oh(\log n)$ term.
\end{proof}

\section{Partitioning}\label{s:partitioning}
Even though ShockHash demonstrates significant speedups, by itself, it still needs exponential running time.
As mentioned in the introduction, real world MPHF constructions usually do not search for a function for the entire input set directly.
Instead, they partition the input of size $N$ and then search on smaller subproblems of size $n$.
In this section, we now give details on how to partition the input set efficiently before then using ShockHash as a building block.

\subsection{ShockHash-RS = ShockHash + RecSplit}\label{s:shockhashInRecSplit}
A first option is to integrate ShockHash as a base case into the highly space efficient RecSplit framework (see \cref{s:recsplit}) and obtain ShockHash-RS.
We keep the general structure of RecSplit intact.
Only in the leaves, we use ShockHash instead of brute-force.
We store the mapping from its keys to their hash function indices in one large retrieval data structure.
Finally, after all leaves are processed, we construct the 1-bit retrieval data structure with all the $N$ entries together.

\myparagraph{Fanouts.}
RecSplit tries to balance the difficulty between the splittings and the bijections.
ShockHash improves the performance of the bijections significantly but does not modify the way that the splittings are calculated.
In this paper, we focus only on the bijections.
To balance the amount of work done between splittings and bijections, we need to adapt the splitting parameters using the same techniques as the RecSplit paper.
The RecSplit paper proves and uses optimal fanouts $\lceil 0.35 n + 0.5 \rceil$ and $\lceil 0.21 n + 0.9 \rceil$ for the two last splitting levels (see \cite[Section 5.4]{esposito2020recsplit}).
For ShockHash-RS, we can adapt their formulas accordingly and get fanouts of $\lfloor 0.10 n + 0.5\rfloor$ and $\lfloor 0.073 n + 0.9 \rfloor$.
However, preliminary experiments show that this is not optimal in practice.
ShockHash is so much faster that the additional time invested into the splittings does not pay off.
We find experimentally that setting the lowest splitting level to 4 and the second lowest to 3 achieves much better results in practice.
To also provide faster and space-inefficient configurations, we set all fanouts to~$2$ when selecting leaf size $n \leq 24$.

\subsection{ShockHash-Flat = ShockHash + $k$-Perfect Hashing}
An additional way to partition the input keys is to use $k$-perfect hashing.
A minimal $k$-perfect hash function maps $N$ keys to $N/k$ output values, where each output value is hit exactly $k$ times (assuming $N$ divides $k$).
This has applications in external memory data structures \cite{kurpicz2023pachash,larson1985external} and there are existing constructions \cite{belazzougui2009hash}.
The idea how to integrate ShockHash with $k$-perfect hashing is straightforward and similar to what we do in \ShockHashRS{} (see \cref{s:shockhashInRecSplit}).
We simply run a two-step process of first determining a $k$-perfect hash function, and then we construct small ShockHash data structures for the $k$ keys hitting each output value.
In contrast to ShockHash-RS (see \cref{s:shockhashInRecSplit}), where some base cases could be smaller than $n$, here all of them have the same size.

\begin{figure}[t]
  \centering
  \includegraphics[scale=0.85]{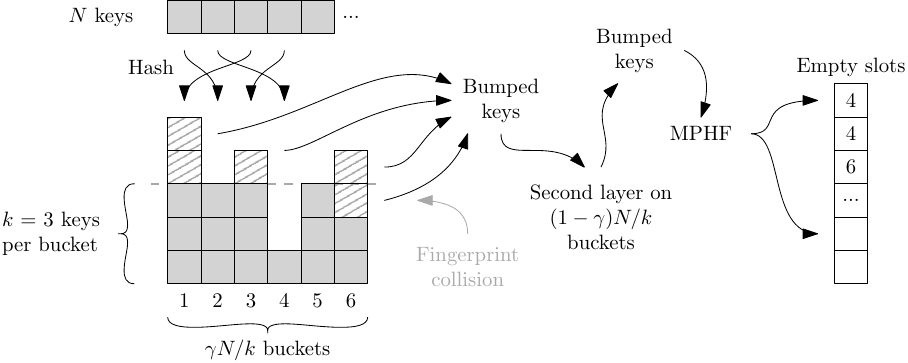}
  \caption{Illustration for our bumped $k$-perfect hash function.}
  \label{fig:kperfect}
\end{figure}

\myparagraph{Bumped $k$-Perfect Hashing.}
We now briefly describe a new $k$-perfect hash function.
This function is focused on fast queries while still having rather small space consumption.
Let us take $N$ keys and hash them uniformly at random to a set of $\gamma N/k$ buckets, $\gamma \in (0, 1]$.
By choosing $\gamma<1$, we can overload the buckets to ensure that most buckets receive at least $k$ keys.
In the experiments, we use $\gamma=0.9$.
We handle overflowing buckets by determining a fingerprint of each key.
Each bucket then stores a threshold value using $\log_2(k)$ bits that indicates which of the keys to bump from the bucket.
This idea of bumping keys based on a threshold is inspired by bumped ribbon retrieval (BuRR) \cite{dillinger2022burr}.
Separator hashing \cite{gonnet1988external} uses a similar idea, however, without bumping keys completely and only supporting non-minimal perfect hash functions.
We use a second level of the same data structure for the bumped keys, mapping them to the remaining $(1 - \gamma) N/k$ buckets.
Finally, we have a small number of keys that still get bumped in the second level.
We first enumerate them by constructing a minimal perfect hash function.
In our implementation, we use ShockHash-RS.
With this, we then index an Elias-Fano coded sequence \cite{fano71number,elias74efficient} storing all empty slots in the output range.
\Cref{fig:kperfect} gives an illustration of the idea.

The advantage of this technique is that the majority of queries need a single access to the array of thresholds and a comparison with the key's threshold.
Few need two accesses to evaluate the second level, and only a tiny fraction of the queries needs to evaluate the explicit re-mapping.

\myparagraph{ShockHash-Flat.}
From the bumped $k$-perfect hash function, we derive an MPHF that has a significantly more flat structure than ShockHash-RS.
Instead of traversing a tree structure, it can perform a simple comparison with the threshold value for a majority of the input keys.
Because we need to access both the threshold and then (usually) the seed of that same bucket, ShockHash-Flat stores thresholds and ShockHash seeds in an interleaved way.

\section{Variants and Refinements}\label{s:refinements}
In the following section, we describe variants and implementation details of ShockHash.
In \cref{s:isolatedKeys}, we first explain how to achieve significant improvements in practice by using a bit-parallel filter.
We then describe two techniques to come up with hash function candidates more efficiently, rotation fitting \cite{bez2023high} and quad split, in \cref{s:quadsplit,s:rotationFitting}.
To use bipartite ShockHash with uneven input sizes $n$, only small tweaks are necessary, which we describe in \cref{ss:unevenN}.
We then continue with practical implementation tricks in \cref{ss:engineering}.
Finally, we describe ideas for parallelization in \cref{ss:parallelization}.

\begin{figure}[t]
  \centering
  \includegraphics[scale=0.85]{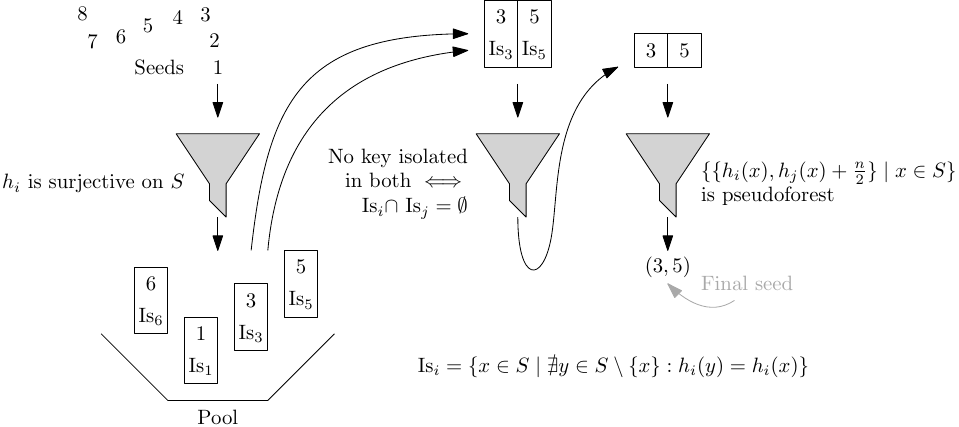}
  \caption{Filtering for isolated keys.
    Each seed in the pool stores a bit vector of isolated keys, here annotated with Is$_i$.
    A seed combination can only work if no key is isolated in both partitions, which can be efficiently checked using bit operations.
    Only if the filter is passed, we need to do the full orientability check.}
  \label{fig:isolatedKeys}
\end{figure}

\subsection{Isolated Keys Filter}\label{s:isolatedKeys}
In \cref{s:shockhash2}, we have already described how making the graph bipartite enables efficient filtering of hash function candidates.
In the following section, we describe an additional way of filtering seeds.
Bipartite ShockHash generates a set of surjective hash function candidates and then tests all combinations for orientability.
By using an additional filter, we can speed up this test for orientability.
The idea is to look at the case that a key is the only one mapping to a position.
We refer to this key as \emph{isolated} for that hash function seed.
More formally, a key $x$ is isolated using a hash function candidate $h$, if $\{y\in S \mid h(x) = h(y)\} = \{x\}$.
If a key is isolated in \emph{both} of the candidate hash functions, then in graph terminology this corresponds to a connected component with two nodes and one edge.
Since each connected component of the final graph must have the same number of nodes and edges, there is then no need to perform the full test for orientability.
We can determine bit patterns for each seed, indicating which of the keys are isolated.
Then seed combinations can be ruled out using simple bit-parallel operations checking if the bit patterns are orthogonal.
Note that the bit patterns used here refer not to the output positions but to the input keys (and therefore have size $n$).
\Cref{fig:isolatedKeys} illustrates the process.

A key is isolated in a partition if none of the other keys hash to its position, which happens with probability $(1-1/(n/2))^{n-1} \rightarrow e^{-2}$.
A seed combination passes the filter if it has no key that is isolated in both partitions.
This is approximately $(1-e^{-4})^{n} \approx 0.98^n$, so the filter makes it possible to avoid the full check for a vast majority of seed combinations.
Note that we apply the filter conditioned on the case that both functions are surjective, which should only make the filter more effective.

What makes this method interesting from a theoretical point of view is that we can be even smarter about filtering here.
As stated before, if one of the hash functions has an isolated key at a position, we can skip testing it with all other hash functions that have an isolated key at the same position.
We can organize all candidate hash functions in a binary trie data structure based on the isolated keys.
Testing a new candidate hash function now boils down to traversing the tree.
In theory, this gives additional exponential improvements in the construction time.
However, preliminary experiments show that it is not helpful for the values of $n$ we use in practice.

\subsection{Rotation Fitting}\label{s:rotationFitting}
A technique to speed up brute-force search for perfect hash functions is \emph{rotation fitting} \cite{bez2023high} (see \cref{s:recsplit}).
The same idea can be used in ShockHash to accelerate the search.
We distribute the keys to two sets using an ordinary and unseeded 1-bit hash function.
We then determine the bit mask of output values that are hit in both of the sets.
Like in the bit mask filter, which we use before checking for orientability (see \cref{s:bitmaskFilter}), only if the logical \texttt{OR} of both masks has all bits set, it is worth testing the seed more closely.
If we now cyclically rotate one of the bit masks and try again, we basically get a new chance of all output values being hit, without having to hash each key again.
We then consider the distance to rotate the keys as part of the hash function seed.
This corresponds to an addition modulo $n$ to all hash values of the second set.
We conjecture that -- as in Ref. \cite{bez2023high} -- this reduces the number of hash function evaluations by a factor of $n$, while the space overhead tends to zero.

For bipartite ShockHash, rotation fitting can be applied as well, though in a slightly different way.
Rotating one of the partitions of the bipartite graph within itself is not useful because it generates isomorphic graphs.
Rotating the two partitions into each other would violate the bipartite condition, thus preventing to use the hash function pools.
Instead, we can use rotation fitting to find seed candidates within each partition.
More specifically, when looking for seed candidates for one partition, we distribute the keys to two subsets using an unseeded 1-bit hash function.
We can then rotate one of the sets (modulo $n/2$) to get additional hash function candidates.
Each rotation can be tested for surjectivity using simple bit shifts that can happen in registers.
In practice, this significantly improves the construction time because fewer hash functions need to be evaluated.
However, the quad split technique described in the following section even enables exponential speedups.

\begin{figure}[t]
  \centering
  \includegraphics[scale=0.85]{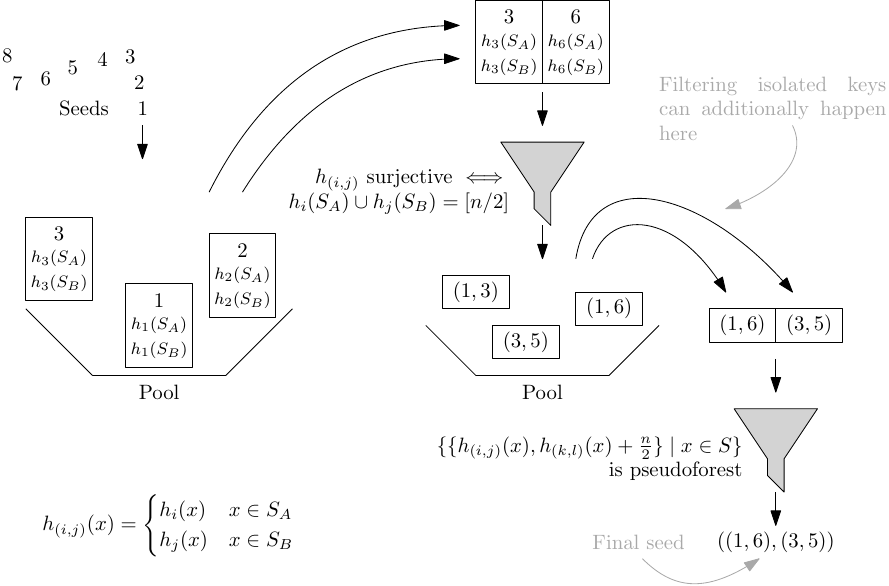}
  \caption{Additional filtering opportunities in quad split.
    We create a pool of all seeds without filtering but annotate each seed a bit vector containing hash function output values.
    Then we combine two seeds to form a seed for one partition.
    Surjectivity can be checked efficiently using bit operations.
    Therefore, the resulting seed is a tuple of four seeds -- one for each half of the keys and each partition.
    In our implementation, we also add the filter for isolated keys on top (see \cref{s:isolatedKeys,fig:isolatedKeys}).
  }
  \label{fig:quadsplit}
\end{figure}

\subsection{Quad Split}\label{s:quadsplit}
The construction is dominated by the time spent evaluating hash function candidates (see \cref{thm:bipartiteShockHashConstruction}), so it is natural to look at this step for improvements.
For bipartite ShockHash, the quad split technique reduces the amount of time spent on finding surjective seed candidates.
It basically applies the idea of bipartite ShockHash on another level of the same data structure.
Like in rotation fitting, we split the input set into two sets $S_A$ and $S_B$ using a constant 1-bit hash function.
Now we can hash each of the two sets using \emph{independent} hash functions.
In particular, we can test all combinations of assigning some hash function to each of the two sets.
This reduces the number of hash function evaluations significantly.

For a seed $i$, let $h_i(S_A)$ and $h_i(S_B)$ indicate the sets of hash function output values of the two subsets $S_A$ and $S_B$.
A seed $i$ for $S_A$ can be used together with a seed $j$ for $S_B$ if $h_i(S_A) \cup h_i(S_B) = [n/2]$.
By storing $h_i(S_A)$ and $h_i(S_B)$ as bit vectors indicating the output values, this compatibility check is a simple and efficient \texttt{OR} operation.
In the quad split technique, we therefore annotate each hash function seed with this bit vector to enable fast search for a possible combination of seeds that is surjective.
This process is illustrated in \cref{fig:quadsplit}.

Like with the isolated keys filter described in \cref{s:isolatedKeys}, we can again use a trie structure again to avoid testing all combinations.
We believe that this enables exponential improvements in the construction time, which could be implemented in future work.
Quad split is orthogonal to the isolated keys filter, so we can actually combine both optimizations.

To encode the combined seed, we use a pairing function again.
In contrast to the bipartite tries in ShockHash, the hash functions cannot be exchanged, so we cannot assume that one seed is larger than the other.
We therefore need a more general pairing function.
The most fitting pairing function here is Szudzik's pairing function (see \cref{s:pairing}),
which enumerates, for all $k \in \mathbb{N}$, all pairs in $[k]\times[k]$ before moving on to pairs involving numbers bigger than $k$.
This means that we can test all combinations of previous hash functions before having to evaluate the next one.
In our implementation, we make sure to try hash function combinations in linear order in the value of the pairing function.

\begin{figure}[t]
  \centering
  \includegraphics[scale=0.85]{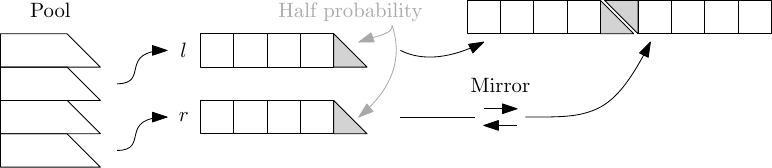}
  \caption{Supporting uneven $n$.}
  \label{fig:unevenN}
\end{figure}

\subsection{Supporting Uneven $n$}\label{ss:unevenN}
To support uneven numbers $n$ of input keys, we can relax the bipartite property.
The idea is that the output value $\lceil n/2 \rceil$ can be hit by both hash functions, but each with half the probability.
When combining the two halves, the value then gets the same probability as all other output values.
For filtering candidate hash functions for surjectivity, the corresponding bit needs to be ignored -- a seed candidate can be valid both if the bit is set or not set.
Now, in order to use hash functions from a single pool for both the left and the right part, we have to mirror the functions used for the right part.%
\footnote{For simplicity, our current implementation uses plain shifting, accepting a doubled probability for the middle bit.
  Supporting uneven $n$ then just boils down to rounding compile time constants to the right direction.
  When integrated into RecSplit, we expect uneven $n$ to happen only once in every two buckets, so it has a negligible overhead of about $1/2b$ bits per key.}
Refer for \cref{fig:unevenN} for an illustration.

\subsection{Engineering}\label{ss:engineering}
In the following section, we explain implementation tricks that we add to make our construction even faster in practice.

\myparagraph{Orientability Check.}
Determining whether a given graph is a pseudoforest can be achieved in linear time using a connected components algorithm.
However, this incurs significant overheads for building a graph data structure.
To avoid this, our implementation therefore uses incremental cuckoo hash table construction with near linear time.
For this, we keep an array representing the graph nodes.
To insert a key, we calculate the \texttt{XOR} of both hash function values and try to assign it into one of the two candidate array cells.
If the cell already contains another key, we evict the old one and insert it using its other candidate position, which we can efficiently retrieve using the \texttt{XOR} operation.
This idea is also known as the \texttt{XOR} trick in cuckoo hashing.
We abort the insertion as soon as we detect a cycle.

\myparagraph{Partial Hash Calculation.}\label{s:partialHashCalculation}
As described in \cref{s:rotationFitting}, we can use rotation fitting in plain ShockHash.
There we make the observation that hashing the first set of keys almost always yields a graph that, by itself, is a pseudoforest.
This is not surprising because the load factor is usually close to the load threshold $c=0.5$ and $n$ is small (which enables higher load \cite{lehmann2023sichash}).
We make use of this fact and reduce the number of hash function evaluations by keeping the hashes for the first set the same and just retrying hash functions for the second set.
More precisely, if $x$ is the hash function seed, we hash each key in the first set with seed $x-(x\textrm{ mod }k)$, where $k$ is a tuning parameter, and the keys of the second set with seed $x$.
Therefore, the hash values of the first set can be cached over multiple iterations.
In preliminary experiments, we find a value of $k=8$ to be a good fit -- values much larger than that have diminishing returns in performance improvement and start to influence the space consumption.
At $k=8$, however, the influence on the space consumption is negligible when $n$ is large.
Given that hashing the keys is a bottleneck during construction, this reduces the number of keys that need to be hashed by a factor of close to $2$.
We only apply this optimization for large $n > 32$.

\myparagraph{Hash Cache.}
In bipartite ShockHash, we regularly combine two hash function candidates from our pool to see if the resulting graph is a pseudoforest.
While we skip that test for many of the candidates using the simple bit parallel filter described in \cref{s:isolatedKeys}, there is still a large number of candidates to compare.
Re-evaluating the hash functions for these candidates can be a bottleneck depending on the input size $n$.
An obvious idea is to cache the hash function output values of the seed candidates.
Because the input sets and therefore the hash values are very small, we can store each hash value in a single byte.
This makes the amount of space needed for each seed candidate relatively small.

\myparagraph{Sentinels.}
For large $n$, the quad split technique spends most of its construction time calculating the logical \texttt{OR} of bit patterns looking for a result that has all bits set.
This inner loop consists of only a very small number of assembler instructions.
We can achieve considerable speedups here by adding a sentinel element to the end of the array that already has all bits set.
Then we no longer need the repeated bounds check for the array.
When using SIMD parallelization (see \cref{s:simd}), we use multiple sentinels depending on the number of SIMD lanes.

\subsection{Parallelization}\label{ss:parallelization}
The main computational load behind ShockHash looks for seeds yielding pseudoforests and can be parallelized on multiple levels:
Over buckets when using RecSplit for partitioning, ShockHash building blocks, seeds, hash-function evaluations and bit-parallel filters.
The remaining operations are also well parallelizable: Hashing of keys to buckets can be split between processors.
Parallel construction of the retrieval data structure can be done in a similar way \cite{dillinger2022burr}.
In the following we explain one possible parallelization with respect to SIMD instructions and multi-threading.
We also outline a hybrid CPU/GPU implementation.

\myparagraph{SIMD.}
\label{s:simd}
In plain ShockHash, we use SIMD parallelism in two locations.
First, we use SIMD to determine the two candidate positions of all keys and to determine the bit mask for filtering.
A key point here is to collect the bitwise \texttt{OR} of individual lanes and to only add the lanes together after all keys are done.
Second, we use SIMD to evaluate the bit mask filter (see \cref{s:bitmaskFilter}) with different rotations in parallel.
Our implementation uses AVX-512 (8 64-bit values) if available and AVX2 (4 64-bit values) otherwise.

Bipartite ShockHash can also be parallelized using SIMD instructions.
When using the quad split technique, we parallelize the test for candidate functions, which involves iterating over long lists of bit patterns, calculating the logical \texttt{OR} with each, and looking for a result that has all bits set.
However, the other more involved data structures are harder to parallelize using SIMD because of more complex control flows.
Therefore, we use SIMD only for checking lists of bit patterns, which is the main bottleneck for large $n$.

\myparagraph{Multi-Threading.}
Because ShockHash is intended to be integrated into a partitioning framework, we can naively parallelize over the different ShockHash base cases.
A simple coarse-grained source of parallelism are the RecSplit buckets.
We can use any kind of load balancing to split them between threads.
Even static load balancing may work because variances in construction time will average out.
However, some kind of dynamic load balancing is likely to be more efficient and also works with cores of different speed that are now becoming standard in many multi-core processors.
The retrieval data structure we use, BuRR \cite{dillinger2022burr}, can be parallelized as well.

\myparagraph{GPUs.}
A full GPU parallelization might be difficult and inefficient for cuckoo hashing as it has irregular control flow and memory access.
Since filtering asymptotically dominates the computations for highly space-efficient variants, one might look at a hybrid implementation where a GPU produces a stream of seeds defining random graphs that cover all nodes and where a multicore CPU performs further stages of computation.
For bipartite ShockHash with quad split, for example, the majority of the construction time is spent on comparing bit patterns to check if two hash function candidates are compatible.
Here we can use the massive parallelism of GPUs to compare many patterns in parallel.
Then the CPU can perform the less frequent and more complex checks for orientability.

\section{Experiments}\label{s:experiments}
We run our experiments on an Intel i7 11700 processor with 8 cores and a base clock speed of 2.5 GHz.
The machine runs Ubuntu 22.04 with Linux 5.15.0 and supports AVX-512 instructions.
We use the GNU C++ compiler version 11.2.0 with optimization flags \texttt{-O3 -march=native}.
For the competitors written in Rust, we compile in release mode with \texttt{target-cpu=native}.

Our implementation of ShockHash uses the BuRR retrieval data structure \cite{dillinger2022burr} with 128-bit ribbon width and 2-bit bumping information.
For ShockHash-RS, we use partitioning based on RecSplit \cite{esposito2020recsplit}, in particular the SIMD-parallel implementation \cite{bez2023high}.
For partitioning keys in ShockHash-Flat, we sort them using IPS$^2$Ra \cite{axtmann2020engineering}.

As input data, we use short strings of uniform random length $\in [10, 50]$ containing random characters except for the zero byte.
The reason for this is that all competitors natively support strings, while some only support integer keys when using their internal structs, which could give them an unfair advantage.
Note that, as a first step, almost all compared codes generate a \emph{master hash code} of each key using a high quality hash function.
Any possible additional hash function can then be evaluated on the master hash code in constant time, independent of the input distribution.
All experiments use a single thread.
While almost all compared codes have a multi-threaded implementation, and perfect hashing can be parallelized trivially by partitioning, this is not the focus here.
The code and scripts needed to reproduce our experiments are available on GitHub under the GNU General Public License \cite{sourceCode,sourceCodeCompetitors}.

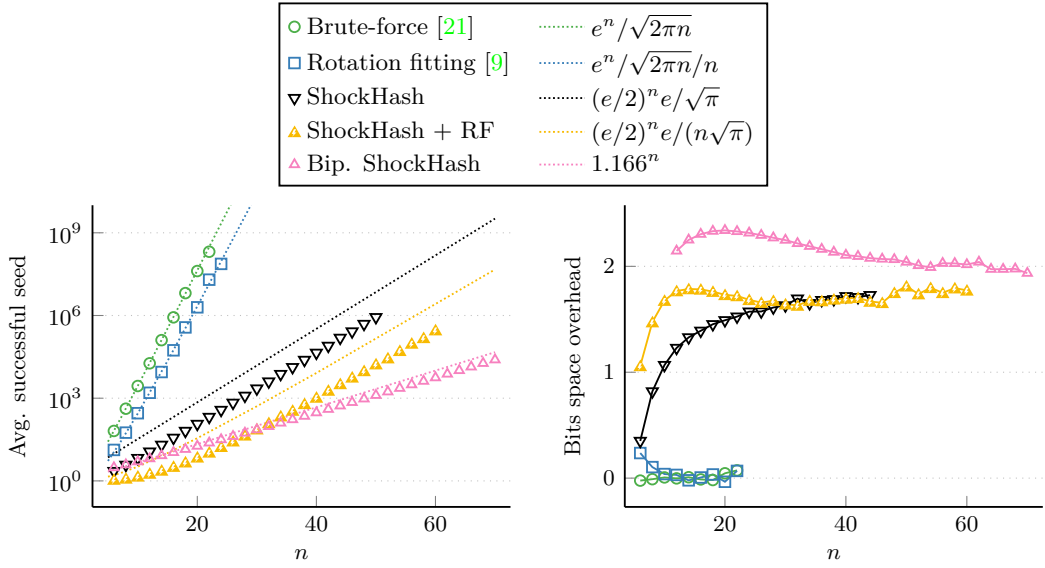
\begin{figure}[t]
  \centering
  \begin{tikzpicture}
    \ref*{legendHashEvals}
  \end{tikzpicture}
  \vspace{2mm}

  \begin{subfigure}[t]{0.48\textwidth}
    \centering
    \input{fig/hashFunctionEvals1}
    \caption{
      Average successful seed.
      For bipartite ShockHash, we use the implementation without quad split and plot the average of the largest seed that was evaluated before pairing up seeds.
      The dotted functions are upper bounds.
      Note that for bipartite ShockHash, we only know an upper bound of $\Oh(1.166^n)$ and not $1.166^n$.
    }
  \end{subfigure}
  \hspace{2mm}
  \begin{subfigure}[t]{0.48\textwidth}
    \centering
    \input{fig/hashFunctionEvals2}
    \caption{
      Idealized space overhead over the lower bound $\log_2(n^n/n!)$ in bits.
      If the average seed is $s$ we charge $\log_2(s)$ bits, plus $n$ bits for retrieval (if applicable).
    }
  \end{subfigure}

  \caption{Hash function evaluations and space overhead of ShockHash compared with more simple brute-force techniques.}
  \label{fig:hashFunctionEvals}
\end{figure}

\subsection{Number of Trials in Theory and Practice}
In \cref{fig:hashFunctionEvals}, we compare the average number of hash function trials for each bijection search technique.
From the different slopes of the curves, it is clearly visible that rotation fitting \cite{bez2023high} saves a polynomial factor compared to plain brute-force, while ShockHash saves an exponential factor.
Additionally, we plot the shown upper bounds for the number of trials of brute-force and ShockHash.
For the rotation fitting variants, we plot the base variants divided by $n$, which is not formally shown to be a theoretical bound, but is an obvious conjecture.
The plot shows that brute-force and rotation fitting are close to the given functions.
For plain ShockHash, the measurements are even better than the theory, which suggests that our proof in \cref{lem:constructionTries} is not tight.
Surprisingly, ShockHash seems to match the function we get when dividing our analysis by $\sqrt{n}$.
We conjecture that the expected number of 1-orientations of a random pseudoforest might actually not be $e\cdot\sqrt{2n}$, but close to constant.
This makes ShockHash an even better replacement for the brute-force technique.
Bipartite ShockHash matches the slope of our analysis as well, but note that our analysis only shows an upper bound of $\Oh(1.166^n)$, not an exact value.

\Cref{fig:hashFunctionEvals} also gives the difference between the idealized space consumption and the space lower bound $\log_2(n^n/n!)$.
It indicates that ShockHash loses space close to constant, which becomes negligible for larger $n$.
This explains why we need to select larger $n$ in ShockHash-RS compared to RecSplit with brute-force to achieve the same space consumption per key.
Even with these larger $n$, ShockHash construction is significantly faster than brute-force.
Bipartite ShockHash appears to have a small constant space overhead over plain ShockHash.

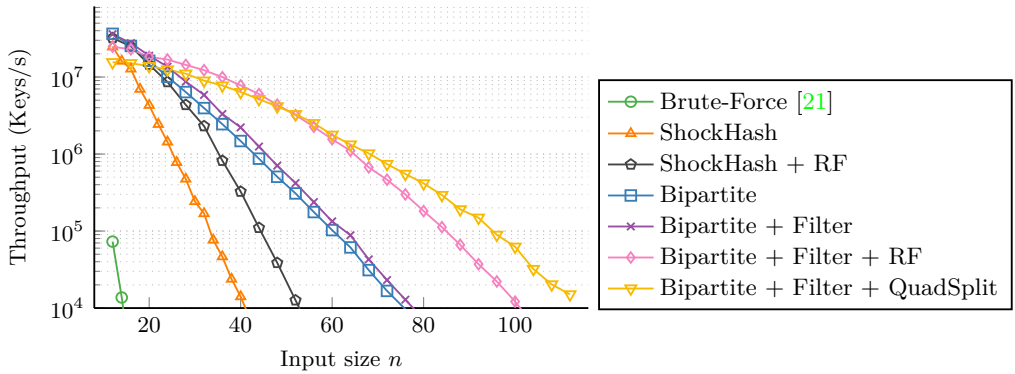
\begin{figure}[t]
  \centering
  \input{fig/differentFilters}
  \caption{Construction throughput using different methods to come up with seed candidates.
      For comparison, the plot also includes Brute-Force search.
      Variants annotated with \emph{RF} use rotation fitting.
      The filtered variants of bipartite ShockHash use the filter for isolated keys.}
  \label{fig:differentFilters}
\end{figure}

\subsection{Seed Candidate Generation}
\cref{fig:differentFilters} shows different methods to generate seed candidates.
For comparison, the plot also includes brute-force search.
It is clearly visible that ShockHash is significantly faster than brute-force.
Also, the bipartite version shows clear speedups compared to the plain version.
Filtering based on isolated keys (see \cref{s:isolatedKeys}) makes the construction about two times faster.
While rotation fitting already gives impressive speedups, our quad split technique (see \cref{s:quadsplit}) is even faster for large~$n$.
Starting with about $n=60$, the quad split technique is up to one order of magnitude faster than the basic bipartite ShockHash implementation.
Note that the comparison here needs to be taken with a grain of salt because different methods have different space overheads based on~$n$.
However, for larger $n$, the methods have almost the same space usage.
For a plot that takes space consumption into account, refer to \cref{fig:pareto}.

\subsection{Partitioning}
\begin{figure}[t]
  \centering
  \input{fig/partitioning}
  \caption{Space usage, construction performance and query performance for different partitioning schemes and base case sizes $n$.
    The total number of input keys is $N=100$ million.
    As a base case we use bipartite ShockHash with quad split.}
  \label{fig:partitioning}
\end{figure}
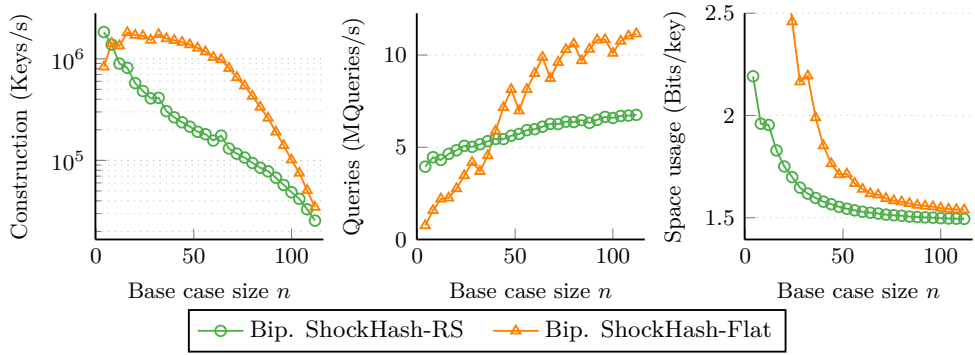

In \cref{fig:partitioning}, we give the construction time, query time, and space consumption for the two different partitioning schemes, ShockHash-RS and ShockHash-Flat.
For small $n$, the ShockHash-RS construction time is dominated by the splittings, which can be seen by the gap between ShockHash-RS and ShockHash-Flat.
For large $n$, the base case dominates and both techniques have a similar throughput.
Looking at the query performance, ShockHash-Flat is faster for $n>48$, which is the most interesting range for good space consumption.
The query throughput increases for larger base cases because we need to spend less time in the partitioning step.
The jumps in the query throughput in ShockHash-Flat are caused by the fixed-width coding of ShockHash seeds, which needs a fallback data structure when a seed does not fit.
For the same base case size, the space consumption of the flat partitioning scheme is higher than with RecSplit.
Compared to ShockHash-RS, ShockHash-Flat trades space consumption for faster queries.

\subsection{Comparison with Competitors}
We now compare ShockHash-RS and ShockHash-Flat with competitors from the literature.
Competitors include CHD \cite{belazzougui2009hash}, SicHash \cite{lehmann2023sichash}, PTHash \cite{pibiri2021pthash}, FMPHGO \cite{beling2023fingerprinting}, RecSplit \cite{esposito2020recsplit}, and SIMDRecSplit \cite{bez2023high}.
We do not plot BBHash \cite{limasset2017fast} because it is significantly outperformed by FMPH \cite{beling2023fingerprinting}, another implementation of the same technique.
While SIMDRecSplit also includes a fast GPU implementation, we leave it out from most plots, as it would be unfair because of the different hardware architecture.
While we measure ShockHash itself with $N=100$ million keys (see \cref{fig:partitioning}), we perform the comparison with competitors with only $N=10$ million keys.
This is because competitor configurations that achieve space usage close to ShockHash would otherwise need unreasonably long to compute.

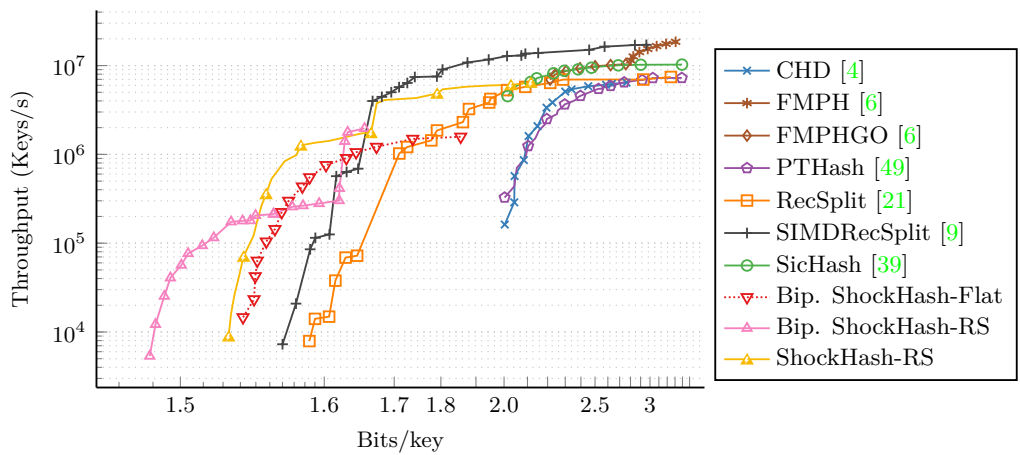
\begin{figure}[t]
  \centering \input{fig/pareto}
  \caption{
    Pareto fronts of the space usage of different competitors.
    $N=10$ million keys.
    Note that both axes are logarithmic.
    For ShockHash-RS, we use ShockHash with SIMD and rotation fitting inside RecSplit.
    For ShockHash-RS, SicHash and PTHash, we plot all Pareto optimal data points but only show markers for every fourth point to increase readability.
    Therefore, the lines might bend on positions without markers.}
  \label{fig:pareto}
\end{figure}

\myparagraph{Construction.}
\Cref{fig:pareto} gives Pareto fronts comparing the construction throughput of different competitors.
Each Pareto front only gives data points that are not dominated by another configuration of the same method regarding both space and construction time.
Note that we are mainly interested in configurations close to the space lower bound.
Therefore, the figure uses an x-axis that is logarithmic to the space lower bound $\log_2(e)$.
Only the RecSplit based competitors and ShockHash-Flat achieve space usage below 1.9 bits per key.
At configurations with less than 1.65 bits per key, ShockHash-RS can significantly outperform SIMDRecSplit.
ShockHash-RS is focused even more on the space efficient configurations than SIMDRecSplit.
For the less space efficient configurations, it does not achieve the same throughput as SIMDRecSplit.
This is not surprising because with these configurations, searching for bijections is fast, so constructing the retrieval data structure has a significant performance penalty.

\begin{table}[t]
  \caption{Query and construction performance of typical configurations of ShockHash and competitors.
    $N=10$ million keys.
    We use SIMD instructions where available.
    For the bipartite versions, we use the quad split technique.
    The table also shows GPURecSplit on an Nvidia RTX 3090.
    }
  \label{tab:queries}
  \centering \input{fig/queries}

\end{table}

\input{fig/measurementMacros}

\Cref{tab:queries} gives a selection of typical configurations.
For SicHash \cite{lehmann2023sichash}, PTHash \cite{pibiri2021pthash} and FMPHGO \cite{beling2023fingerprinting}, we use the configurations given in the original papers, where FMPHGO is configured to use the hash cache.
For RecSplit-based techniques, we mainly use space-efficient configurations with $b=2000$ and select the leaf size $n$ such that a similar space consumption is achieved.

Comparing the configurations with a space consumption of 1.56 bit per key, \ShockHashRS{} is about \speedupSIMD156{} times faster than SIMDRecSplit, which the next competitor not based on ShockHash.
For space consumptions below $1.53$ bits per key, bipartite ShockHash-RS is more space efficient and simultaneously three orders of magnitude faster than SIMDRecSplit.
Comparing different methods where each is given about half an hour of construction time, RecSplit is able to produce a perfect hash function with $1.58$ bits per key.
During the course of three years, SIMDRecSplit started a chain of work, first improving the space consumption to $1.56$ bits per key.
ShockHash-RS is then able to achieve $1.52$ bits per key, reducing the gap to the lower space bound of $\approx 1.442$ bits per key by about $30\%$.
Finally, bipartite ShockHash-RS then reduces the space usage to just $1.489$ bits per key, which is within $3.3$\% of the lower bound with practically feasible construction time.
Using a single CPU thread, bipartite ShockHash-RS achieves a space usage better than what was previously only achieved using thousands of threads on a GPU \cite{bez2023high}.
To construct an MPHF with $1.53$ bits per key, SIMDRecSplit takes more than $15$ hours, while bipartite ShockHash-RS achieves a better space consumption in just 57 seconds.
Bipartite ShockHash-Flat trades space usage for better query speed.
However, at a configuration achieving the same space usage as SIMDRecSplit, it is still about $32$ times faster to construct.

\myparagraph{Queries.}
\Cref{tab:queries} also shows the query throughput of typical configurations.
RecSplit-based techniques have slower queries than the other techniques as they have to traverse several levels of a tree, decoding variable-bitlength data in each step.
ShockHash-RS additionally needs to access a retrieval data structure.
However, when comparing configurations that achieve a similar space efficiency, the query performance of ShockHash-RS is similar to competitors.
This shows that the overhead of the retrieval operation is small compared to the work for traversing the heavily compressed tree.

Compared to plain ShockHash-RS, our bipartite implementation loses about 10\% of the query performance at the same space usage.
This is due to un-pairing seeds and having to do case distinctions based on the partitions and on the subset during quad split.
However, the larger leaves in bipartite ShockHash-RS can reduce the time spent on traversing RecSplit's splitting tree data structure, which is a major bottleneck for the queries.
Bipartite ShockHash-Flat can be constructed about $32$ times faster than the previously most space-efficient competitor SIMDRecSplit for the same space requirement.
Simultaneously, it achieves 30\% faster queries, which brings the query performance of very space efficient MPHFs much closer to competitors that are not focused on space consumption.

When query performance is the main concern, PTHash \cite{pibiri2021pthash} trades much higher space usage for much faster queries.
SicHash takes the middle-ground between PTHash and bipartite ShockHash-Flat both with respect to query time and space consumption while allowing faster construction than both approaches.

\section{Conclusion and Future Work}\label{s:conclusion}
ShockHash is a new way to compute minimal perfect hash functions on small sets.
By combining trial-and-error search with cuckoo hashing and retrieval data structures, ShockHash achieves an exponential speedup over plain brute-force (almost a factor~$2^n$).
While the plain brute-force technique samples functions and hopes for one to be an MPHF, ShockHash samples graphs and hopes for one to be a pseudoforest.
With bipartite ShockHash, we present an extension that samples bipartite graphs and performs aggressive filtering to achieve additional exponential speedups.
These improvements enable the currently fastest way to achieve near space-optimal minimal perfect hash functions and breaks the dominance of the previous best methods that relied on pure brute-force for their base-case subproblems.

We integrate ShockHash as a base case into different partitioning frameworks.
When using ShockHash inside RecSplit, we get ShockHash-RS, which can be constructed up to three orders of magnitude faster than the previous state of the art when comparing sequential codes.
Constructing with a single thread, ShockHash-RS is even faster than a tuned GPU implementation of the brute-force technique.
Using ShockHash inside a newly developed $k$-perfect hash function, we get ShockHash-Flat, which can be constructed about 32 times faster than the previous state of the art, while simultaneously having 30\% faster queries.
This starts to close the gap to constructions that are way less space efficient and brings space efficient perfect hash functions closer to practical applications.

\myparagraph{Future Work.}
In the future, additional filtering techniques would be interesting that could give additional speedups.
We already describe one such filter, using a trie to skip incompatible hash function seeds before trying to combine them.
To support even larger building blocks, a hybrid parallelization of the technique using a GPU would be an interesting future direction.
Finally, we believe that the $k$-perfect hash function we presented is useful on its own.
A next step would be to analyze the function and compare it with other $k$-perfect hash functions.

\bibliography{paper}

\end{document}

%% file: pgf_header.tex
\usepackage{pgfplots}
\pgfplotsset{compat=newest}

\usepgfplotslibrary{groupplots}
\pgfplotsset{every axis/.style={scale only axis}}

\definecolor{veryLightGrey}{HTML}{F2F2F2}
\definecolor{lightGrey}{HTML}{DDDDDD}
\definecolor{colorSimdRecSplit}{HTML}{444444}
\definecolor{colorChd}{HTML}{377EB8}
\definecolor{colorRustFmph}{HTML}{A65628}
\definecolor{colorRustFmphGo}{HTML}{A65628}
\definecolor{colorSicHash}{HTML}{4DAF4A}
\definecolor{colorPthash}{HTML}{984EA3}
\definecolor{colorRecSplit}{HTML}{FF7F00}
\definecolor{colorBbhash}{HTML}{F781BF}
\definecolor{colorShockHash}{HTML}{F8BA01}
\definecolor{colorBipartiteShockHash}{HTML}{F781BF}
\definecolor{colorBipartiteShockHashFlat}{HTML}{E41A1C}

\colorlet{colorBruteForce}{colorSicHash}
\colorlet{colorRotationFitting}{colorChd}

\pgfdeclareimage[interpolate,height=1.85mm,width=1.85mm]{shockhashMark}{fig/shockhash}
\pgfdeclareplotmark{shockhash}{\pgftext[at=\pgfpointorigin]{\pgfuseimage{shockhashMark}}}

\pgfdeclareplotmark{flippedTriangle}{%
  \pgfpathmoveto{\pgfqpointpolar{-90}{1.2\pgfplotmarksize}}%
  \pgfpathlineto{\pgfqpointpolar{30}{1.2\pgfplotmarksize}}%
  \pgfpathlineto{\pgfqpointpolar{150}{1.2\pgfplotmarksize}}%
  \pgfpathclose%
  \pgfusepath{stroke}
}

\pgfplotscreateplotcyclelist{myColorList}{%
  colorSicHash,mark=o\\%
  colorRecSplit,mark=triangle\\%
  colorSimdRecSplit,mark=pentagon\\%
  colorChd,mark=square\\%
  colorPthash,mark=x\\%
  colorBbhash,mark=diamond\\%
  colorShockHash,mark=flippedTriangle\\%
}
\pgfplotscreateplotcyclelist{transparentHeatmap}{%
  colorSicHash,mark=*,fill=colorSicHash\\
}

\pgfplotsset{
  mark repeat*/.style={
    scatter,
    scatter src=x,
    scatter/@pre marker code/.code={
      \pgfmathtruncatemacro\usemark{
        or(mod(\coordindex,#1)==0, (\coordindex==(\numcoords-1))
      }
      \ifnum\usemark=0
        \pgfplotsset{mark=none}
      \fi
    },
    scatter/@post marker code/.code={}
  },
  major grid style={thin,dotted},
  minor grid style={thin,dotted},
  ymajorgrids,
  yminorgrids,
  every axis/.append style={
    line width=0.7pt,
    tick style={
      line cap=round,
      thin,
      major tick length=4pt,
      minor tick length=2pt,
    },
    mark options={solid},
  },
  legend cell align=left,
  legend style={
    line width=0.7pt,
    /tikz/every even column/.append style={column sep=3mm,black},
    /tikz/every odd column/.append style={black},
    mark options={solid},
  },
  legend style={font=\small},
  title style={yshift=-2pt},
  enlarge x limits=0.04,
  every tick label/.append style={font=\footnotesize},
  every axis label/.append style={font=\small},
  every axis y label/.append style={yshift=-1ex},
  /pgf/number format/1000 sep={},
  axis lines*=left,
  xlabel near ticks,
  ylabel near ticks,
  axis lines*=left,
  label style={font=\footnotesize},
  tick label style={font=\footnotesize},
  cycle list name=myColorList,
  plotParameters/.style={
    width=35.0mm,
    height=30.0mm,
  },
  plotScaling/.style={
    width=38.0mm,
    height=30.0mm,
  },
  plotScalingConfigs/.style={
    width=35.0mm,
    height=30.0mm,
  },
  plotPareto/.style={
    width=80.0mm,
    height=50.0mm,
  },
  plotHfEvals/.style={
    width=55.0mm,
    height=40.0mm,
  },
  plotLeafMethods/.style={
    width=29.0mm,
    height=30.0mm,
  },
  plotPartitioning/.style={
    width=30.0mm,
    height=30.0mm,
  },
  plotProbabilities/.style={
    width=35.0mm,
    height=30.0mm,
    only marks,
    mark size=.75pt,
    cycle list name=transparentHeatmap,
  },
  plotFilters/.style={
    width=65.0mm,
    height=40.0mm,
  },
}

%% file: unicode.tex
\usepackage{newunicodechar}
\usepackage{silence}
\newunicodechar{♢}{\tikz \node[inner sep=1.5,draw,diamond] {};}
\newunicodechar{☆}{\tikz \node[inner sep=1,draw,star,star point ratio=2] {};}
\newunicodechar{△}{\triangle}
\newunicodechar{⬜}{\kern 0.5pt\tikz \node[inner sep=1.7,draw,regular polygon,regular polygon sides=4] {};\kern 0.5pt}
\newunicodechar{◯}{\tikz[baseline=-3pt] \node[inner sep=1.7,draw,cloud,cloud puffs=4,cloud puff arc=190] {};}

\newunicodechar{⊥}{\bot}
\newunicodechar{•}{\item}
\newunicodechar{✓}{\checkmark}
\newunicodechar{✗}{\xmark}
\newunicodechar{…}{\dots}
\newunicodechar{≔}{\coloneqq}
\newunicodechar{⁻}{^-}
\newunicodechar{⁺}{^+}
\newunicodechar{₋}{_-}
\newunicodechar{₊}{_+}
\newunicodechar{ℓ}{\ell}
\newunicodechar{•}{\item}
\newunicodechar{…}{\dots}
\newunicodechar{≔}{\coloneqq}
\newif\ifnormopen\normopenfalse
\newunicodechar{‖}{\ifnormopen\rVert\normopenfalse\else\rVert\normopentrue\fi}
\newunicodechar{≤}{\leq}
\newunicodechar{≥}{\geq}
\newunicodechar{≰}{\nleq}
\newunicodechar{≱}{\ngeq}
\newunicodechar{⊕}{\oplus}
\newunicodechar{⊗}{\otimes}
\newunicodechar{≠}{\neq}
\newunicodechar{¬}{\neg}
\newunicodechar{≡}{\equiv}
\newunicodechar{₀}{_0}
\newunicodechar{₁}{_1}
\newunicodechar{₂}{_2}
\newunicodechar{₃}{_3}
\newunicodechar{₄}{_4}
\newunicodechar{₅}{_5}
\newunicodechar{₆}{_6}
\newunicodechar{₇}{_7}
\newunicodechar{₈}{_8}
\newunicodechar{₉}{_9}
\newunicodechar{ₚ}{_p}
\newunicodechar{ₙ}{_n}
\newunicodechar{ₐ}{_a}
\newunicodechar{ₑ}{_e}
\newunicodechar{ₕ}{_h}
\newunicodechar{ₖ}{_k}
\newunicodechar{ₗ}{_l}
\newunicodechar{ₘ}{_m}
\newunicodechar{ₛ}{_s}
\newunicodechar{ₜ}{_t}
\newunicodechar{ₓ}{_x}
\newunicodechar{⁰}{^0}
\newunicodechar{¹}{^1}
\newunicodechar{²}{^2}
\newunicodechar{³}{^3}
\newunicodechar{⁴}{^4}
\newunicodechar{⁵}{^5}
\newunicodechar{⁶}{^6}
\newunicodechar{⁷}{^7}
\newunicodechar{⁸}{^8}
\newunicodechar{⁹}{^9}
\newunicodechar{∈}{\in}
\newunicodechar{∉}{\notin}
\newunicodechar{⊂}{\subset}
\newunicodechar{⊃}{\supset}
\newunicodechar{⊆}{\subseteq}
\newunicodechar{⊇}{\supseteq}
\newunicodechar{⊄}{\nsubset}
\newunicodechar{⊅}{\nsupset}
\newunicodechar{⊈}{\nsubseteq}
\newunicodechar{⊉}{\nsupseteq}
\newunicodechar{∪}{\cup}
\newunicodechar{∩}{\cap}
\newunicodechar{∀}{\forall}
\newunicodechar{∃}{\exists}
\newunicodechar{∄}{\nexists}
\newunicodechar{∨}{\vee}
\newunicodechar{∧}{\wedge}
\newunicodechar{ℝ}{\mathbb{R}}
\newunicodechar{ℙ}{\mathbb{P}}
\newunicodechar{ℕ}{\mathbb{N}}
\newunicodechar{𝔼}{\mathbb{E}}
\newunicodechar{𝔽}{\mathbb{F}}
\newunicodechar{ℤ}{\mathbb{Z}}
\newunicodechar{⌊}{\lfloor}
\newunicodechar{⌋}{\rfloor}
\newunicodechar{⌈}{\lceil}
\newunicodechar{⌉}{\rceil}
\newunicodechar{·}{\cdot}
\newunicodechar{∘}{\circ}
\newunicodechar{×}{\times}
\newunicodechar{↑}{\uparrow}
\newunicodechar{↓}{\downarrow}
\newunicodechar{→}{\rightarrow}
\newunicodechar{←}{\leftarrow}
\newunicodechar{⇒}{\Rightarrow}
\newunicodechar{⇐}{\Leftarrow}
\newunicodechar{↔}{\leftrightarrow}
\newunicodechar{⇔}{\Leftrightarrow}
\newunicodechar{↦}{\mapsto}
\newunicodechar{∅}{\varnothing}
\newunicodechar{∞}{\infty}
\newunicodechar{≅}{\cong}
\newunicodechar{≈}{\approx}
\newunicodechar{ℓ}{\ell}
\newunicodechar{𝟙}{\mathds{1}}
\newunicodechar{𝟘}{\mathds{0}}
\newunicodechar{↪}{\hookrightarrow}

\newunicodechar{α}{\alpha}
\newunicodechar{β}{\beta}
\newunicodechar{γ}{\gamma}
\newunicodechar{Γ}{\Gamma}
\newunicodechar{δ}{\delta}
\newunicodechar{Δ}{\Delta}
\newunicodechar{ε}{\varepsilon}
\newunicodechar{ζ}{\zeta}
\newunicodechar{η}{\eta}
\newunicodechar{θ}{\theta}
\newunicodechar{Θ}{\Theta}
\newunicodechar{ι}{\iota}
\newunicodechar{κ}{\kappa}
\newunicodechar{λ}{\lambda}
\newunicodechar{Λ}{\Lambda}
\newunicodechar{μ}{\mu}
\newunicodechar{ν}{\nu}
\newunicodechar{ξ}{\xi}
\newunicodechar{Ξ}{\Xi}
\newunicodechar{π}{\pi}
\newunicodechar{Π}{\Pi}
\newunicodechar{ρ}{\rho}
\newunicodechar{σ}{\sigma}
\newunicodechar{Σ}{\Sigma}
\newunicodechar{τ}{\tau}
\newunicodechar{υ}{\upsilon}
\newunicodechar{ϒ}{\Upsilon}
\newunicodechar{φ}{\varphi}
\newunicodechar{ϕ}{\phi}
\newunicodechar{Φ}{\Phi}
\newunicodechar{χ}{\chi}
\newunicodechar{ψ}{\psi}
\newunicodechar{Ψ}{\Psi}
\newunicodechar{ω}{\omega}
\newunicodechar{Ω}{\Omega}

\newunicodechar{𝒜}{\mathcal{A}}
\newunicodechar{ℬ}{\mathcal{B}}
\newunicodechar{𝒞}{\mathcal{C}}
\newunicodechar{𝒟}{\mathcal{D}}
\newunicodechar{ℰ}{\mathcal{E}}
\newunicodechar{ℱ}{\mathcal{F}}
\newunicodechar{𝒢}{\mathcal{G}}
\newunicodechar{ℋ}{\mathcal{H}}
\newunicodechar{ℐ}{\mathcal{I}}
\newunicodechar{𝒥}{\mathcal{J}}
\newunicodechar{𝒦}{\mathcal{K}}
\newunicodechar{ℒ}{\mathcal{L}}
\newunicodechar{ℳ}{\mathcal{M}}
\newunicodechar{𝒩}{\mathcal{N}}
\newunicodechar{𝒪}{\mathcal{O}}
\newunicodechar{𝒫}{\mathcal{P}}
\newunicodechar{𝒬}{\mathcal{Q}}
\newunicodechar{ℛ}{\mathcal{R}}
\newunicodechar{𝒮}{\mathcal{S}}
\newunicodechar{𝒯}{\mathcal{T}}
\newunicodechar{𝒰}{\mathcal{U}}
\newunicodechar{𝒱}{\mathcal{V}}
\newunicodechar{𝒲}{\mathcal{W}}
\newunicodechar{𝒳}{\mathcal{X}}
\newunicodechar{𝒴}{\mathcal{Y}}
\newunicodechar{𝒵}{\mathcal{Z}}
\newunicodechar{𝒶}{\mathcal{a}}
\newunicodechar{𝒷}{\mathcal{b}}
\newunicodechar{𝒸}{\mathcal{c}}
\newunicodechar{𝒹}{\mathcal{d}}
\newunicodechar{ℯ}{\mathcal{e}}
\newunicodechar{𝒻}{\mathcal{f}}
\newunicodechar{ℊ}{\mathcal{g}}
\newunicodechar{𝒽}{\mathcal{h}}
\newunicodechar{𝒾}{\mathcal{i}}
\newunicodechar{𝒿}{\mathcal{j}}
\newunicodechar{𝓀}{\mathcal{k}}
\newunicodechar{𝓁}{\mathcal{l}}
\newunicodechar{𝓂}{\mathcal{m}}
\newunicodechar{𝓃}{\mathcal{n}}
\newunicodechar{ℴ}{\mathcal{o}}
\newunicodechar{𝓅}{\mathcal{p}}
\newunicodechar{𝓆}{\mathcal{q}}
\newunicodechar{𝓇}{\mathcal{r}}
\newunicodechar{𝓈}{\mathcal{s}}
\newunicodechar{𝓉}{\mathcal{t}}
\newunicodechar{𝓊}{\mathcal{u}}
\newunicodechar{𝓋}{\mathcal{v}}
\newunicodechar{𝓌}{\mathcal{w}}
\newunicodechar{𝓍}{\mathcal{x}}
\newunicodechar{𝓎}{\mathcal{y}}
\newunicodechar{𝓏}{\mathcal{z}}

%% file: fig/hashFunctionEvals1.tex
    \begin{tikzpicture}
        \begin{axis}[
            xlabel={$n$},
            ylabel={Avg. successful seed},
            plotHfEvals,
            ymode=log,
            ymax=1e10,
            legend columns=5,
            transpose legend,
            legend to name=legendHashEvals,
          ]
          \addplot+[color=colorBruteForce,only marks] coordinates { (6,64.8549) (8,416.091) (10,2767.63) (12,18417.6) (14,127722.0) (16,856746.0) (18,6.48792e+06) (20,4.09977e+07) (22,2.05908e+08) };
          \addlegendentry{Brute-force \cite{esposito2020recsplit}};
          \addplot+[mark=square,color=colorRotationFitting,only marks] coordinates { (6,13.3014) (8,56.3617) (10,285.461) (12,1573.24) (14,8911.15) (16,54628.5) (18,369627.0) (20,1.9798e+06) (22,1.99377e+07) (24,7.45176e+07) };
          \addlegendentry{Rotation fitting \cite{bez2023high}};
          \addplot+[mark=flippedTriangle,color=black,only marks] coordinates { (6,2.2757) (8,3.8287) (10,6.73595) (12,11.5547) (14,20.3844) (16,36.5247) (18,64.9892) (20,116.499) (22,208.758) (24,374.636) (26,680.764) (28,1221.88) (30,2238.19) (32,4023.07) (34,7519.32) (36,13354.2) (38,24472.3) (40,44311.0) (42,79172.2) (44,154603.0) (46,263326.0) (48,492942.0) (50,868759.0) };
          \addlegendentry{ShockHash};
          \addplot+[mark=shockhash,color=colorShockHash,only marks] coordinates { (6,1.12525) (8,1.23685) (10,1.4658) (12,1.85955) (14,2.40425) (16,3.3074) (18,4.78105) (20,7.2033) (22,10.88855) (24,17.4039) (26,27.6061) (28,45.1996) (30,75.3199) (32,127.268) (34,223.582) (36,357.877) (38,625.271) (40,1065.71) (42,1908.39) (44,3303.09) (46,5731.75) (48,9796.6) (50,17251.6) (52,28482.7) (54,52678.4) (56,97198.0) (58,154722.0) (60,295654.0) };
          \addlegendentry{ShockHash + RF};
          \addplot[mark=triangle,color=colorBipartiteShockHash,only marks] coordinates { (6,3.03015) (8,3.87401) (10,4.97847) (12,6.42104) (14,8.31067) (16,10.79041) (18,14.0589) (20,18.3211) (22,23.9634) (24,31.5383) (26,41.3931) (28,54.5764) (30,72.0575) (32,95.2126) (34,126.487) (36,167.987) (38,223.363) (40,298.241) (42,398.312) (44,529.578) (46,707.522) (48,949.309) (50,1276.0) (52,1709.08) (54,2301.27) (56,3099.2) (58,4155.41) (60,5554.71) (62,7558.61) (64,10148.5) (66,13711.7) (68,18469.5) (70,24832.8) };
          \addlegendentry{Bip. ShockHash};

          \addplot[mark=none,densely dotted,color=colorBruteForce] coordinates { (5,26.4787) (7,165.357) (10,2778.78) (15,336730) (20,4.32797e+07) (30,7.78366e+11) (40,1.48477e+16) (70,1.19943e+29) };
          \addlegendentry{$e^n/\sqrt{2 \pi n}$};
          \addplot[mark=none,densely dotted,color=colorRotationFitting] coordinates { (5,5.29575) (7,23.6224) (10,277.878) (15,22448.7) (20,2.16399e+06) (30,2.59455e+10) (40,3.71193e+14) (70,1.71347e+27) };
          \addlegendentry{$e^n/\sqrt{2 \pi n}/n$};
          \addplot[mark=none,densely dotted,color=black] coordinates { (5,7.11282) (7,13.1393) (10,32.9886) (15,152.998) (20,709.593) (30,15263.5) (40,328321) (70,3.26764e+09) };
          \addlegendentry{$(e/2)^ne/\sqrt{\pi}$};
          \addplot[mark=none,densely dotted,color=colorShockHash] coordinates { (5,1.42256) (7,1.87704) (10,3.29886) (15,10.1999) (20,35.4796) (30,508.783) (40,8208.03) (70,4.66805e+07) };
          \addlegendentry{$(e/2)^ne/(n\sqrt{\pi})$};
          \addplot[mark=none,densely dotted,color=colorBipartiteShockHash] coordinates { (5,2.15523) (7,2.93015) (10,4.645) (15,10.011) (20,21.576) (30,100.22) (40,465.524) (70,46655) };
          \addlegendentry{$1.166^n$};
        \end{axis}
    \end{tikzpicture}

%% file: fig/hashFunctionEvals2.tex
    \begin{tikzpicture}
        \begin{axis}[
            plotHfEvals,
            title={},
            xlabel={$n$},
            ylabel={Bits space overhead}, %
            cycle list name=myColorList,
          ]
          \addplot+[color=colorBruteForce] coordinates { (6,-0.0234508) (8,-0.0100292) (10,0.00662729) (12,-0.00246884) (14,0.00832896) (16,-0.0123481) (18,-0.0181847) (20,0.0455501) (22,0.0731805) };
          \addlegendentry{Brute-Force  \cite{esposito2020recsplit}};
          \addplot+[mark=square,color=colorRotationFitting] coordinates { (6,0.236825) (8,0.103994) (10,0.0384379) (12,0.0303382) (14,-0.0208516) (16,0.00583817) (18,0.0344676) (20,-0.0344609) (22,0.0682468) };
          \addlegendentry{Rotation Fitting \cite{bez2023high}};
          \addplot+[mark=flippedTriangle,color=black] coordinates { (6,0.351081) (8,0.818735) (10,1.06764) (12,1.22911) (14,1.32902) (16,1.39139) (18,1.45315) (20,1.49068) (22,1.52505) (24,1.57209) (26,1.57358) (28,1.61109) (30,1.63094) (32,1.69517) (34,1.65146) (36,1.68107) (38,1.68893) (40,1.71764) (42,1.70878) (44,1.72791) };
          \addlegendentry{ShockHash};
          \addplot+[mark=triangle,color=colorBipartiteShockHash] coordinates { (12,2.14548) (14,2.24975) (16,2.30365) (18,2.33233) (20,2.33966) (22,2.3294) (24,2.31568) (26,2.29397) (28,2.26973) (30,2.24778) (32,2.21629) (34,2.18955) (36,2.1621) (38,2.13457) (40,2.10602) (42,2.0947) (44,2.07744) (46,2.06926) (48,2.0649) (50,2.04046) (52,2.00809) (54,1.99055) (56,2.02893) (58,2.02773) (60,2.0171) (62,2.04118) (64,1.97305) (66,1.97289) (68,1.97629) (70,1.93544) };
          \addlegendentry{ShockHash2Filter};
          \addplot+[mark=shockhash,color=colorShockHash] coordinates { (6,1.06005) (8,1.47376) (10,1.67663) (12,1.76716) (14,1.7863) (16,1.78234) (18,1.76139) (20,1.73096) (22,1.72061) (24,1.68402) (26,1.6575) (28,1.675) (30,1.64282) (32,1.63045) (34,1.67673) (36,1.67297) (38,1.68878) (40,1.6969) (42,1.70341) (44,1.66812) (46,1.65585) (48,1.7503) (50,1.81236) (52,1.73831) (54,1.79739) (56,1.75005) (58,1.79413) (60,1.77358) };
          \addlegendentry{ShockHash + RF};

          \legend{};
        \end{axis}
    \end{tikzpicture}

%% file: fig/differentFilters.tex
    \centering
    \begin{tikzpicture}
        \begin{axis}[
            xlabel={Input size $n$},
            ylabel={Throughput (Keys/s)},
            plotFilters,
            ymode=log,
            ymin=1e4,
            legend to name=differentFiltersLegend,
            legend columns=1,
          ]

          \addplot coordinates { (12,72954.1) (14,13712) (16,1804.67) (18,258.165) (20,30.9706) (22,18.5795) (24,0.305102) };
          \addlegendentry{Brute-Force \cite{esposito2020recsplit}};
          \addplot coordinates { (12,2.46852e+07) (14,1.61179e+07) (16,1.27499e+07) (18,6.96077e+06) (20,4.30233e+06) (22,2.4352e+06) (24,1.43743e+06) (26,779082) (28,474338) (30,242158) (32,168671) (34,76508.5) (36,46466.8) (38,23780.8) (40,14076.1) (42,7302.42) (44,4388.28) (46,2397.72) };
          \addlegendentry{ShockHash};
          \addplot coordinates { (12,3.21106e+07) (16,2.49918e+07) (20,1.44941e+07) (24,8.61837e+06) (28,4.34824e+06) (32,2.30607e+06) (36,819098) (40,325415) (44,110245) (48,38643.6) (52,12624) (56,4028.57) (60,1524.02) };
          \addlegendentry{ShockHash + RF};
          \addplot coordinates { (12,3.64201e+07) (16,2.53469e+07) (20,1.62569e+07) (24,1.00866e+07) (28,6.37979e+06) (32,3.94932e+06) (36,2.4417e+06) (40,1.47519e+06) (44,870815) (48,506554) (52,308068) (56,176472) (60,102804) (64,61283.3) (68,31095.4) (72,16704.2) (76,9666.53) (80,5862.87) (84,3376.21) (88,1737.9) (92,983.151) (96,552.843) (100,279.71) (104,140.275) (108,76.2389) (112,42.5768) };
          \addlegendentry{Bipartite};
          \addplot coordinates { (12,3.60051e+07) (16,2.75746e+07) (20,1.89344e+07) (24,1.36095e+07) (28,8.71083e+06) (32,5.85406e+06) (36,3.30934e+06) (40,2.21734e+06) (44,1.25075e+06) (48,706951) (52,419426) (56,237742) (60,133464) (64,88280.9) (68,42861.9) (72,23105.2) (76,12844.6) (80,7164.41) (84,4047.02) (88,2155.46) (92,1156.52) (96,647.188) (100,332.908) (104,167.862) (108,88.3739) (112,46.8627) };
          \addlegendentry{Bipartite + Filter};
          \addplot coordinates { (12,2.47656e+07) (16,2.28857e+07) (20,1.83184e+07) (24,1.68556e+07) (28,1.43833e+07) (32,1.2275e+07) (36,9.95818e+06) (40,7.79776e+06) (44,6.01491e+06) (48,4.37409e+06) (52,3.20406e+06) (56,2.25674e+06) (60,1.55756e+06) (64,1.10376e+06) (68,671864) (72,463762) (76,299266) (80,181679) (84,112386) (88,66194.3) (92,37164.2) (96,22146.7) (100,12123) (104,6065.13) (108,5919.82) (112,2518.49) };
          \addlegendentry{Bipartite + Filter + RF};
          \addplot coordinates { (12,1.55641e+07) (16,1.51147e+07) (20,1.38537e+07) (24,1.25171e+07) (28,1.10488e+07) (32,8.97648e+06) (36,7.74943e+06) (40,6.37761e+06) (44,5.15994e+06) (48,4.14812e+06) (52,3.2887e+06) (56,2.51739e+06) (60,1.77738e+06) (64,1.32896e+06) (68,1.01784e+06) (72,741740) (76,552165) (80,415909) (84,293239) (88,191343) (92,148909) (96,89241.7) (100,62518) (104,32120.6) (108,20279.7) (112,15039) };
          \addlegendentry{Bipartite + Filter + QuadSplit};

        \end{axis}
    \end{tikzpicture}
    \begin{tikzpicture}[baseline=-3cm]
        \ref*{differentFiltersLegend}
    \end{tikzpicture}

%% file: fig/partitioning.tex
    \centering
    \begin{tikzpicture}
        \begin{axis}[
            xlabel={Base case size $n$},
            ylabel={Construction (Keys/s)},
            plotPartitioning,
            legend to name=partitioningLegend,
            legend columns=2,
            ymode=log,
          ]

          \addplot coordinates { (4,1.83184e+06) (8,1.37906e+06) (12,895688) (16,815116) (20,576342) (24,481176) (28,407558) (32,412821) (36,305721) (40,265189) (44,237143) (48,214774) (52,191266) (56,181085) (60,156710) (64,175208) (68,131050) (72,116493) (76,106718) (80,94363.9) (84,84760.8) (88,78083.9) (92,67605.3) (96,57318.1) (100,48616.4) (104,41855.5) (108,32998.9) (112,25559.8) };
          \addlegendentry{Bip. ShockHash-RS};
          \addplot coordinates { (4,832771) (8,1.42286e+06) (12,1.34286e+06) (16,1.81133e+06) (20,1.705e+06) (24,1.67726e+06) (28,1.52774e+06) (32,1.73509e+06) (36,1.57938e+06) (40,1.51955e+06) (44,1.45758e+06) (48,1.37671e+06) (52,1.27392e+06) (56,1.17214e+06) (60,1.03231e+06) (64,982830) (68,805860) (72,654463) (76,541589) (80,429356) (84,334814) (88,260736) (92,190476) (96,140246) (100,101325) (104,75050.7) (108,50694.8) (112,34618.4) };
          \addlegendentry{Bip. ShockHash-Flat};

        \end{axis}
    \end{tikzpicture}
    \begin{tikzpicture}
        \begin{axis}[
            xlabel={Base case size $n$},
            ylabel={Queries (MQueries/s)},
            plotPartitioning,
            legend to name=partitioningLegend2,
            legend columns=1,
            ymin=0,
          ]

          \addplot coordinates { (4,3.94649) (8,4.4603) (12,4.3109) (16,4.63886) (20,4.83465) (24,5.07846) (28,5.02412) (32,5.15943) (36,5.33049) (40,5.44929) (44,5.4407) (48,5.62746) (52,5.72082) (56,5.92909) (60,5.97836) (64,6.11583) (68,6.26841) (72,6.25586) (76,6.38855) (80,6.3674) (84,6.47165) (88,6.32231) (92,6.48971) (96,6.64673) (100,6.58848) (104,6.69748) (108,6.71953) (112,6.75584) };
          \addlegendentry{RecSplit};
          \addplot coordinates { (4,0.755955) (8,1.57421) (12,2.19829) (16,2.25012) (20,2.75786) (24,3.46452) (28,4.18673) (32,3.69495) (36,4.54711) (40,5.88512) (44,7.15564) (48,8.1367) (52,6.97691) (56,8.13935) (60,9.00252) (64,9.8824) (68,8.74049) (72,9.60615) (76,10.3018) (80,10.6033) (84,9.7012) (88,10.3146) (92,10.8213) (96,10.8401) (100,10.1031) (104,10.7585) (108,11.0327) (112,11.1532) };
          \addlegendentry{$k$-perfect};

        \end{axis}
    \end{tikzpicture}
    \begin{tikzpicture}
        \begin{axis}[
            xlabel={Base case size $n$},
            ylabel={Space usage (Bits/key)},
            plotPartitioning,
            legend to name=partitioningLegend3,
            legend columns=1,
            ymax=2.5,
          ]

          \addplot coordinates { (4,2.19151) (8,1.95973) (12,1.95337) (16,1.82904) (20,1.75034) (24,1.69838) (28,1.64706) (32,1.61786) (36,1.59658) (40,1.57846) (44,1.56609) (48,1.55303) (52,1.54385) (56,1.53607) (60,1.52918) (64,1.52451) (68,1.52036) (72,1.51498) (76,1.51194) (80,1.50962) (84,1.50555) (88,1.50325) (92,1.5017) (96,1.50021) (100,1.49751) (104,1.49611) (108,1.49525) (112,1.49412) };
          \addlegendentry{RecSplit};
          \addplot coordinates { (4,16.6252) (8,6.20831) (12,4.02398) (16,3.53059) (20,2.87134) (24,2.45883) (28,2.1651) (32,2.19361) (36,1.98962) (40,1.852) (44,1.76436) (48,1.71077) (52,1.71404) (56,1.66814) (60,1.63968) (64,1.61802) (68,1.61181) (72,1.59429) (76,1.58348) (80,1.57765) (84,1.56961) (88,1.56156) (92,1.55683) (96,1.55392) (100,1.54695) (104,1.54242) (108,1.53978) (112,1.53835) };
          \addlegendentry{$k$-perfect};

        \end{axis}
    \end{tikzpicture}

    \begin{tikzpicture}
        \ref*{partitioningLegend}
    \end{tikzpicture}

%% file: fig/pareto.tex
\centering
\def\diff{1.443}
    \begin{tikzpicture}
        \begin{axis}[
            plotPareto,
            xlabel={Bits/key},
            ylabel={Throughput (Keys/s)},
            legend to name=paretoLegend,
            legend columns=1,
            xmin=1.48-\diff,
            xmode=log,
            ymode=log,
            minor xtick={1.48-\diff,1.49-\diff,1.51-\diff,1.52-\diff,1.53-\diff,1.54-\diff,1.55-\diff,1.56-\diff,1.57-\diff,1.58-\diff,1.59-\diff},
            xtick=      {1.5-\diff, 1.6-\diff, 1.7-\diff, 1.8-\diff, 1.9-\diff, 2.0-\diff, 2.1-\diff, 2.2-\diff, 2.3-\diff, 2.4-\diff, 2.5-\diff, 2.6-\diff, 2.7-\diff,
                         2.8-\diff, 2.9-\diff, 3-\diff, 3.1-\diff, 3.2-\diff, 3.3-\diff, 3.4-\diff, 3.5-\diff},
            xticklabels={1.5,       1.6,       1.7,       1.8,       ~,         2.0,       ~,         ~,         ~,         ~,         2.5,       ~,         ~,
                         ~,         ~,         3,       ~,         ~,         ~,         ~,         ~},
          ]
          \addplot[mark=x,color=colorChd,solid] coordinates { (0.55975,162161) (0.59852,288484) (0.59988,569930) (0.64022,863036) (0.66231,1.60953e+06) (0.70496,2.07943e+06) (0.75118,3.35683e+06) (0.78332,3.8373e+06) (0.85723,5.09424e+06) (0.89593,5.40833e+06) (1.00738,5.85138e+06) (1.14262,6.08273e+06) (1.19009,6.23053e+06) (1.31893,6.42674e+06) };
          \addlegendentry{CHD \cite{belazzougui2009hash}};
          \addplot[mark=asterisk,color=colorRustFmph,solid] coordinates { (1.36055,1.14025e+07) (1.38392,1.26582e+07) (1.44465,1.41844e+07) (1.53219,1.53846e+07) (1.63141,1.65563e+07) (1.744,1.75439e+07) (1.86478,1.85185e+07) };
          \addlegendentry{FMPH \cite{beling2023fingerprinting}};
          \addplot[mark=diamond,color=colorRustFmphGo,solid] coordinates { (0.77008,6.94927e+06) (0.79816,7.93651e+06) (0.86224,8.69565e+06) (0.95067,9.32836e+06) (1.05783,9.82318e+06) (1.17955,1.01215e+07) (1.31284,1.03734e+07) };
          \addlegendentry{FMPHGO \cite{beling2023fingerprinting}};
          \addplot[mark=pentagon,color=colorPthash,solid,mark repeat*=4] coordinates { (0.55915,326755) (0.5987,437178) (0.6095,706564) (0.64676,903914) (0.66001,1.241e+06) (0.69441,1.51883e+06) (0.71228,1.86289e+06) (0.74449,2.18914e+06) (0.75265,2.50125e+06) (0.80817,2.8393e+06) (0.81253,3.10752e+06) (0.8415,3.44116e+06) (0.85396,3.65764e+06) (0.90678,3.98089e+06) (0.92326,4.12882e+06) (0.93972,4.42478e+06) (0.95373,4.55996e+06) (0.99091,4.82859e+06) (1.00548,4.91642e+06) (1.05668,5.31067e+06) (1.08328,5.47046e+06) (1.10345,5.48246e+06) (1.12563,5.70451e+06) (1.14476,5.70776e+06) (1.17896,5.90667e+06) (1.18915,5.95238e+06) (1.23149,6.14628e+06) (1.25906,6.33312e+06) (1.29759,6.48088e+06) (1.34228,6.58328e+06) (1.3909,6.72495e+06) (1.4456,6.83527e+06) (1.47356,6.86342e+06) (1.49946,6.97837e+06) (1.52879,7.02741e+06) (1.56381,7.09723e+06) (1.58804,7.24113e+06) (1.94732,7.26216e+06) };
          \addlegendentry{PTHash \cite{pibiri2021pthash}};
          \addplot[mark=square,color=colorRecSplit,solid] coordinates { (0.1414,7902.7) (0.14714,14056.1) (0.16251,14894.7) (0.16981,37885.4) (0.18248,68818.9) (0.19795,72320.2) (0.26638,1.0227e+06) (0.28155,1.20963e+06) (0.33362,1.43699e+06) (0.34748,1.85908e+06) (0.41687,2.3084e+06) (0.43497,3.24254e+06) (0.50151,3.8373e+06) (0.50591,4.2123e+06) (0.56893,5.29101e+06) (0.64679,5.75705e+06) (0.77108,6.38162e+06) (0.84398,6.95894e+06) (1.48486,6.96379e+06) (1.80348,7.44048e+06) };
          \addlegendentry{RecSplit \cite{esposito2020recsplit}};
          \addplot[mark=+,color=colorSimdRecSplit,solid] coordinates { (0.11705,7278.14) (0.12868,20867.6) (0.14219,85321.3) (0.14751,114911) (0.16291,125518) (0.17096,568150) (0.18379,633955) (0.19939,690369) (0.22072,3.99202e+06) (0.23552,4.44444e+06) (0.25176,5.00751e+06) (0.26602,5.7241e+06) (0.28184,6.36943e+06) (0.29715,7.43494e+06) (0.34713,7.5358e+06) (0.36115,9.01713e+06) (0.43057,1.08342e+07) (0.49994,1.16959e+07) (0.56897,1.28041e+07) (0.62915,1.2987e+07) (0.64728,1.36054e+07) (0.7085,1.38696e+07) (1.01447,1.50376e+07) (1.13051,1.63399e+07) (1.39963,1.70358e+07) (1.51742,1.71233e+07) };
          \addlegendentry{SIMDRecSplit \cite{bez2023high}};
          \addplot[mark=o,color=colorSicHash,solid,mark repeat*=4] coordinates { (0.57121,4.52694e+06) (0.60093,5.79039e+06) (0.63208,6.28931e+06) (0.64081,6.43501e+06) (0.67132,6.52316e+06) (0.67136,6.75219e+06) (0.6893,6.77048e+06) (0.697,7.00771e+06) (0.70143,7.18907e+06) (0.7277,7.53012e+06) (0.75819,7.69823e+06) (0.75906,8.1103e+06) (0.78904,8.19001e+06) (0.78973,8.56164e+06) (0.82048,8.62813e+06) (0.84592,8.67303e+06) (0.85121,8.73362e+06) (0.87731,8.81057e+06) (0.88136,8.88099e+06) (0.90725,8.99281e+06) (0.93843,9.06618e+06) (0.94304,9.07441e+06) (0.96894,9.27644e+06) (0.97379,9.44287e+06) (1.03052,9.46074e+06) (1.03459,9.61538e+06) (1.0918,9.72763e+06) (1.18379,9.94036e+06) (1.24485,1.00604e+07) (1.2759,1.01317e+07) (1.30593,1.0142e+07) (1.36769,1.01626e+07) (1.45941,1.02145e+07) (1.95003,1.02564e+07) };
          \addlegendentry{SicHash \cite{lehmann2023sichash}};
          \addplot[mark=flippedTriangle,color=colorBipartiteShockHashFlat,densely dotted] coordinates { (0.08862,14634.7) (0.09578,23336.1) (0.09658,42359.2) (0.09797,63433.2) (0.10439,104247) (0.11095,143940) (0.11635,223623) (0.12218,299652) (0.13458,434122) (0.1417,551568) (0.159,756143) (0.18351,907605) (0.1965,1.05685e+06) (0.22686,1.21447e+06) (0.29277,1.49298e+06) (0.41049,1.57878e+06) };
          \addlegendentry{Bip. ShockHash-Flat};
          \addplot[mark=triangle,color=colorBipartiteShockHash,solid] coordinates { (0.04594,5354.34) (0.04785,12183) (0.05091,25325.1) (0.05314,40679) (0.05739,56677.8) (0.0602,76601.9) (0.06654,93980.5) (0.07211,115334) (0.08149,173892) (0.08831,178651) (0.09335,179469) (0.09681,205373) (0.10969,212062) (0.11003,214110) (0.12538,257050) (0.13526,264606) (0.15146,279728) (0.17435,301905) (0.17476,414662) (0.18141,1.40233e+06) (0.18524,1.77999e+06) (0.20851,1.9604e+06) };
          \addlegendentry{Bip. ShockHash-RS};
          \addplot[mark=shockhash,color=colorShockHash,solid,mark repeat*=4] coordinates { (0.08006,8961.08) (0.08143,15933.7) (0.08344,26258.2) (0.08619,42659.9) (0.08895,70807.5) (0.09543,121448) (0.09795,177528) (0.10089,262715) (0.10425,358474) (0.10835,530560) (0.11891,838715) (0.12912,985902) (0.13314,1.26279e+06) (0.14667,1.35044e+06) (0.16329,1.42268e+06) (0.20457,1.72652e+06) (0.21827,1.74703e+06) (0.2272,3.7594e+06) (0.23985,4.09333e+06) (0.30007,4.3122e+06) (0.34752,4.84966e+06) (0.3615,5.40249e+06) (0.42801,5.78369e+06) (0.50226,5.97015e+06) (0.58441,6.05694e+06) (0.67273,6.43501e+06) };
          \addlegendentry{ShockHash-RS};
        \end{axis}
    \end{tikzpicture}
    \begin{tikzpicture}[baseline=-4.5cm]
        \ref*{paretoLegend}
    \end{tikzpicture}

%% file: fig/queries.tex
\addtolength\tabcolsep{-2pt}
\small
\begin{centering}
\begin{tabular}[t]{l rrr}
    \toprule
    Method & Space & Construction & Query \\
           & bits/key & ns/key & ns/query \\ \midrule

                                     FMPH, $\gamma$=$2.15$ & 3.529 &          51 &  54 \\
                                      FMPH, $\gamma$=$1.0$ & 2.804 &          89 &  69 \\ \midrule
                    FMPHGO, $\gamma$=$2.65, s$=$4, b$=$16$ & 3.547 &          86 &  51 \\
                     FMPHGO, $\gamma$=$1.0, s$=$4, b$=$16$ & 2.212 &         135 &  69 \\ \midrule
                   PTHash, $c$=$7.0$, $\alpha$=$0.99$, C-C & 3.524 &         198 &  20 \\
                    PTHash, $c$=$6.0$, $\alpha$=$0.99$, EF & 2.345 &         247 &  34 \\ \midrule
           SicHash, $\alpha$=$0.9$, $p_1$=$21$, $p_2$=$78$ & 2.412 &         116 &  40 \\
          SicHash, $\alpha$=$0.97$, $p_1$=$45$, $p_2$=$31$ & 2.082 &         169 &  41 \\ \midrule
                              RecSplit, $n$=$8$, $b$=$100$ & 1.792 &         713 &  74 \\
                            RecSplit, $n$=$14$, $b$=$2000$ & 1.585 &    125\,521 &  97 \\ \midrule
                          SIMDRecSplit, $n$=$8$, $b$=$100$ & 1.808 &         117 &  80 \\
                        SIMDRecSplit, $n$=$14$, $b$=$2000$ & 1.585 &     11\,749 & 108 \\
                        SIMDRecSplit, $n$=$16$, $b$=$2000$ & 1.560 &    137\,902 & 100 \\
                        SIMDRecSplit, $n$=$18$, $b$=$2000$ & 1.547 &    271\,524 &  99 \\
                        SIMDRecSplit, $n$=$20$, $b$=$2000$ & 1.535 & 5\,569\,394 &  97 \\ \midrule
                       (GPURecSplit, $n$=$20$, $b$=$2000$) & 1.536 &     55\,988 & 100 \\
                       (GPURecSplit, $n$=$24$, $b$=$2000$) & 1.496 &    508\,768 &  94 \\ \midrule
     \textbf{Bipartite ShockHash-RS}, $n$=$64$, $b$=$2000$ & 1.524 &      5\,724 & 131 \\
    \textbf{Bipartite ShockHash-RS}, $n$=$104$, $b$=$2000$ & 1.496 &     24\,406 & 121 \\
    \textbf{Bipartite ShockHash-RS}, $n$=$128$, $b$=$2000$ & 1.489 &    188\,041 & 113 \\ \midrule
              \textbf{Bipartite ShockHash-Flat}, $n$=$100$ & 1.547 &      9\,620 &  62 \\
              \textbf{Bipartite ShockHash-Flat}, $n$=$128$ & 1.537 &    174\,366 &  58 \\ \midrule
               \textbf{ShockHash-RS}, $n$=$30$, $b$=$2000$ & 1.582 &         797 & 121 \\
               \textbf{ShockHash-RS}, $n$=$39$, $b$=$2000$ & 1.556 &      1\,813 & 123 \\
               \textbf{ShockHash-RS}, $n$=$58$, $b$=$2000$ & 1.523 &    112\,072 & 121 \\

    \bottomrule
\end{tabular}
\end{centering}

%% file: fig/measurementMacros.tex
\def\speedupSIMD156{76}

\def\maxSpeedupNonSimdPlainRotate{195}

\def\maxSpeedupNonSimdPlain{25}